\documentclass{ws-ijmpe}
\usepackage[super,compress]{cite}
\usepackage{amsmath,bm,ascmac,calc}
\usepackage{url}
\begin{document}

\markboth{H. Nakada}{Properties of exotic nuclei}

\catchline{}{}{}{}{}

\newcommand{\be}{\begin{equation}}
\newcommand{\ee}{\end{equation}}
\newcommand{\bea}{\begin{eqnarray}}
\newcommand{\eea}{\end{eqnarray}}
\newcommand{\bra}{\langle}
\newcommand{\ket}{\rangle}
\newcommand{\vect}[1]{\boldsymbol{#1}}

\title{Properties of exotic nuclei
  and their linkage to the nucleonic interaction}

\author{H. Nakada
}

\address{Department of Physics, Graduate School of Science,
 Chiba University\\
 Yayoi-cho 1-33, Inage, Chiba 263-8522, Japan\\
nakada@faculty.chiba-u.jp}



\maketitle

\begin{history}
\received{Day Month Year}
\revised{Day Month Year}
\end{history}


\begin{abstract}
The structure of exotic nuclei sheds new light
on the linkage of the nuclear structure to the nucleonic interaction.
The self-consistent mean-field (SCMF) theories are useful
to investigate this linkage,
which are applicable to many nuclei
covering almost the whole range of the nuclear chart
without artificial truncation of model space.
For this purpose, it is desired to develop
effective interaction for the SCMF calculations
well connected to the bare nucleonic interaction.
Focusing on ground-state properties,
I show results of the SCMF calculations
primarily with the M3Y-type semi-realistic interaction,
M3Y-P6 and M3Y-P6a to be precise,
and discuss in detail how the nucleonic interaction affects
structure of nuclei including those far off the $\beta$-stability.

The central channels of the effective interaction
are examined by the properties of the infinite nuclear matter
up to the spin- and the isospin-dependence.
While experimental information of the infinite matter is obtained
by extrapolating systematic data on finite nuclei in principle,
it is not easy to constrain the spin- and the isospin-dependence
without connection to the bare nucleonic interaction.
The non-central channels play important roles
in the shell structure of the finite nuclei.
The tensor force is demonstrated to affect $Z$- or $N$-dependence
of the shell structure and the magic numbers,
on which the spin-isospin channel in the central force often acts cooperatively.
By using the M3Y-P6 interaction,
the prediction of magic numbers is given in a wide range of the nuclear chart,
which is consistent with almost all the available data.
In relation to the erosion of magic numbers in unstable nuclei,
effects of the tensor force on the nuclear deformation are also argued,
being opposite between nuclei at the $\ell s$- and the $jj$-closed magicities.
Qualitatively consistent with the $3N$-force effect on the $\ell s$-splitting
suggested based on the chiral effective field theory,
the density-dependent LS channel,
which is newly introduced in M3Y-P6a,
reproduces the observed kinks in the differential charge radii
at the $jj$-closed magic numbers
and predicts anti-kinks at the $\ell s$-closed magic numbers.
The pairing correlation has significant effects
on the halos near the neutron drip line.
A new mechanism called `unpaired-particle haloing' is disclosed.
\end{abstract}


\ccode{PACS numbers: 21.30.Fe, 21.10.Pc, 21.10.Ft, 21.10.Gv,
  21.65.Cd, 21.65.Mn}


\section{Introduction}\label{sec:intro}

Atomic nuclei are quantum many-body systems
in which a finite number of nucleons are bound by themselves,
showing a variety of interesting and non-trivial properties.
In particular, exotic phenomena have been disclosed
since the invention of the secondary beams:
for instance, the advent of nuclear halos
near the drip lines~\cite{Riisager_1994,Tanihata_1995},
disappearances of the known magic numbers
and appearances of new magic numbers~\cite{Sorlin-Porquet_2008},
and clusters glued by excess neutrons~\cite{KanadaEnyo-Kimura-Ono_2012}.
They have supplied an opportunity to perceive the structure of nuclear systems
from a more general and profound perspective than before.
As rich physics has been revealed so far and is further expected,
experimental facilities of radioactive nuclear beams have been
developed worldwide and upgraded.
More abundant data will be accumulated in the next few decades.

Nuclei far off the $\beta$ stability accessed by the radioactive beams
supply an indispensable laboratory
for investigating the effects of the nucleonic interaction.
First of all, they have isospin values
substantially distant from those near the $\beta$ stability.
I here give a simple argument.
Denoting the averaged proton-proton, neutron-neutron and proton-neutron
interaction by $v_{pp}$, $v_{nn}$ and $v_{pn}$
and assuming $v_{pp}\approx v_{nn}$ for the sake of simplicity,
we can assess the total interaction energy as
\be\begin{split}
\bra V\ket &\sim \Big[\frac{Z(Z-1)}{2}+\frac{N(N-1)}{2}\Big]\bra v_{nn}\ket
+ ZN\bra v_{pn}\ket \nonumber\\
&= \Big(\frac{A}{2}\Big)^2 \bigg[\big(\bra v_{nn}\ket+\bra v_{pn}\ket\big)
  + \eta_t^2\big(\bra v_{nn}\ket-\bra v_{pn}\ket\big)
  -\frac{2}{A}\bra v_{nn}\ket\bigg]\,,
\end{split}\ee
where $Z$ ($N$) denoted the proton (neutron) number, $A=Z+N$
and $\eta_t=(Z-N)/A$.
Whereas $\eta_t^2\lesssim 0.05$ in the stable nuclei,
$\eta_t^2$ exceeds $0.1$ in the nuclei with $N\approx 2Z$,
illustrating the sizable contribution
of the $\big(\bra v_{nn}\ket-\bra v_{pn}\ket\big)$ term.
Moreover, quantum effects such as the shell structure
are greatly influenced by the nucleonic interaction,
particularly the non-central channels and the three-nucleon ($3N$) interaction,
as will be discussed in later sections.

Toward an unambiguous description of nuclear properties,
it is desirable to construct a quantum-mechanical model
based on the nucleonic interaction.
However, there have been obstacles.
First, the nucleonic interaction has not been established completely,
even at the bare level.
Though it originates from the quantum chromo-dynamics (QCD),
it is a difficult task to derive the nucleonic interaction from the QCD.
Despite the progress based on the chiral effective field theory
($\chi$EFT)~\cite{Epelbaum-HammerMeissner_2009,Machleidt-Entem_2011}
and the progress in the lattice QCD calculations~\cite{Aoki-etal_2012},
there is still a gap between the precision
which is required for the nuclear structure studies
and that can be obtained directly from the QCD.
Second, even if the bare nucleonic interaction is well established,
it does not mean that nuclear structure theory provides reliable results
because of the complication of the quantum many-body correlations.
While a part of this problem is shared with the electronic systems,
nuclei have additional complications due to the strong short-range repulsion
and the non-central channels in the bare interaction.
Br\"{u}ckner's $G$-matrix~\cite{Fetter-Walecka} and its extension
(\textit{e.g.} the hole-line expansion~\cite{Hufner-Mahaux_1972})
have provided a theoretical framework to handle
the influence of the short-range repulsion and the non-central channels.
The $G$-matrix may work as an effective two-nucleon ($2N$) interaction.
It is still questioned
whether these many-body calculations are fully convergent,
although it seems close~\cite{Lu-etal_2017}.
These problems prevent us from fully understanding
basic properties concerning nuclear structure, \textit{e.g.} the saturation,
from microscopic standpoints,
despite significant progress~\cite{Akmal-Pandharipande_1997,Mueller-etal_2000,
  Hebeler-etal_2011,Sammarruca-etal_2012,Coraggio-etal_2014,Kohno_2015,
  Hu-etal_2017,Drischler-Hebeler-Schwenk_2019}.
In atomic systems, which are composed of electrons,
an approach using the energy-density functional (EDF)
has been established to be a standard theoretical framework
in describing their ground-state (g.s.) properties.
The Hohenberg-Kohn (HK) theorem guarantees the existence of an EDF
that gives exact g.s. energy and density~\cite{Hohenberg-Kohn}.
In the atomic case, a singularity originating from quantum effects
is circumvented via the Kohn-Sham (KS) method~\cite{Kohn-Sham},
and the local-density approximation (LDA) has successfully been applied
to the regular part of the EDF~\cite{Sahni-Bohnen-Harbola}.
On the contrary,
it is not evident in nuclei if the KS method works well,
because the non-central channels can be additional sources of the singularities
and the LDA is not established so well as in the atomic cases.

Among practical models of nuclear structure,
those within the self-consistent mean-field (SCMF) theory
combined with effective interactions
are suited to the global description of the nuclear structure
covering almost the whole range of the nuclear chart.~\footnote{
  The effective interactions applied to the SCMF calculations so far
  include terms depending on the nucleon density $\rho(\vect{r})$,
  linked to the saturation.
  This density-dependence may motivate to reinterpret
  the SCMF approaches as a sort of the EDF approaches,
  in which EDF is comprised of the one-body density matrix
  containing non-local currents.
  However, the words SCMF and effective interaction are used in this review,
  in order to stress connection to the nucleonic interaction.}
The SCMF approaches are also advantageous
in the respect that they do not need artificial truncation of model space.
In this review, I shall discuss the properties of exotic nuclei
based on the SCMF approaches,
constraining to g.s. properties
and mainly focusing on their linkage to the nucleonic interaction.
For this purpose, I shall employ
the M3Y-type semi-realistic nucleonic interaction~\cite{
  Nakada_2003,Nakada_2008b},
which is based on the $G$-matrix but is partly modified
from phenomenological standpoints.
Though it cannot be exhaustive,
I expect that this review sheds light on their relationship
and stimulate future studies in this line.

\section{New aspects of nuclear structure disclosed by radioactive beams}
\label{sec:exotica}

In this section, I discuss some of the g.s. properties of nuclei
observed in experiments,
which are characteristic to those far off the $\beta$-stability.

\subsection{Neutron skins}

Because large asymmetry between protons and neutrons gives rise to energy loss,
protons and neutrons tend to distribute according to the ratio of their numbers
at any position, in the first approximation.
This distribution yields $\sqrt{\bra r^2\ket_p}\approx\sqrt{\bra r^2\ket_n}$,
where $\sqrt{\bra r^2\ket_\tau}$ represents root-mean-square (r.m.s.) radius
for the distribution of the point-particle $\tau(=p,n)$.
However, to be more precise,
$\sqrt{\bra r^2\ket_n}$ is somewhat larger than $\sqrt{\bra r^2\ket_p}$
under neutron excess.
This broader distribution of neutrons than protons,
called `neutron skin'~\cite{Tanihata_1995},
becomes conspicuous far off the $\beta$-stability,
as observed in $^{6,8}$He~\cite{Tanihata-etal_1992}.
It is customary to represent the thickness of the neutron skin
by $\sqrt{\bra r^2\ket_n}-\sqrt{\bra r^2\ket_p}$.
Experimentally, $\sqrt{\bra r^2\ket_n}$ has mostly been extracted
via hadronic probes~\cite{Greenlees-Pyle-Tang_1968,
  GarciaRecio-Nieves-Oset_1992,Clark-Kerr-Hama_2002},
while $\sqrt{\bra r^2\ket_p}$ can be obtained from electromagnetic probes
to excellent precision~\cite{Angeli_2004,Angeli-Marinova_2013}.
Experiments via the neutral weak current
have been proposed and are proceeding~\cite{Donnelly-Dubach-Sick_1989,
  Horowitz-etal_2001,PREX_2012,Horowitz-Kumar-Michaels_2014},
to extract $\sqrt{\bra r^2\ket_n}$
without ambiguity originating from the nucleonic interaction.
Since the thickness of neutron skins is governed
by the energy loss due to the proton-neutron asymmetry
around the nuclear surface,
they depend on the density-dependence of the symmetry energy.
Indeed, as far as the SCMF approaches are concerned,
the neutron-skin thickness of individual nuclide
well correlates to the slope parameter of the symmetry energy
$\mathcal{L}_{t0}$~\cite{Centelles-RocaMaza-Vinas-Warda_2009,
  Chen-Ko-Li-Xu_2010,Agrawal-De-Samaddar_2012,Inakura-Nakada_2015},
which will be defined by Eq.~(\ref{eq:L_t0}) in Sec.~\ref{subsec:matter}.
The neutron-skin thickness has attracted great interest,
as $\mathcal{L}_{t0}$ is a crucial parameter to the equation-of-state (EoS)
of the neutron-star matter~\cite{Lattimer-Prakash_2007},

\subsection{Nuclear halos}\label{subsec:halo}

Immediately after the invention of the radioactive beam technology,
an exotic structure of $^{11}$Li was discovered~\cite{Tanihata-etal_1985};
this nucleus has extraordinarily large interaction cross-section ($\sigma_I$).
Since $\sigma_I$ well correlates
to the nuclear matter radius~\cite{Tanihata-etal_1985,Karol_1975},
the anomalous $\sigma_I$ suggests deviation
from the simple rule of the nuclear radius ($\propto A^{1/3}$)
which is linked to the saturation,
\textit{i.e.} stability of matter consisting of nucleons
at a certain density $\rho_0(\approx 0.16\,\mathrm{fm}^{-3})$
(see Sec.~\ref{subsec:matter}).
This anomalous $\sigma_I$ was interpreted and eventually confirmed
as an effect of the loosely bound last two neutrons
distributing in a broad spatial region~\cite{Tanihata_1995},
called `neutron halo'.
The presence of such broad density distribution was also established
in $^{11}$Be~\cite{Fukuda-etal_1991},
to which the last one neutron contribute.
Since then one- and two-neutron halos have been observed
in a number of nuclei in $A\lesssim 40$~\cite{
  Tanihata-Savajols-Kanungo_2013,Takechi-etal_2012,Takechi-etal_2014}.
Neutron halos have been predicted in some heavier nuclei as well~\cite{
  Hamamoto_2017},
and possibility of halos comprised of several neutrons,
called `giant halo', has been argued in \textit{e.g.} Zr~\cite{Meng-Ring_1998}.
The neutron halos may be regarded as a particular form of the neutron skins.

In the vicinity of the neutron drip line,
the last one or two neutrons are bound with very small separation energy.
In the case of the one-neutron halo in a nucleus with mass number $A$,
the last neutron is bound in the field produced by the other $(A-1)$ nucleons.
Whereas the $A$-body wave function (w.f.) $|\Psi_A\ket$ depends
on interactions among all the $A$ nucleons,
its asymptotic form in the distance should be dominated
by the two-body channel comprised of $|\Psi_{A-1}\ket$,
the nucleus with mass number $(A-1)$, and the last neutron,
$|\Psi_A\ket\sim|\Psi_{A-1}\ket\otimes|\varphi_A\ket$.
Here $|\varphi_A\ket$ is the w.f. of the last neutron
as a function of the coordinate relative to $|\Psi_{A-1}\ket$,
$\vect{r}'_A=\vect{r}_A-[1/(A-1)]\sum_{i=1}^{A-1}\vect{r}_i$,
apart from the spin variable.
Because of the short-range nature of the nuclear force,
the asymptotic Schr\"{o}dinger equation for $|\varphi_A\ket$ is given as
\be -\frac{1}{2M'_A}\frac{\partial^2}{\partial r_A^{\prime 2}}
[r'_A\,\varphi_A(\vect{r}'_A)]
= \epsilon [r'_A\,\varphi_A(\vect{r}'_A)]\,, \label{eq:asymp-rel} \ee
where $M'_A=[(A-1)/A]M$ is the reduced mass
between the nucleus with mass number $(A-1)$ and the last neutron
with $M$ standing for the nucleon mass, $-\epsilon$ is the separation energy,
and $r=|\vect{r}|$.
This equation yields the asymptotic form of the w.f. as
\be
\varphi_A(\vect{r}'_A) \approx \xi'\,
\frac{\exp(-\sqrt{2M'_A|\epsilon|}\,r'_A)}{r'_A}\mathcal{Y}(\hat{\vect{r}}'_A)
  = \xi \,\frac{\exp(-\sqrt{2M|\epsilon'|}|\,\vect{r}_A-\vect{R}|)}
  {|\vect{r}_A-\vect{R}|}\mathcal{Y}(\widehat{\vect{r}_A-\vect{R}})\,,
 \label{eq:asymp-phi} \ee
where $\vect{R}=(1/A)\sum_{i=1}^A \vect{r}_i$ denotes
the center-of-mass (c.m.) coordinate of the whole nucleus,
$\epsilon'=[A/(A-1)]\epsilon$,
$\mathcal{Y}(\hat{\vect{r}})$ represents the spin-angular part
with $\hat{\vect{r}}=\vect{r}/r$.
It should be noted that the amplitude $\xi$ (or $\xi'$)
depends on the structure inside.
Because of the centrifugal barrier,
this amplitude is suppressed when the neutron has
a high $\ell$ (orbital angular momentum).
If the w.f. of the last neutron is dominated by that of the asymptotic region,
the r.m.s. radius of this state diverges
according to $|\epsilon|^{-1}$ for the $s$-wave
and to $|\epsilon|^{-1/2}$ for the $p$-wave,
as $\epsilon\to 0$.
As it converges for the higher partial waves,
this argument illustrates the importance of the $s$- and $p$-waves
in the halos~\cite{Riisager-Jensen-Moller_1992}.

On the other hand, the Coulombic effect cannot be ignored
in the asymptotics for a proton.
The Coulomb barrier reduces the amplitude of the asymptotic function
in the distance.
Moreover, since the Coulomb force has long-range,
its effects should be incorporated into the asymptotic form
in terms of the Coulomb w.f.~\cite{HandbookOfMathematicalFunctions}.
Therefore proton halos should be hindered
even in the vicinity of the proton drip line,
though possibilities remain in light-mass regions~\cite{Warner-etal_1995,
  Ozawa-etal_1994,Cai-etal_2002}.
For a two-neutron halo,
a correlation between the last two neutrons could influence the asymptotics.

\subsection{Appearance and disappearance of magic numbers}

The shell structure, which is manifested by the magic numbers,
is fundamental to nuclear structure physics.
Although the well-known magic numbers, $Z,N=2,8,20,28,50,82$ and $N=126$,
had once seemed rigorous,
experiments using the radioactive beams revealed
that nuclear shell structure varies depending on $Z$ and $N$,
eventually giving rise to appearance and disappearance of magic numbers.
It is difficult to exaggerate the importance of these phenomena.

The earliest evidence was found in the disappearance
of the $N=8$ magic number at $^{11}$Be,
whose g.s. has $J^\pi=(1/2)^+$ rather than $(1/2)^-$~\cite{
  Wilkinson-Alburger_1959}.
This $J^\pi$ value is hard to be accounted for
without including an orbit in the $1s0d$-shell~\cite{Millener-Kurath},
indicating a breakdown of the $N=8$ magic nature (`magicity').
However, it had not been quick
that magic numbers were recognized to appear and disappear in other regions
as well.

Magic numbers are experimentally identified
via irregularities of binding energies,
kinks in the separation energies,
and high excitation energies~\cite{Heyde_2004}.
Measured binding energies and excitation energies
indicated loss of the $N=20$ magicity at $^{31}$Na and $^{32}$Mg~\cite{
  Thibault-etal_1975,GuillemaudMueller-etal_1984},
and the region involving these nuclei is later called
`island of inversion'~\cite{Warburton-Becker-Brown_1990}.
The collapse of the $N=28$ magicity has also been established
at $^{42}$Si and $^{40}$Mg~\cite{Bastin-etal_2007,Takeuchi-etal_2012,
  Crawford-etal_2019}.
On the contrary, it has been pointed out
that $N=16$ behaves like a magic number around $^{24}$O
in the neutron separation energy~\cite{Ozawa-etal_2000}
and the excitation energy~\cite{Kanungo-Tanihata-Ozawa_2002}.
In the neutron-rich region around Ca,
$N=32$ and $34$ have been suggested to be the new magic numbers~\cite{
  Kanungo-Tanihata-Ozawa_2002,Steppenbeck-etal_2013}.
These magicities remind us of several submagic numbers
near the $\beta$-stability line~\cite{deShalit-Feshbach,Kleinheinz_1979},
\textit{e.g.} $Z=40$ around $^{90}$Zr and $Z=64$ at $^{146}$Gd.
As neutron magicity is responsible for the waiting points
of the synthesizing process of heavy elements~\cite{Iliadis_2007},
it is significant to comprehend and predict magic numbers correctly
for elucidating the origin of matters as well as the nuclear structure itself.
Proton magic numbers in extreme neutron excess
are essential also in understanding the structure
of the neutron-star crust~\cite{Negele-Vautherin_1973,Onsi-etal_2008}.

The appearance and the disappearance of magic numbers
are ascribed to variation of the shell structure,
sometimes called `shell evolution'~\cite{Otsuka-etal_2005},
as departing from the $\beta$-stability line.
For the shell evolution,
roles of the centrifugal barrier that gives rise to $\ell$-dependence
in the single-particle (s.p.) energies~\cite{Ozawa-etal_2000}
and of specific channels of effective interactions~\cite{Otsuka-etal_2005,
  Otsuka-etal_2001}
have been argued.
The nuclear shell structure emerges under the nuclear mean field (MF),
which is produced by the interaction among constituent nucleons.
Constructing s.p. orbitals self-consistently
without artificial truncation of model space,
the SCMF theory supplies the desired framework to study shell evolution
including appearance and disappearance of magic numbers,
if an appropriate effective interaction is applied.

\section{Self-consistent mean-field theory
  and effective interactions}\label{sec:SCMF}

The theoretical framework,
\textit{i.e.} the self-consistent mean-field (SCMF) theory,
is briefly reviewed in this section.
Effective interactions,
which are in principle the only input to the SCMF theory,
are argued in some detail.
For the broad applicability of the SCMF theory and its extension
to nuclear structure problems,
see Ref.~\refcite{Bender-Heenen-Reinhard_2003} for instance.

\subsection{Variational aspects}\label{subsec:variation}

The SCMF theory relies on the variational principle.
The variational principle for a single Slater determinant
derives the Hartree-Fock (HF) theory~\cite{Ring-Schuck}.
For a single Slater determinant $|\Phi\ket$,
expectation values of any operators represented in the second quantized form
are decomposed into the product of the one-body density matrix
$\varrho_{\mu\mu'}=\bra\Phi|a_{\mu'}^\dagger a_\mu|\Phi\ket$,
where $\mu$ and $\mu'$ are indices of the s.p. bases,
owing to Wick's theorem~\cite{Ring-Schuck,Fetter-Walecka}.
By applying this consequence to the Hamiltonian $H$,
the total energy expectation value $E=\bra\Phi|H|\Phi\ket$ is expressed
as a functional of $\varrho_{\mu\mu'}$.
Therefore the HF theory can be formulated
in terms of the variation of the energy with respect to $\varrho_{\mu\mu'}$.
Extension of $|\Phi\ket$ so as to include the pair condensate
yields the Hartree-Fock-Bogolyubov (HFB) theory,
in which the pairing tensor $\kappa_{\mu\mu'}=\bra\Phi|a_{\mu'} a_\mu|\Phi\ket$
and $\kappa_{\mu\mu'}^\ast$ enter in addition to $\varrho_{\mu\mu'}$.
Throughout this review, $|\Phi\ket$ denotes the MF state.

The saturation is a fundamental property in nuclei.
In applying the SCMF theory to nuclei,
it is necessary to adopt an effective interaction,
which masks distortion of the w.f.
due to the high-momentum components,
instead of the bare nucleonic interaction.
It seems unavoidable that the effective interaction depends
on the local density $\rho(\vect{r})$
as derived from the $G$-matrix to acquire the saturation,
although a many-body pseudo-potential may do a similar job~\cite{
  Sadoudi-etal_2013,Lacroix-Bennaceur_2015}.
Since $\varrho_{\mu\mu'}$ determines $\rho(\vect{r})$,
it is straightforward to incorporate the $\rho$-dependent effective interaction
into the SCMF framework.
It should be noted, however, that $\rho$-dependent interaction
is not allowed in the full quantum-mechanical respect,
as they are not represented in the second-quantized form.
As a result, it gives rise to a serious problem
when applied to the calculations beyond MF
including the quantum-number projections~\cite{
  Dobaczewski-Stoitsov-Nazarewicz-Reinhard_2007,Duguet-etal_2009,Robledo_2010}.

\subsection{History of effective interactions for SCMF calculations:
  biased overview}

The SCMF calculations have been progressed with the development of computers.
It was Vautherin and Brink~\cite{Vautherin-Brink}
who first implemented a SCMF calculation in nuclei
without artificial truncation of model space.
They assumed the zero-range form for effective interactions
as proposed by Skyrme~\cite{Skyrme_1959},
by which the resultant EDF is represented by only local variables
(`quasi-local' currents),
facilitating computation.
The quasi-local currents involve low-order derivatives of the s.p. w.f.'s,
in addition to the local density.
In the original Skyrme interaction,
the $3N$ contact term was used to realize the saturation,
which is convertible at the HF level to the $2N$ interaction
with a coupling coefficient proportional to $\rho(\vect{r})$.
Since the $3N$ contact interaction gives rise to the instability
for the spin excitation~\cite{Chang_1975},
the corresponding term (sometimes called $t_3$ term)
is usually handled as the $\rho$-dependent $2N$ interaction,
in which fractional power of $\rho$ is extensively allowed~\cite{Kohler_1976}.
The zero-range attraction suffers the divergence problem
when applied to the pairing field.
While a cut-off scheme was introduced in later studies~\cite{
  Esbensen-Bertsch-Hencken_1997,Bulgac-Yu_2002,Goriely-etal_2002},
Gogny and his collaborators developed a finite-range effective interaction
to avoid this problem
and applied it to the HFB calculations~\cite{Gogny_D1}.

The nucleonic interaction has various ranges.
Apart from the Coulomb force among protons,
the one-pion exchange potential (OPEP) provides the longest-range component,
since the pions are the lightest meson mediating the nucleonic interaction,
linked to the chiral symmetry breaking~\cite{Hosaka-Toki_2001}.
The range of the OPEP is $\sim 1.4\,\mathrm{fm}$,
not sufficiently short to justify the momentum expansion of the interaction
that immediately derives quasi-local currents,
as the nucleons distribute with momentum up to $\sim 1.4\,\mathrm{fm}^{-1}$.
Negele and Vautherin proposed a density-matrix expansion (DME)
and applied it to a finite-range interaction derived from the $G$-matrix~\cite{
  Negele-Vautherin_1972}.
The DME seems to validate describing a nuclear structure
with the quasi-local currents,
more extensively than the simple momentum expansion.
Campi and Sprung also developed an effective interaction
derived from the $G$-matrix via the LDA~\cite{Campi-Sprung_1972}.
However, it was not easy to attain high accuracy
with effective interactions derived in a purely theoretical manner.
A part of the reason should be attributed to the limitation of the accuracy
in the bare nucleonic interaction available at that time.
It had become popular to obtain effective interactions for the SCMF approaches
by assuming a certain functional form
(\textit{e.g.} the EDF form based on the Skyrme interaction)
and fitting the parameters to nuclear structure data.
A recent example is found in the UNEDF project~\cite{UNEDF1,UNEDF2},
in which parameters of the Skyrme EDF were searched
by using a great number of experimental data known to date.
Still, it was concluded,
``the standard Skyrme energy density has reached its limits,
and significant changes to the form of the functional are needed.'' 
For further recent progress,
see Ref.~\refcite{Grasso_2019} and references therein.

Another approach came from applying the SCMF to the relativistic Lagrangian
composed of nucleons and mesons~\cite{Serot-Walecka_1986},
which is called relativistic mean-field (RMF) theory.
While the exchange terms due to the Pauli principle are ignored
in the usual RMF approaches,
relativistic Hartree-Fock calculations were implemented
relatively recently~\cite{Ban-etal_2006,Long-etal_2009}.
Despite advantages and progress,
the anti-nucleon degrees of freedom (d.o.f.)
bring complication in the RMF approaches and their extensions.
In the following, the arguments are restricted to the SCMF approaches
with non-relativistic effective Hamiltonians,
besides quoting RMF results from literature in a few cases.
Though relativistic effects should become sizable as the density grows,
a part of them may be incorporated in the effective interactions.

The Michigan-three-range-Yukawa (M3Y) interaction was explored,
based on the $G$-matrix around the nuclear surface
($\rho\approx \rho_0/3$)~\cite{M3Y}.
Originally intended to apply to nuclear reactions,
the M3Y interaction provides matrix elements among the valence orbitals
similar to those determined empirically~\cite{USD}.
However, while having reliability for properties
dominated around the nuclear surface,
it is not suitable for describing phenomena associated with density variation,
\textit{e.g.} the saturation,
since the M3Y interaction was obtained at the fixed density.
Although overall coefficients depending on density
were introduced for reaction problems~\cite{Farid-Satchler_1985,
  Khoa-Oertzen-Ogloblin_1996},
such density-dependence is questionable in microscopic respect;
while in reality the short-range repulsion becomes dominant at high density,
repulsion would be produced via the long-range channels
by the overall coefficients with the inverted sign.
Another cure was adding density-dependent contact terms,
instead of the overall coefficients,
as proposed in Ref.~\refcite{Nakada_2003}.
Along this line, several parameter-sets of the M3Y-type interaction
have been developed for SCMF studies~\cite{Nakada_2008b,Nakada_2008b_erratum,
  Nakada_2010,Nakada_2010_erratum,Nakada_2013},
in which some of the parameters have been adjusted
to nuclear structure data
while others kept unchanged from the $G$-matrix result.
For this reason, they are called `semi-realistic' interactions.

The bare nucleonic interaction contains the tensor force.
Theoretically, the tensor force naturally appears in the meson exchange picture.
The existence of the tensor force is experimentally proven
by the finite quadrupole moment of the deuteron.
However, the tensor force has not been included
in the conventional SCMF calculations,
both with the Skyrme and the Gogny interactions.
Although there were a few attempts to include tensor force
in the SCMF calculations~\cite{Stancu-Brink-Flocard_1977,Co-Lallena_1996}
and a certain effect of the tensor force was found,
its significance in the nuclear structure had not been recognized widely.
As the $Z$- and $N$-dependence of the shell structure has been confirmed,
it was pointed out~\cite{Otsuka-etal_2005}
that the tensor force plays an important role in it,
and its effects have become a hot topic~\cite{Sagawa-Colo_2014}.
Tensor-force effects will be discussed also in later sections of this review.

\subsection{Effective Hamiltonian containing semi-realistic interaction}

Throughout this paper, the following Hamiltonian is employed,
\be\begin{split} H =& K + V_N + V_C - H_\mathrm{c.m.}\,;\\
& K = \sum_i \frac{\vect{p}_i^2}{2M}\,,\quad
V_N = \sum_{i<j} v_{ij}\,,\quad
V_C = \alpha_\mathrm{em} \sum_{i<j(\in p)} \frac{1}{r_{ij}}\,,\\
& H_\mathrm{c.m.} = \frac{\vect{P}^2}{2AM}
= \frac{1}{A}\bigg[\sum_i \frac{\vect{p}_i^2}{2M}
  + \sum_{i<j} \frac{\vect{p}_i\cdot\vect{p}_j}{M}\bigg]\quad
\Big(\vect{P}=\sum_i \vect{p}_i\Big)\,,
\end{split}\label{eq:Hamil}\ee
where $i,j$ are the indices of individual nucleons,
and $\vect{r}_{ij}= \vect{r}_i - \vect{r}_j$.
I set $M=(M_p+M_n)/2$, where $M_p$ ($M_n$) is the mass
of a proton (a neutron)~\cite{PDG_2006}.
The fine structure constant is denoted by $\alpha_\mathrm{em}$.
The effective $2N$ interaction $v_{ij}$
is comprised of the following terms,
holding the rotational and the isospin symmetry,
together with the translational symmetry
except in the $\rho$-dependent coupling coefficients,
\be\begin{split} v_{ij} =& v_{ij}^{(\mathrm{C})}
 + v_{ij}^{(\mathrm{LS})} + v_{ij}^{(\mathrm{TN})}
 + v_{ij}^{(\mathrm{C}\rho)} + v_{ij}^{(\mathrm{LS}\rho)}\,;\\
& v_{ij}^{(\mathrm{C})} = \sum_n \big\{t_n^{(\mathrm{SE})} P_\mathrm{SE}
+ t_n^{(\mathrm{TE})} P_\mathrm{TE} + t_n^{(\mathrm{SO})} P_\mathrm{SO}
+ t_n^{(\mathrm{TO})} P_\mathrm{TO}\big\}
 f_n^{(\mathrm{C})} (r_{ij})\,,\\
& v_{ij}^{(\mathrm{LS})} = \sum_n \big\{t_n^{(\mathrm{LSE})} P_\mathrm{TE}
 + t_n^{(\mathrm{LSO})} P_\mathrm{TO}\big\}
 f_n^{(\mathrm{LS})} (r_{ij})\,\vect{L}_{ij}\cdot
(\vect{s}_i+\vect{s}_j)\,,\\
& v_{ij}^{(\mathrm{TN})} = \sum_n \big\{t_n^{(\mathrm{TNE})} P_\mathrm{TE}
 + t_n^{(\mathrm{TNO})} P_\mathrm{TO}\big\}
 f_n^{(\mathrm{TN})} (r_{ij})\, r_{ij}^2 S_{ij}\,,\\
& v_{ij}^{(\mathrm{C}\rho)} = \Big\{C_\mathrm{SE}[\rho(\vect{R}_{ij})]\,P_\mathrm{SE}
 + C_\mathrm{TE}[\rho(\vect{R}_{ij})]\,P_\mathrm{TE}\Big\}\,\delta(\vect{r}_{ij})\,,
 \\
& v_{ij}^{(\mathrm{LS}\rho)} = 2i\,D[\rho(\vect{R}_{ij})]\,
 \vect{p}_{ij}\times\delta(\vect{r}_{ij})\,\vect{p}_{ij}\cdot
 (\vect{s}_i+\vect{s}_j) \\
&\qquad~ = D[\rho(\vect{R}_{ij})]\,\{-\nabla_{ij}^2\delta(\vect{r}_{ij})\}\,
 \vect{L}_{ij}\cdot(\vect{s}_i+\vect{s}_j)\,.
\end{split}\label{eq:effint}\ee
Here $\vect{s}_i$ is the spin operator,
$\vect{R}_{ij}=(\vect{r}_i+\vect{r}_j)/2$,
$\vect{p}_{ij}= (\vect{p}_i - \vect{p}_j)/2$,
$\vect{L}_{ij}= \vect{r}_{ij}\times \vect{p}_{ij}$,
$S_{ij}= 4\,[3(\vect{s}_i\cdot\hat{\vect{r}}_{ij})
(\vect{s}_j\cdot\hat{\vect{r}}_{ij})
- \vect{s}_i\cdot\vect{s}_j ]$,
and $\rho(\vect{r})$ is the isoscalar nucleon density.
$P_\mathrm{Y}$ ($\mathrm{Y}=\mathrm{SE},\mathrm{TE},\mathrm{SO},\mathrm{TO}$)
are the projection operators on the singlet-even (SE), triplet-even (TE),
singlet-odd (SO) and triplet-odd (TO) $2N$ states,
which are expressed by the the spin- and isospin-exchange operators
$P_\sigma$ and $P_\tau$ as
\be\begin{split} P_\mathrm{SE} = \frac{1-P_\sigma}{2}\,\frac{1+P_\tau}{2}\,,
\quad& P_\mathrm{TE} = \frac{1+P_\sigma}{2}\,\frac{1-P_\tau}{2}\,,\\
P_\mathrm{SO} = \frac{1-P_\sigma}{2}\,\frac{1-P_\tau}{2}\,.
\quad& P_\mathrm{TO} = \frac{1+P_\sigma}{2}\,\frac{1+P_\tau}{2}\,.
\end{split}\label{eq:proj_T}\ee
The subscript $n$ in (\ref{eq:effint}) corresponds to the range parameter,
whose inverse is denoted by $\mu_n^{(\mathrm{X})}$
($\mathrm{X}=\mathrm{C},\mathrm{LS},\mathrm{TN}$),
and the associated coupling constants by $t_n^{(\mathrm{Y})}$.
The Skyrme interaction is obtained
by setting $f_1^{(\mathrm{C})}(r)=\delta(\vect{r})$,
$f_2^{(\mathrm{C})}(r)=f^{(\mathrm{LS})}(r)=f^{(\mathrm{TN})}(r)
=\nabla^2\delta(\vect{r})$,
and the Gogny interaction by $f_n^{(\mathrm{C})}(r)=e^{-(\mu_n^{(\mathrm{C})} r)^2}$,
$f^{(\mathrm{LS})}(r)=\nabla^2\delta(\vect{r})$.
The Skyrme EDF was extended to the form
not representable by the interaction~\cite{Reinhard-Flocard_1995}.
The Gogny interaction was extended so as to contain $v^{(\mathrm{TN})}$
with $f^{(\mathrm{TN})}(r)=e^{-(\mu^{(\mathrm{TN})} r)^2}/r^2$~\cite{Otsuka-Matsuo-Abe_2006,
  Anguiano-etal_2012}.
In the M3Y interaction,
the Yukawa function $f_n^{(\mathrm{X})}(r)=e^{-x}/x$ with $x=\mu_n^{(\mathrm{X})} r$
is used for all of $\mathrm{X}=\mathrm{C},\mathrm{LS},\mathrm{TN}$.
The $\rho$-dependent channel $v_{ij}^{(\mathrm{C}\rho)}$
is essential to realize the saturation.
Physically, $v^{(\mathrm{C}\rho)}$ may carry
effects of the $3N$ interaction and of the $\rho$-dependence
that originates from many-body effects.
For the functional $C_\mathrm{Y}[\rho(\vect{r})]$
($\mathrm{Y}=\mathrm{SE},\mathrm{TE}$),
it is customary to assume
\be C_\mathrm{Y}[\rho]=t_\rho^{(\mathrm{Y})}\,\rho^{\alpha^{(\mathrm{Y})}}\,,
\label{eq:CinC}\ee
with the parameters $t_\rho^{(\mathrm{Y})}$ and $\alpha^{(\mathrm{Y})}$.
By taking $\alpha^{(\mathrm{TE})}=1/3$, a reasonable value
of the nuclear incompressibility is obtained (see Subsec.~\ref{subsec:matter}).
It is interesting that $\alpha^{(\mathrm{Y})}=1/3$ is compatible
with the low-density limit of the energy of the infinite Fermi gas~\cite{
  Lee-Yang_1957}.
The channel $v_{ij}^{(\mathrm{LS}\rho)}$ has recently been introduced~\cite{
  Nakada-Inakura_2015}
and will be discussed in Sec.~\ref{subsec:r_c}.
For the functional $D[\rho(\vect{r})]$,
the following form is employed,~\footnote{
  The sign of $D[\rho]$ was wrong
  in Refs.~\protect\refcite{Nakada-Inakura_2015} and \refcite{Nakada_2015}.
See Ref.~\refcite{Nakada_2019}.}
\be D[\rho(\vect{r})] = w_1\,\frac{\rho(\vect{r})}
 {1+d_1\rho(\vect{r})}\,, \label{eq:DinLS}
 \ee
where $w_1$ and $d_1$ are constants.

For $\rho$-dependent channels,
it is a question at which point $\rho$ in the coupling coefficient
should be evaluated.
This ambiguity is avoided in (\ref{eq:effint})
since all the $\rho$-dependent channels contain $\delta(\vect{r}_{ij})$,
by which $\vect{R}_{ij}$ in $C_\mathrm{Y}[\rho]$ or $D[\rho]$
is replaced by $\vect{r}_i$ or $\vect{r}_j$ without any difference.
The Hamiltonian is translationally invariant
if all the variables depend only on the relative coordinates.
However, the $\rho$-dependent coefficient $C_\mathrm{Y}[\rho]$ and $D[\rho]$
can break the translational invariance
because of the dependence on $\vect{R}_{ij}$.
Although it is reasonable to consider
as $\rho$ in $C_\mathrm{Y}[\rho]$ and $D[\rho]$ should depend on
$\vect{R}_{ij}-\vect{R}$ ($\vect{R}$ is the c.m. coordinate),
it is practical to approximate it as
\be C_\mathrm{Y}[\rho(\vect{R}_{ij}-\vect{R})] \approx
C_\mathrm{Y}[\rho(\vect{R}_{ij}-\bra\vect{R}\ket)]
= C_\mathrm{Y}[\rho(\vect{R}_{ij})]\,,
\ee
and likewise for $D[\rho]$.

A class of the M3Y-P$n$ parameter-sets ($n$ represents an integer)
of the semi-realistic interaction have been developed~\cite{
  Nakada_2003,Nakada_2008b,Nakada_2008b_erratum,Nakada_2010,Nakada_2010_erratum,
  Nakada_2013}
by modifying the M3Y-Paris interaction~\cite{M3Y-Paris},
the M3Y interaction derived from the Paris $2N$ potential~\cite{
  Paris-potential}.
The range parameters $\mu_n^{(\mathrm{X})}$ of the M3Y-Paris interaction
are maintained,
and the longest-range part in $v^{(\mathrm{C})}$ (\textit{i.e.} the $n=3$ term)
is kept identical to the central channels of the OPEP,
$v^{(\mathrm{C})}_\mathrm{OPEP}$.
In this review, I shall mainly present SCMF results
using the M3Y-P6 parameter-set and its variant M3Y-P6a~\cite{
  Nakada_2013,Nakada-Inakura_2015}.
In M3Y-P6,
the TE channel of $v^{(\mathrm{C}\rho)}$ is responsible for the saturation,
consistent with the indication
that the second-order effect of the bare tensor force is the dominant source
of the saturation~\cite{Bethe_1971}.
The SE channel of $v^{(\mathrm{C}\rho)}$,
which is relevant to the structure of the neutron stars,
is fitted to the microscopically calculated EoS of the pure neutron matter
in Ref.~\refcite{Friedman-Pandharipande_1981}.
As $v^{(\mathrm{C}\rho)}$ is added,
the coupling constants in the $n=1,2$ terms of $v^{(\mathrm{C})}$
have been modified from those of Ref.~\refcite{M3Y-Paris}
so as to reproduce the binding energies
of several doubly magic nuclei~\cite{AtomicMass_2003}
and the matter radii of $^{208}$Pb~\cite{Alkhazov-Belostotsky-Vorobyov_1978}
(see Sec.~\ref{subsec:doubly&pairing}).
For the tensor force, $v^{(\mathrm{TN})}$ of Ref.~\refcite{M3Y-Paris}
is maintained without any modification,
and in this respect the tensor force is regarded as realistic,
directly connected to the $G$-matrix.
Owing to this connection to the microscopic theory,
the tensor-force effects can be investigated with little ambiguity.
In M3Y-P6, $v^{(\mathrm{LS}\rho)}$ is not used
and $v^{(\mathrm{LS})}$ in Ref.~\refcite{M3Y-Paris} is enhanced
by an overall factor $2.2$,
to reproduce the observed level sequence around $^{208}$Pb~\cite{
  TableOfIsotopes}.
In M3Y-P6a,
$v^{(\mathrm{LS}\rho)}$ is introduced instead of enhancing $v^{(\mathrm{LS})}$,
with all the other parameters are kept identical to M3Y-P6.
The parameters in M3Y-P6 and P6a are tabulated
in Table~\ref{tab:param_M3Y},
together with the original M3Y-Paris interaction.

\begin{table}
  \caption{Parameters of M3Y-type interactions.
    See Eqs.~(\protect\ref{eq:effint},\protect\ref{eq:CinC},\protect\ref{eq:DinLS})
    for the definition.
  \label{tab:param_M3Y}}
\centerline
{\begin{tabular}{ccr@{.}lr@{.}lr@{.}lr@{.}l}
\toprule 
\multicolumn{2}{c}{parameters} & \multicolumn{2}{c}{M3Y-Paris~} &
 \multicolumn{2}{c}{~~M3Y-P6~~} & \multicolumn{2}{c}{~~M3Y-P6a~} \\
\colrule 
$1/\mu_1^{(\mathrm{C})}$ &(fm)& $0$&$25$ & $0$&$25$ & $0$&$25$ \\
$t_1^{(\mathrm{SE})}$ &(MeV)& $11466$& & $10766$& & $10766$& \\
$t_1^{(\mathrm{TE})}$ &(MeV)& $13967$& & $8474$& & $8474$& \\
$t_1^{(\mathrm{SO})}$ &(MeV)& $-1418$& & $-728$& & $-728$& \\
$t_1^{(\mathrm{TO})}$ &(MeV)& $11345$& & $12453$& & $12453$& \\
$1/\mu_2^{(\mathrm{C})}$ &(fm)& $0$&$40$ & $0$&$40$ & $0$&$40$ \\
$t_2^{(\mathrm{SE})}$ &(MeV)& $-3556$& & $-3520$& & $-3520$& \\
$t_2^{(\mathrm{TE})}$ &(MeV)& $-4594$& & $-4594$& & $-4594$& \\
$t_2^{(\mathrm{SO})}$ &(MeV)& $950$& & $1386$& & $1386$& \\
$t_2^{(\mathrm{TO})}$ &(MeV)& $-1900$& & $-1588$& & $-1588$& \\
$1/\mu_3^{(\mathrm{C})}$ &(fm)& $1$&$414$ & $1$&$414$ & $1$&$414$ \\
$t_3^{(\mathrm{SE})}$ &(MeV)& $-10$&$463$ & $-10$&$463$ & $-10$&$463$ \\
$t_3^{(\mathrm{TE})}$ &(MeV)& $-10$&$463$ & $-10$&$463$ & $-10$&$463$ \\
$t_3^{(\mathrm{SO})}$ &(MeV)& $31$&$389$ & $31$&$389$ & $31$&$389$ \\
$t_3^{(\mathrm{TO})}$ &(MeV)& $3$&$488$ & $3$&$488$ & $3$&$488$ \\
$1/\mu_1^{(\mathrm{LS})}$ &(fm)& $0$&$25$ & $0$&$25$ & $0$&$25$ \\
$t_1^{(\mathrm{LSE})}$ &(MeV)& $-5101$& & $-11222$&$2$ & $-5101$& \\
$t_1^{(\mathrm{LSO})}$ &(MeV)& $-1897$& & $-4173$&$4$ & $-1897$& \\
$1/\mu_2^{(\mathrm{LS})}$ &(fm)& $0$&$40$ & $0$&$40$ & $0$&$40$ \\
$t_2^{(\mathrm{LSE})}$ &(MeV)& $-337$& & $-741$&$4$ & $-337$& \\
$t_2^{(\mathrm{LSO})}$ &(MeV)& $-632$& & $-1390$&$4$ & $-632$& \\
$1/\mu_1^{(\mathrm{TN})}$ &(fm)& $0$&$40$ & $0$&$40$ & $0$&$40$ \\
$t_1^{(\mathrm{TNE})}$ &(MeV$\cdot$fm$^{-2}$)& $-1096$& & $-1096$& & $-1096$& \\
$t_1^{(\mathrm{TNO})}$ &(MeV$\cdot$fm$^{-2}$)& $244$& & $244$& & $244$& \\
$1/\mu_2^{(\mathrm{TN})}$ &(fm)& $0$&$70$ & $0$&$70$ & $0$&$70$ \\
$t_2^{(\mathrm{TNE})}$ &(MeV$\cdot$fm$^{-2}$)& $-30$&$9$ & $-30$&$9$ & $-30$&$9$ \\
$t_2^{(\mathrm{TNO})}$ &(MeV$\cdot$fm$^{-2}$)& $15$&$6$ & $15$&$6$ & $15$&$6$ \\
$\alpha^{(\mathrm{SE})}$ && \multicolumn{2}{c}{---} &
 \multicolumn{2}{c}{$1$} & \multicolumn{2}{c}{$1$} \\
$t_\rho^{(\mathrm{SE})}$ &(MeV$\cdot$fm$^6$)& $0$& & $384$& & $384$& \\
$\alpha^{(\mathrm{TE})}$ && \multicolumn{2}{c}{---} &
 \multicolumn{2}{c}{$1/3$} & \multicolumn{2}{c}{$1/3$} \\
$t_\rho^{(\mathrm{TE})}$ &(MeV$\cdot$fm$^4$)& $0$& & $1930$& & $1930$& \\
$w_1$ &(MeV$\cdot$fm$^8$)& $0$& & $0$& & $742$& \\
$d_1$ &(fm$^3$)& \multicolumn{2}{c}{---} & \multicolumn{2}{c}{---} & $1$&$0$ \\
\botrule 
\end{tabular}}
\end{table}

For comparison,
the SLy5 parameter-set~\cite{Skyrme-Lyon} of the Skyrme interaction
and the D1S, D1M parameter-sets~\cite{Gogny_D1S,Gogny_D1M}
of the Gogny interaction are employed as well.
There are plenty of parameter-sets of the Skyrme interactions in the market,
and the SLy5 parameter-set is no more than a single example,
not necessarily a typical one.
It is worth noting, however, that SLy5 was adjusted
to an \textit{ab initio} result of the neutron-matter EoS,
as well as to some nuclear structure data.
The D1S parameter-set has been among those most widely applied
to nuclear structure calculations.
The D1M parameter-set is relatively new,
which takes care of nuclear masses and neutron-matter EoS.

$V_C$ and $H_\mathrm{c.m.}$ are treated without additional approximation
up to the exchange and the pairing terms,
unlike many of the SCMF calculations.
Effects of $V_C$ on the pairing are not negligible~\cite{
  Anguiano-Egido-Robledo_2001,Lesinski-Duguet-Bennaceur-Meyer_2009,
  Nakada-Yamagami_2011}.

In the HFB calculations with the Gogny or the M3Y-type interaction,
the same effective interaction is applied to the pairing channels
as to the particle-hole ($ph$) channels.
The $\rho$-dependent coupling coefficients
$C_\mathrm{Y}[\rho]$ and $D[\rho]$
are assumed not to depend on the pairing tensor.
In the spirit of the EDF approaches,
there is no need for the $ph$ and the pairing channels to hold this consistency.
From the Br\"{u}ckner-Hartree-Fock results in the infinite nuclear matter,
effective interaction in the pairing channel is not identical
to that in the $ph$ channel~\cite{Ring-Schuck}.
However, it seems unnatural in finite nuclei that interaction between nucleons
occupying specific orbitals suddenly changes its character
from the $ph$ channel to the pairing channel.
The consistent interaction between the $ph$ and the pairing channels
enables us to avoid this unnaturalness.

\subsection{Properties of infinite nuclear matter}\label{subsec:matter}

It is convenient to consider the hypothetical matter
which is comprised of an infinite number of nucleons
interacting only via the strong interaction.
Experimental information of such nuclear matter
can be obtained from the $A\to\infty$ limit of relevant quantities
of finite nuclei
if a sufficient number of systematic data are collected.
In nature, matters inside the neutron stars may correspond
well to infinite nuclear matter with a particular combination
of the density and the isospin ratio.
Whether actualized or not,
properties of the nuclear matter are useful,
providing information on specific parts of the effective interaction.

Although stable clusterization or spinodal instability~\cite{
  Chomaz-Colonna-Randrup_2004,LG-transition_2019}
can take place at low densities,
spatially homogeneous nuclear-matter is considered here,
with effective interactions in which the influence of temporary fluctuation
is incorporated.
Owing to the translational symmetry,
the s.p. state is expressed by the plane wave,
\be \varphi_{\vect{k}\sigma\tau}(\vect{r})
= \frac{1}{\sqrt\Omega}\,e^{i\vect{k}\cdot\vect{r}}\,\chi_\sigma \chi_\tau\quad
(\sigma=\uparrow,\downarrow\,;~\tau=p,n)\,.
\label{eq:NM-spwf}\ee
Here $\chi_\sigma$ ($\chi_\tau$) denotes the spin (isospin) w.f.
For the volume of the system $\Omega$,
the $\Omega\rightarrow\infty$ limit will be taken afterward
with keeping the nucleon density $\rho=A/\Omega$ finite.
The homogeneity prevents the non-central channels of the interaction
from contributing.
The energy of the system can be obtained only
from $v^{(\mathrm{C})}+v^{(\mathrm{C}\rho)}$ in Eq.~(\ref{eq:effint}),
as well as from $K$ in Eq.~(\ref{eq:Hamil}).
At zero temperature, the nucleons occupy the s.p. states
up to the Fermi momentum $k_{\mathrm{F}\tau\sigma}$,
which may depend on the spin and isospin.
The energy of the nuclear matter is then a function of $k_{\mathrm{F}\tau\sigma}$.
Although the Bardeen-Cooper-Schrieffer (BCS) state
can be lower in energy at low density,
the energy gain due to the pairing is small and is neglected here.
With densities of each spin-isospin component,
\be \rho_{\tau\sigma} = \frac{1}{6\pi^2} k_{\mathrm{F}\tau\sigma}^3\,,
\ee
the spin- and the isospin-asymmetry parameters as well as the total density
are defined as
\be\begin{split}
\rho &= \sum_{\sigma\tau} \rho_{\tau\sigma}
= \rho_{p\uparrow}+\rho_{p\downarrow}+\rho_{n\uparrow}+\rho_{n\downarrow}\,,\\
\eta_s &= \frac{{\displaystyle\sum_{\sigma\tau}}\sigma\rho_{\tau\sigma}}{\rho}
= \frac{\rho_{p\uparrow}-\rho_{p\downarrow}+\rho_{n\uparrow}-\rho_{n\downarrow}}{\rho}\,,\\
\eta_t &= \frac{{\displaystyle\sum_{\sigma\tau}}\tau\rho_{\tau\sigma}}{\rho}
= \frac{\rho_{p\uparrow}+\rho_{p\downarrow}-\rho_{n\uparrow}-\rho_{n\downarrow}}{\rho}\,,\\
\eta_{st} &= \frac{{\displaystyle\sum_{\sigma\tau}}\sigma\tau\rho_{\tau\sigma}}{\rho}
= \frac{\rho_{p\uparrow}-\rho_{p\downarrow}-\rho_{n\uparrow}+\rho_{n\downarrow}}{\rho}\,,
\end{split}\ee
where $\sigma$ ($\tau$) in the summation takes $\pm 1$,
corresponding to $\sigma=\uparrow,\downarrow$ ($\tau=p,n$).
The spin-saturated symmetric nuclear matter is defined
by $\eta_s=\eta_t=\eta_{st}=0$,
giving $k_{\mathrm{F}p\uparrow}=k_{\mathrm{F}p\downarrow}
=k_{\mathrm{F}n\uparrow}=k_{\mathrm{F}n\downarrow}$
which is denoted simply by $k_\mathrm{F}$,
and $\rho_{p\uparrow}=\rho_{p\downarrow}=\rho_{n\uparrow}
=\rho_{n\downarrow}=\rho/4$.
The minimum of the energy per nucleon $\mathcal{E}=E/A=E/\rho\Omega$
with respect to all of the four variables
corresponds to the saturation point,
which should lie along the $\eta_s=\eta_t=\eta_{st}=0$ line.
The minimum of $\mathcal{E}$, which satisfies
\be \frac{\partial\mathcal{E}}{\partial\rho}\Big\vert_\mathrm{sat.}=0\,,
\label{eq:sat-point}\ee
yields the saturation density $\rho_0$
(equivalently, $k_{\mathrm{F}0}$) and energy $\mathcal{E}_0$.
The expression $\vert_\mathrm{sat.}$ stands for evaluation
at the saturation point.

The total energy of the nuclear matter is calculated as~\cite{Nakada_2003}
\be\begin{split}
E = \bra K\ket + \Big\bra&\sum v_{ij}^{(\mathrm{C})}\Big\ket
+ \Big\bra\sum v_{ij}^{(\mathrm{C}\rho)}\Big\ket\,;\\
\bra K\ket &= \frac{\Omega}{(2\pi)^3} \sum_{\sigma_1\tau_1}
  \int_{k_1\leq k_{\mathrm{F}\tau_1\sigma_1}}d^3k_1
  \frac{\vect{k}_1^2}{2M}
  = \frac{3}{5}\Omega\sum_{\sigma\tau}\frac{k_{\mathrm{F}\tau\sigma}^2}{2M}\,
  \rho_{\tau\sigma}\,,\\
  \Big\bra\sum v_{ij}^{(\mathrm{C})}\Big\ket
  &= \frac{\Omega^2}{2(2\pi)^6} \sum_{\sigma_1\sigma_2\tau_1\tau_2}
  \int_{k_1\leq k_{\mathrm{F}\tau_1\sigma_1}}d^3k_1 \int_{k_2\leq k_{\mathrm{F}\tau_2\sigma_2}}d^3k_2 \\
&\hspace*{4.5cm}\times\bra\vect{k}_1\sigma_1\tau_1,\vect{k}_2\sigma_2\tau_2
  |v_{12}^{(\mathrm{C})}|\vect{k}_1\sigma_1\tau_1,\vect{k}_2\sigma_2\tau_2\ket \\
&= \frac{\Omega}{2(2\pi)^6} \sum_n \sum_{\sigma_1\sigma_2\tau_1\tau_2} \\
&\quad\times \Big[\big(t_n^{(\mathrm{W})} + t_n^{(\mathrm{B})} \delta_{\sigma_1\sigma_2}
- t_n^{(\mathrm{H})} \delta_{\tau_1\tau_2}
- t_n^{(\mathrm{M})} \delta_{\sigma_1\sigma_2} \delta_{\tau_1\tau_2}\big)
\mathcal{W}_n^\mathrm{H}(k_{\mathrm{F}\tau_1\sigma_1},k_{\mathrm{F}\tau_2\sigma_2}) \\
&\quad + \big(t_n^{(\mathrm{M})}
+ t_n^{(\mathrm{H})} \delta_{\sigma_1\sigma_2} - t_n^{(\mathrm{B})} \delta_{\tau_1\tau_2}
- t_n^{(\mathrm{W})} \delta_{\sigma_1\sigma_2} \delta_{\tau_1\tau_2}\big)
\mathcal{W}_n^\mathrm{F}(k_{\mathrm{F}\tau_1\sigma_1},k_{\mathrm{F}\tau_2\sigma_2})\Big]\,,\\
\Big\bra\sum v_{ij}^{(\mathrm{C}\rho)}\Big\ket &= \frac{\Omega^2}{2(2\pi)^6}
  \sum_{\sigma_1\sigma_2\tau_1\tau_2} \int_{k_1\leq k_{\mathrm{F}\tau_1\sigma_1}}d^3k_1
  \int_{k_2\leq k_{\mathrm{F}\tau_2\sigma_2}}d^3k_2 \\
&\hspace*{4.5cm}\times\bra\vect{k}_1\sigma_1\tau_1,\vect{k}_2\sigma_2\tau_2
  |v_{12}^{(\mathrm{C}\rho)}|\vect{k}_1\sigma_1\tau_1,\vect{k}_2\sigma_2\tau_2\ket \\
&= \frac{\Omega}{2(2\pi)^6} \sum_{\sigma_1\sigma_2\tau_1\tau_2}
  \Big[\frac{C_\mathrm{SE}[\rho]+C_\mathrm{TE}[\rho]}{2}
    \big(1 - \delta_{\sigma_1\sigma_2} \delta_{\tau_1\tau_2}\big) \\
  &\hspace*{2cm} + \frac{-C_\mathrm{SE}[\rho]+C_\mathrm{TE}[\rho]}{2}
    \big(\delta_{\sigma_1\sigma_2} -  \delta_{\tau_1\tau_2}\big)\Big]\,
    \frac{16\pi^2}{9} k_{\mathrm{F}\tau_1\sigma_1}^3 k_{\mathrm{F}\tau_2\sigma_2}^3 \,,
\label{eq:NME}\end{split}\ee
where
\be\begin{split}
\mathcal{W}_n^\mathrm{H}(k_{\mathrm{F}1},k_{\mathrm{F}2})
=& \int_{k_1\leq k_{\mathrm{F}1}} d^3k_1 \int_{k_2\leq k_{\mathrm{F}2}} d^3k_2\,
\tilde f_n^{(\mathrm{C})}(0)
= \frac{16\pi^2}{9} k_{\mathrm{F}1}^3 k_{\mathrm{F}2}^3\,\tilde f_n^{(\mathrm{C})}(0)\,,\\
\mathcal{W}_n^\mathrm{F}(k_{\mathrm{F}1},k_{\mathrm{F}2})
=& \int_{k_1\leq k_{\mathrm{F}1}} d^3k_1 \int_{k_2\leq k_{\mathrm{F}2}} d^3k_2\,
\tilde f_n^{(\mathrm{C})}(2k_{12}) \\
=& 8\pi^2 \bigg[ \int_0^{(k_{\mathrm{F}2}-k_{\mathrm{F}1})/2}
  dk_{12}\,\frac{16}{3}k_{\mathrm{F}1}^3 k_{12}^2\,\tilde f_n^{(\mathrm{C})}(2k_{12}) \\
&\quad + \int_{(k_{\mathrm{F}2}-k_{\mathrm{F}1})/2}^{(k_{\mathrm{F}1}+k_{\mathrm{F}2})/2}
dk_{12}\,\Big\{-\frac{1}{2}(k_{\mathrm{F}2}^2-k_{\mathrm{F}1}^2)^2 k_{12}
+\frac{8}{3}(k_{\mathrm{F}1}^3+k_{\mathrm{F}2}^3) k_{12}^2 \\
&\hspace*{4cm} -4(k_{\mathrm{F}1}^2+k_{\mathrm{F}2}^2) k_{12}^3
+\frac{8}{3}k_{12}^5 \Big\}\,\tilde f_n^{(\mathrm{C})}(2k_{12})\bigg]
\end{split}\label{eq:W-func}\ee
with
\be \tilde f_n^{(\mathrm{C})}(q) = \int d^3r\,f_n^{(\mathrm{C})}(r)\,
e^{-i\vect{q}\cdot\vect{r}}
\label{eq:Four1}\ee
and
\be\begin{split}
t_n^{(\mathrm{SE})} = t_n^{(\mathrm{W})}-t_n^{(\mathrm{B})}
-t_n^{(\mathrm{H})}+t_n^{(\mathrm{M})}\,,&\quad
t_n^{(\mathrm{TE})} = t_n^{(\mathrm{W})}+t_n^{(\mathrm{B})}
+t_n^{(\mathrm{H})}+t_n^{(\mathrm{M})}\,,\\
t_n^{(\mathrm{SO})} = t_n^{(\mathrm{W})}-t_n^{(\mathrm{B})}
+t_n^{(\mathrm{H})}-t_n^{(\mathrm{M})}\,,&\quad
t_n^{(\mathrm{TO})} = t_n^{(\mathrm{W})}+t_n^{(\mathrm{B})}
-t_n^{(\mathrm{H})}-t_n^{(\mathrm{M})}\,.
\end{split}\ee

The analytic expression of the $\mathcal{W}$ functions
of (\ref{eq:W-func}) can be obtained for most $f_n^{(\mathrm{C})}(r)$
employed in the SCMF approaches.
Then, by applying Eq.~(\ref{eq:NME}) with (\ref{eq:W-func}),
energy per nucleon $\mathcal{E}$ and their derivatives
are calculated almost analytically.
The saturation density and energy are determined by Eq.~(\ref{eq:sat-point}).
The incompressibility is
\be \mathcal{K}_0 = k_\mathrm{F}^2 \frac{\partial^2\mathcal{E}}
  {\partial k_\mathrm{F}^2}\Big\vert_\mathrm{sat.}
 = 9\rho^2 \frac{\partial^2\mathcal{E}}{\partial\rho^2}\Big\vert_\mathrm{sat.}\,.
\label{eq:K_0}\ee
The volume symmetry energy and analogous curvature
for the spin asymmetry are obtained as a function of $\rho$ as
\be\begin{split}
a_t(\rho) = \frac{1}{2} \frac{\partial^2\mathcal{E}}{\partial\eta_t^2}
	\Big\vert_{\eta_s=\eta_t=\eta_{st}=0}\,,\quad
a_s(\rho) &= \frac{1}{2} \frac{\partial^2\mathcal{E}}{\partial\eta_s^2}
	\Big\vert_{\eta_s=\eta_t=\eta_{st}=0}\,,\\
a_{st}(\rho) &= \frac{1}{2} \frac{\partial^2\mathcal{E}}{\partial\eta_{st}^2}
	\Big\vert_{\eta_s=\eta_t=\eta_{st}=0}\,.
\end{split}\label{eq:a_s&t}\ee
Their values at the saturation point are denoted by
$a_{t0}=a_t(\rho_0)$, $a_{s0}=a_s(\rho_0)$, $a_{st0}=a_{st}(\rho_0)$.
The s.p. energy $\epsilon_{\tau\sigma}(\vect{k})$ is defined
by variation of $E$ with respect to the occupation probability
of the state $|\vect{k}\sigma\tau\ket$~\cite{Baym-Pethick}.
The effective mass ($k$-mass) at the saturation point $M^\ast_0$ is defined
from $\epsilon_{\tau\sigma}(\vect{k})$ by
\be \frac{\partial\epsilon_{\tau\sigma}(\vect{k})}{\partial k}
 \Big\vert_\mathrm{sat.} = \frac{k_{\mathrm{F}0}}{M^\ast_0}\,,
 \label{eq:M*}\ee
and is calculated by the derivative of the $\mathcal{W}$ functions
in Eq.~(\ref{eq:W-func}) with respect to $k_{\mathrm{F}1}$~\cite{Nakada_2003}.
It is noted that these quantities
of Eqs.~(\ref{eq:K_0},\ref{eq:a_s&t},\ref{eq:M*}) are linked
to the Landau-Migdal parameters~\cite{Nakada_2003}.
The slope parameter of $a_t(\rho)$ is
\be \mathcal{L}_{t0} = 3\frac{d}{d\rho} a_t(\rho)\Big\vert_\mathrm{sat.}
 = \frac{1}{2}k_\mathrm{F}\frac{\partial^3\mathcal{E}}
  {\partial k_\mathrm{F}\,\partial\eta_t^2}\Big\vert_\mathrm{sat.}
 = \frac{3}{2}\rho\frac{\partial^3\mathcal{E}}
 {\partial\rho\,\partial\eta_t^2}\Big\vert_\mathrm{sat.}\,.
 \label{eq:L_t0}
\ee
The third derivative of $\mathcal{E}$ with respect to $\rho$
is denoted by $\mathcal{Q}_0$,
\be \mathcal{Q}_0 = k_\mathrm{F}^3
  \frac{\partial^3\mathcal{E}}{\partial k_\mathrm{F}^3}\Big\vert_\mathrm{sat.}
 = 27\rho^3\frac{\partial^3\mathcal{E}}{\partial\rho^3}\Big\vert_\mathrm{sat.}\,.
\ee

\begin{figure}
\centerline{\includegraphics[scale=0.5]{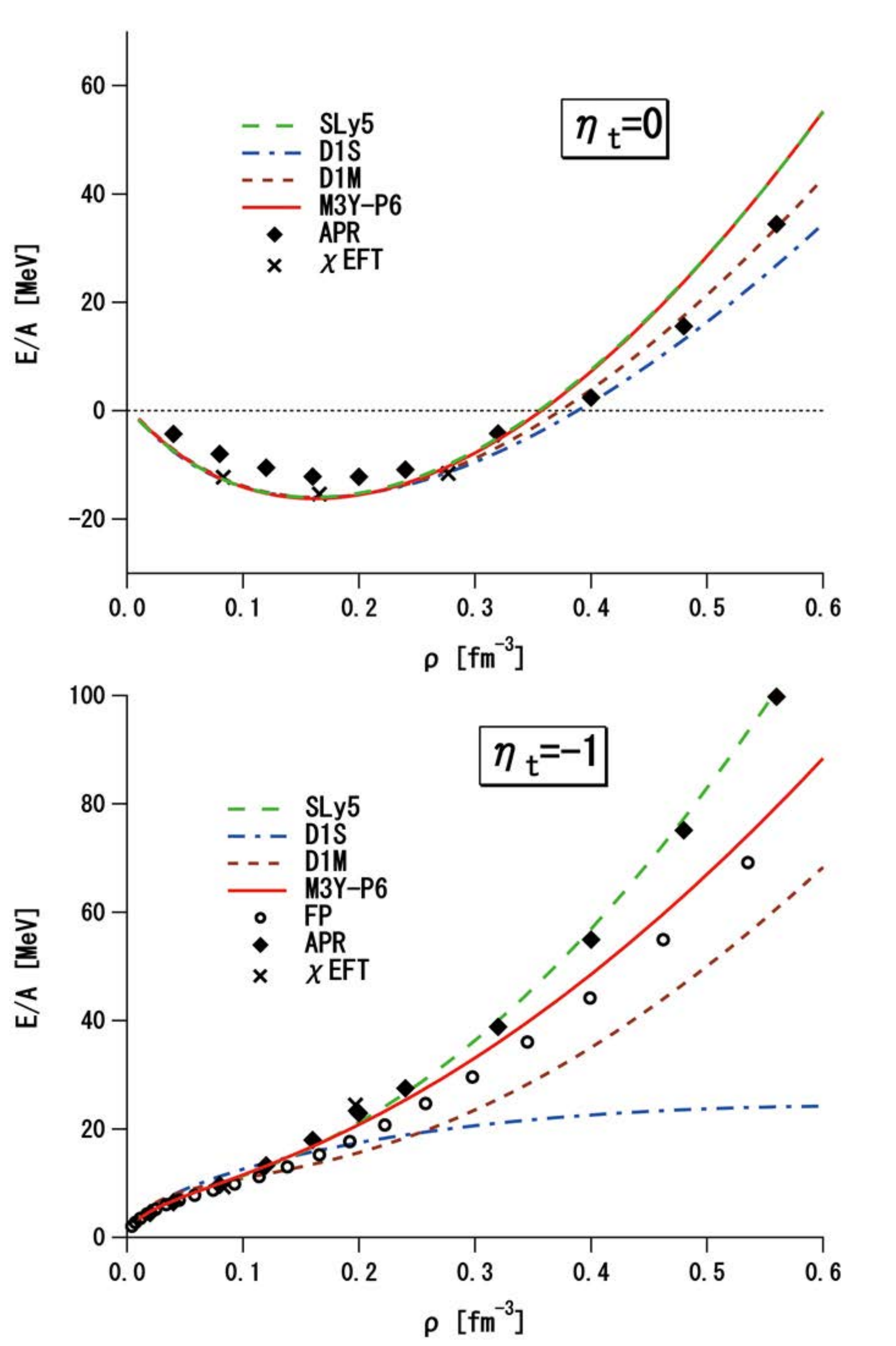}}
\caption{$\mathcal{E}=E/A$ \textit{vs.} $\rho$
  in the symmetric nuclear matter (upper panel)
  and the pure neutron matter (lower panel),
  calculated with SLy5 (green long dashed line), D1S (blue dot-dashed line),
  D1M (brown short dashed line) and M3Y-P6 (red line).
  Circles, diamonds and crosses represent the \textit{ab initio} results
  of FP~\protect\cite{Friedman-Pandharipande_1981},
  APR~\protect\cite{Akmal-Pandharipande-Ravenhall_1998}
  and the $\chi$EFT of Ref.~\protect\refcite{Kohno_2013_erratum}
  with $\Lambda=450\,\mathrm{MeV}$,
  respectively.
\label{fig:NME}}
\end{figure}

In Fig.~\ref{fig:NME}, $\mathcal{E}(\rho)$'s
at $\eta_t=0$ (symmetric nuclear matter)
and at $\eta_t=-1$ (pure neutron matter)
are compared among several effective interactions.
In both cases $\eta_s=\eta_{st}=0$ is assumed.
The \textit{ab initio} results of
Refs.~\refcite{Friedman-Pandharipande_1981} (for the pure neutron matter)
and \refcite{Akmal-Pandharipande-Ravenhall_1998} (for both),
as well as those of the $\chi$EFT with the cut-off energy $450\,\mathrm{MeV}$
in the lowest-order Br\"{u}ckner calculations~\cite{Kohno_2013_erratum},
are also displayed for reference.
Fitted to properties of finite nuclei,
all of SLy5, D1S, D1M and M3Y-P6 give similar $\mathcal{E}(\rho)$
for the symmetric matter in $\rho\lesssim 0.25\,\mathrm{fm}^{-3}$.
In contrast, D1S behaves quite differently for the pure neutron matter
from the others,
since this parameter-set is obtained with no reference
to the \textit{ab initio} EoS of the pure neutron matter.
This illustrates that it is not easy to fix neutron-matter EoS
only from nuclear structure data~\cite{Brown_2000}.

Several quantities at the saturation points are listed in Table~\ref{tab:NMsat},
where those of M3Y-P6, D1S and SLy5 are compared.
$k_{\mathrm{F}0}$, $\mathcal{E}_0$, $\mathcal{K}_0$, $M^\ast_0/M$
and $a_{t0}$ are constrained moderately well
from staple nuclear structure data.
Slightly lower $\mathcal{E}_0$ in M3Y-P6 than the others
is related to the tensor force.
M3Y-P6 includes realistic tensor force,
while SLy5 and D1S do not contain tensor force explicitly,
although some of its effects might be incorporated
into other channels in an effective manner.
Since the tensor force acts on finite nuclei mostly repulsively
(see Sec.~\ref{subsec:def-TNS}),
fitting the parameters to the measured binding energies with the tensor force
gives slightly lower saturation energy.
$M^\ast_0$ is connected to the nuclear excitations
and the energy-dependence of the optical potential.
Experimental information for the latter indicates
$M^\ast_0\approx 0.7M$~\cite{Bohr-Mottelson_1}.
The small $M^\ast_0$ value could be a defect of M3Y-P6,
which could become problematic when applied to excitations,
even if it is not apparent in g.s. properties.
$\mathcal{K}_0$ and $\mathcal{Q}_{0}$ are more or less reflected
in the upper panel of Fig.~\ref{fig:NME}.
The neutron-matter energy is sometimes approximated as
\be \mathcal{E}(\rho,\eta_t=-1) \approx
\mathcal{E}(\rho,\eta_t=0) + a_t(\rho)\,.
\label{eq:E_PNM-approx}\ee
The so-called higher-order symmetry energy governs
the precision of this approximation,
and see Ref.~\refcite{Tsukioka-Nakada_2017} for a detailed analysis.
Because the expansion of $a_t(\rho)$ around $\rho_0$ gives
\be a_t(\rho) \approx a_{t0}
+ \mathcal{L}_{t0}\Big(\frac{\rho-\rho_0}{3\rho_0}\Big)\,,
\ee
the values of $\mathcal{L}_{t0}$ correlate
to the lower panel in Fig.~\ref{fig:NME}.

\begin{table}[pt]
\caption{Nuclear matter properties at the saturation point.
  \label{tab:NMsat}}
\centerline
{\begin{tabular}{ccrrr}
\toprule 
&&~~M3Y-P6 &~~~~D1S~~ &~~~~SLy5~ \\ \colrule 
$k_{\mathrm{F}0}$ & (fm$^{-1}$) & $1.340$~~& $1.342$~~& $1.334$~~\\
$\mathcal{E}_0$ & (MeV) & $-16.24$~~& $-16.01$~~& $-15.98$~~\\
$\mathcal{K}_0$ & (MeV) & $239.7$~~& $202.9$~~& $229.9$~~\\
$M^\ast_0/M$ && $0.596$~~& $0.697$~~& $0.697$~~\\
$a_{t0}$ & (MeV) & $32.14$~~& $31.12$~~& $32.03$~~\\
$a_{s0}$ & (MeV) & $26.47$~~& $26.18$~~& $37.47$~~\\
$a_{st0}$ & (MeV) & $41.00$~~& $29.13$~~& $15.15$~~\\
$\mathcal{Q}_{0}$ & (MeV) & $-378.0$~~& $-515.7$~~& $-363.9$~~\\
$\mathcal{L}_{t0}$ & (MeV) & $44.64$~~& $22.44$~~& $48.27$~~\\
\botrule 
\end{tabular}}
\end{table}

The curvatures $a_s$ and $a_{st}$ are linked to the magnetic susceptibility
of the nuclear matter~\cite{Vidana-Bombaci_2002}.
Interaction-dependence of $a_{st0}$ is obvious in Table~\ref{tab:NMsat}.
Data on the spin-isospin excitation,
which customarily supply results in terms of the Landau-Migdal parameter,
have been reported
in Refs.~\refcite{Suzuki-Sakai_1999,Ichimura-Sakai-Wakasa,Yasuda-etal_2018}.
Converted to $a_{st0}$, they yield $a_{st0}\approx 35\,\mathrm{MeV}$.
Whereas few data on $a_s$ have been obtained so far,
a recent experiment has indicated that isoscalar spin excitations are not
much fragmented over higher energy region~\cite{Matsubara-etal_2015}.
This implicates that $a_{s0}$ is substantially smaller than $a_{st0}$.
As both experimental and theoretical studies on the quantities
involving the spin d.o.f. have been limited,
further efforts on both sides are desired.

\begin{figure}
\centerline{\includegraphics[scale=0.5]{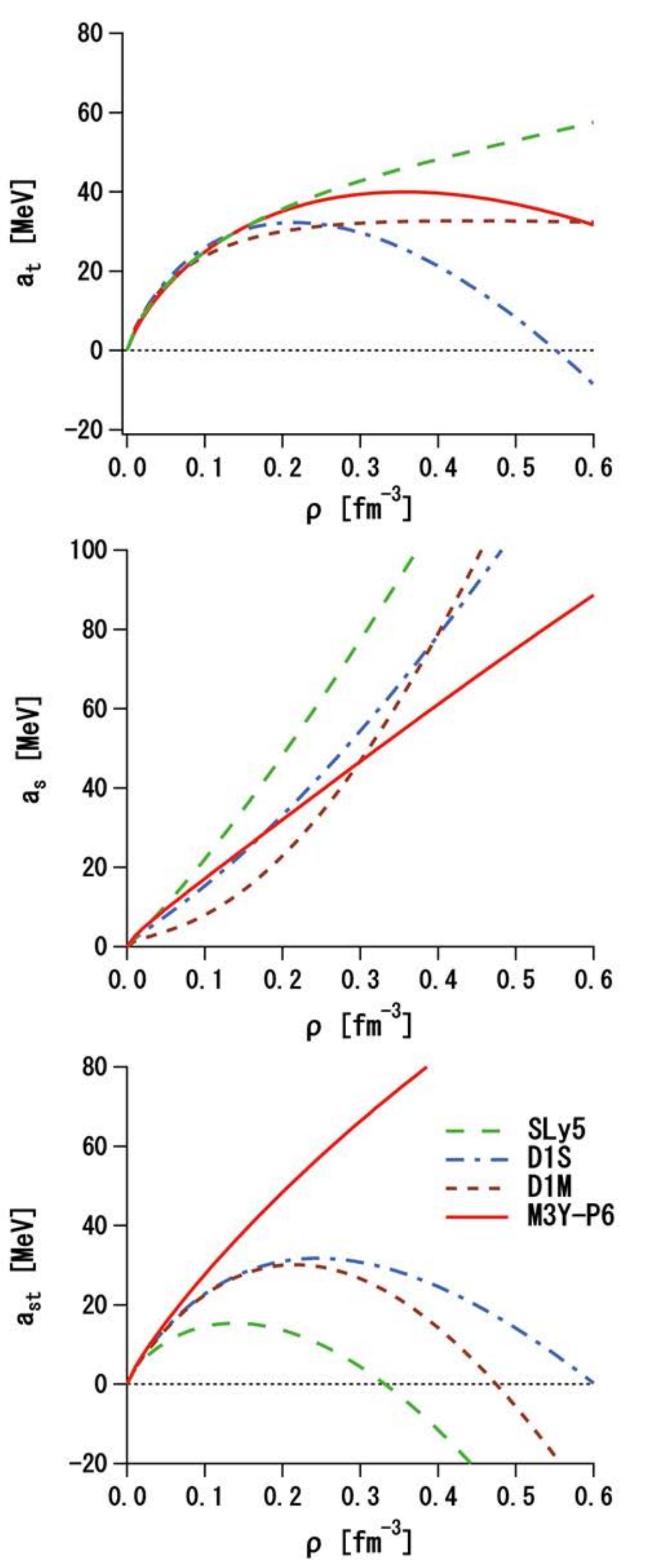}}
\caption{$a_t(\rho)$, $a_s(\rho)$ and $a_{st}(\rho)$
in the symmetric nuclear matter.
See Fig.~\protect\ref{fig:NME} for the conventions.
\label{fig:a_st}}
\end{figure}

Together with $a_t$, $\rho$-dependence of $a_s$ and $a_{st}$
is presented in Fig.~\ref{fig:a_st}.
The spin-saturated symmetric nuclear matter becomes unstable
if any of $a_t$, $a_s$ and $a_{st}$ is negative.
All the interactions under consideration other than M3Y-P6
give rise to instability at $\rho<0.6\,\mathrm{fm}^{-3}\approx 4\rho_0$.
In particular, in contrast to rigidity with M3Y-P6,
the nuclear matter is predicted to be unstable
against the spin-isospin excitation with the other interactions,
as revealed by $a_{st}(\rho)$ in Fig.~\ref{fig:a_st}.
Though beyond the scope of this review,
this instability could have influence
when these effective interactions are applied to the neutron star.
In the rigidity concerning $a_{st}(\rho)$,
$v^{(\mathrm{C})}_\mathrm{OPEP}$ plays a significant role,
which is included explicitly in M3Y-P6 but not in the others.

An interesting extension of the SCMF approaches
is applications to neutron stars,
extremely compact astrophysical objects.
In the neutron stars,
matters comprised of hadrons are actualized, covering a wide range of $\rho$.
The SCMF framework can be a tool to connect
properties of finite nuclei on earth to extreme matters in compact stars
in a unified manner.
At a glance, as the density differs significantly,
properties of high-density matters
might not seem closely connected to the structure of the finite nuclei,
which can be measured by experiments on earth.
However, the nuclear structure is indeed informative
to properties of high-density matters,
and \textit{vice versa}.
Effective interaction underlies this connection.
Applications of the M3Y-type semi-realistic interactions
to the neutron stars and the proto-neutron stars are found
in Refs.~\refcite{Loan-etal_2011,Tan-etal_2016}.

\subsection{Asymptotics of quasiparticle wave functions}\label{subsec:asymp}

The asymptotic behavior of the quasiparticle (q.p.) w.f.'s
at large $r$ is discussed,
which is relevant to halos.
While it is discussed in the HFB frame,
it is easy to reduce the consequence to the s.p. w.f.'s in the HF.

Consider the HFB equation in terms of the spherical coordinate.
At sufficiently large $r$, the nuclear force becomes negligible,
and the following asymptotic equations are derived for a neutron q.p.~\cite{
  Skyrme-P,Nakada_2006},
\be\begin{split}
\Big\{-\frac{1}{2M}\frac{\partial^2}{\partial r^2}
- \lambda\Big\}\,[r\,U_k(\vect{r})]
 &\approx \varepsilon(k)\,[r\,U_k(\vect{r})]\,,\\
\Big\{-\frac{1}{2M}\frac{\partial^2}{\partial r^2}
- \lambda\Big\}\,[r\,V_k(\vect{r})]
 &\approx -\varepsilon(k)\,[r\,V_k(\vect{r})]\,.
 \label{HFBeq-asymp}\end{split}\ee
$V_k$ ($U_k$) is  the w.f. for the occupied (unoccupied) component
of the q.p. state $k$,
and is associated with the coefficients in Eq.~(\ref{eq:Bogolyubov})
in Sec.~\ref{subsec:GEM} as
\be V_k(\vect{r}) = \sum_\mu V_{\mu k}\,\phi_\mu(\vect{r})\,,\quad
U_k(\vect{r}) = \sum_\mu U_{\mu k}\,\phi_\mu(\vect{r})\,,
\ee
where $\{\phi_\mu\}$ is the set of the basis functions.
As long as the nucleus is bound,
the chemical potential $\lambda$ is negative.
Since the q.p. energy $\varepsilon(k)$ is taken to be positive,
Eq.~(\ref{HFBeq-asymp}) derives the asymptotic form,
apart from the amplitude and the spin-angular part,
\be
 r\,V_k(\vect{r})\sim \exp(-\eta_{k+} r)\,,\quad
 r\,U_k(\vect{r})\sim\left\{\begin{array}{ccc}
 \exp(-\eta_{k-} r) & \mbox{for}& \lambda+\varepsilon(k) < 0 \\
 \cos(p_k r + \theta_k) & \mbox{for}& \lambda+\varepsilon(k) > 0 \end{array}
 \right.,
\label{qpwf-asymp}\ee
where $\eta_{k\pm} = \sqrt{2M[-\lambda\pm\varepsilon(k)]}$,
$p_k = \sqrt{2M[\lambda+\varepsilon(k)]}$
and $\theta_k$ is an appropriate real number.
It is noted that the q.p. energies are discrete
only for $0\leq\varepsilon(k)<-\lambda$,
and both the exponential and the oscillatory asymptotics have to be treated
in the HFB calculations.

For proton w.f.'s,
the Coulomb interaction influences the w.f.'s at large $r$,
although it does not affect the criterion
whether or not individual q.p. levels are discrete.
The above arguments for the neutron w.f.'s
can be applied if the asymptotic forms are properly modified.

The asymptotic form of Eq.~(\ref{qpwf-asymp}) is compared
to that of Eq.~(\ref{eq:asymp-phi}).
The SCMF theory gives the product form
$|\Psi_A\ket\sim|\Psi_{A-1}\ket\otimes|\varphi_A\ket$
for the $A$-body w.f.,
as defining the s.p. (or q.p.) orbitals for the individual nucleons.
In the SCMF theory, the coordinate of the individual nucleon $\vect{r}_i$
is measured relative to an origin fixed in the rest frame,
which usually corresponds to the average
of the c.m. position $\bra\vect{R}\ket$,
rather than the relative coordinate like $\vect{r}_i-\vect{R}$.
For sufficiently large $r_A\,(\gg R)$,
$\varphi_A$ in Eq.~(\ref{eq:asymp-phi}) is reduced as
\be
\varphi_A \approx \xi \,\frac{\exp(-\sqrt{2M|\epsilon'|}\,r_A)}
  {r_A}\mathcal{Y}(\hat{\vect{r}}_A)\,.
\label{eq:asymp-phi1} \ee
The form of Eq.~(\ref{qpwf-asymp}) is harmonious
with that of Eq.~(\ref{eq:asymp-phi1}),
if $\lambda\mp\varepsilon(k)$ for the last neutron
is equal to $\epsilon'$,
where the sign is determined by whether $N$ is even or odd,
as discussed in Sec.~\ref{subsec:def-halo}.
The description of neutron halos by the SCMF approaches is feasible
as far as these energies are close.

\section{Numerical methods}\label{sec:numerical}

Numerical calculations in the SCMF frame are often intensive,
and computational methods have significance
in applying the SCMF theory to finite nuclei.
I devote this section to numerical methods of the SCMF calculations.
After general arguments,
the methods that will give the results in the subsequent sections are discussed.

\subsection{General arguments}

In the SCMF theory, the whole nuclear w.f. is expressed
by a product of the s.p. or q.p. functions,
apart from the anti-symmetrization.
The main task in the SCMF calculations is
obtaining the s.p. or q.p. functions.
There are at least three ways to handle the s.p. functions:
(i) representing the s.p. functions
at discretized points in the coordinate space,
(ii) representing them at discretized points in the momentum space,
and (iii) representing them by a linear combination
of a set of basis functions.
Although the method (ii) could have advantages in certain respects,
I do not go into detail,
as it has not been much explored so far.
While the method (i) is suitable when used with zero-range interactions
(or quasi-local EDFs) such as the Skyrme interaction,
it is not easy to implement when applied with finite-range interactions,
though carried out in a few works~\cite{DeDonno-Co-Anguiano-Lallena_2014}.
The method (iii) does not lead to severe difficulties
even when used with finite-range interactions (or non-local EDFs).
However, since results somewhat depend on the basis functions,
properties of the basis functions may limit applicability,
and it is crucial to choose an appropriate set of functions.
Notice that, in practical numerical computations,
one can handle a limited number of basis functions.
The completeness with an infinite number of the bases,
as guaranteed for a set of the harmonic-oscillator (HO) basis-functions,
is not necessarily helpful.

Both for the methods (i) and (iii),
special care is needed when applied to nuclei far off the stability,
in which the w.f.'s may extend in a spatially broad range
with an energy-dependent tail.
In the HFB calculations, even oscillatory asymptotics may come
into bound-state problems [see Eq.~(\ref{qpwf-asymp})].
For the method (i), a standard homogeneous mesh is inefficient
to describe such w.f.'s,
irrespective of the one- to three-dimensional calculations.
To cure this problem, methods using an adaptive mesh have been
exploited~\cite{Modine-Zumbach-Kaxiras,Nakatsukasa-Yabana_2005}.
To improve precision of the kinetic energy term
that contains derivatives of the w.f.'s with respect to the spatial points,
the method using the Lagrange mesh has been developed~\cite{Baye_2015},
which lies at an intersection of the methods (i) and (iii).
For the method (iii),
the most popular basis function in the nuclear structure calculations
has been the HO basis-functions.
However, since the asymptotics of the HO function is given by a Gaussian
with a definite range,
it is not suitable for nuclei with a broad spatial distribution.
To remedy the asymptotics of the total density distribution,
a modification using the transformed HO (THO)
was proposed~\cite{Stoitsov-Dobaczewski-Ring-Pittel}.
Still, it does not mean that the THO set is capable
of describing energy-dependent asymptotics of individual s.p. orbitals;
asymptotics of all the s.p. functions is forced to match
that of the most loosely bound particle.
It is also noted that the THO set leads to a complication
when applied to finite-range interactions.
Another set of basis functions to improve the asymptotics
is given by numerically solving the Woods-Saxon potential~\cite{
  Zhou-Meng-Ring_2003}.

For the algorithm for the SCMF computation,
one may solve the s.p. Schr\"{o}dinger equations
(\textit{i.e.} the HF or the HFB equation) iteratively
or may apply the gradient method~\cite{Ring-Schuck}.
The latter could be extended to
\textit{e.g.} the conjugate gradient method~\cite{ConjugateGradientMethod_1994}.

\subsection{Gaussian expansion method}\label{subsec:GEM}

For numerical results shown in the subsequent sections,
the Gaussian expansion method (GEM) is applied.
First developed for computations
in few-body systems~\cite{Kamimura_1988,Kameyama-Kamimura-Fukushima},
the GEM has been extended to the SCMF calculations~\cite{Nakada-Sato_2002,
  Nakada_2006}.
The basis functions are taken to be
\be \phi_{\nu\ell jm}(\vect{r})
= R_{\nu\ell j}(r)\,[Y^{(\ell)}(\hat{\vect{r}})\chi_\sigma]^{(j)}_m\,; \quad
R_{\nu\ell j}(r) = \mathcal{N}_{\nu\ell j}\,r^\ell\exp(-\nu r^2)\,,
\label{eq:basis} \ee
with an appropriate constant $\mathcal{N}_{\nu\ell j}$.
The isospin index is dropped without confusion.
The radial function is taken to be a Gaussian
with the range parameter $\nu$,
which is a complex number in general~\cite{Hiyama-Kino-Kamimura_2003};
$\nu=\nu_\mathrm{r}+i\nu_\mathrm{i}$ ($\nu_\mathrm{r}>0$).
In the GEM, $\nu$'s belong to a geometric progression.
In Refs.~\refcite{Nakada_2006,Nakada_2006_erratum},
it is found that a combination of the real- and complex-range Gaussian bases
is suitable for nuclear MFmean-field calculations.
In all the following calculations,
we take the basis-set of
\begin{equation}
\nu_\mathrm{r}=\nu_0\,b^{-2n}\,,\quad
\left\{\begin{array}{ll}\nu_\mathrm{i}=0 & (n=0,1,\cdots,5)\\
{\displaystyle\frac{\nu_\mathrm{i}}{\nu_\mathrm{r}}
=\pm\frac{\pi}{2}} & (n=0,1,2)\end{array}\right.\,,
 \label{eq:basis-param}
\end{equation}
with $\nu_0=(2.40\,\mathrm{fm})^{-2}$ and $b=1.25$,
irrespective of $(\ell,j)$.
Namely, 12 bases are employed for each $(\ell,j)$;
6 bases have real $\nu$ and the other 6 have complex $\nu$.
Superposition of these Gaussians efficiently describes
various s.p. or q.p. w.f.'s, as shown below.
Note that the GEM bases are non-orthogonal.
Therefore, the HF (HFB) equation,
the linear equation giving s.p. (q.p.) energies and w.f.'s,
becomes a generalized eigenvalue equation containing a norm matrix.

Because of the non-orthogonality,
the creation and annihilation operators
$a_{\nu\ell jm}^\dagger$ and $a_{\nu\ell jm}$
associated with $\phi_{\nu\ell jm}$
obey the non-canonical commutation relations,
\be \{a_{\nu\ell jm}, a_{\nu'\ell'j'm'}^\dagger\}
 = \delta_{\ell\ell'} \delta_{jj'} \delta_{mm'} N^{(\ell j)}_{\nu\nu'}\,,\quad
 \{a_{\nu\ell jm}, a_{\nu'\ell'j'm'}\} =
 \{a_{\nu\ell jm}^\dagger, a_{\nu'\ell'j'm'}^\dagger\} = 0\,,
\ee
where $\mathsf{N}=(N^{(\ell j)}_{\nu\nu'})$ is the norm matrix.
The HFB equation is modified accordingly,
and the HF equation is straightforwardly deduced from it.
The generalized Bogolyubov transformation giving the q.p. state $k$ is
\be \alpha_k^\dagger = \sum_\mu \big[U_{\mu k} a_\mu^\dagger
  + V_{\mu k} a_\mu\big]\,,
\label{eq:Bogolyubov}\ee
with $\mu=(\nu\ell jm)$.
The matrix properties of $\mathsf{U}=(U_{\mu k})$ and $\mathsf{V}=(V_{\mu k})$,
as well as the HFB equation and the HFB Hamiltonian,
are shown in Appendices of Ref.~\refcite{Nakada_2006}~\footnote{
  In Eq.~(A.10) of Ref.~\refcite{Nakada_2006},
  the variation ${\displaystyle\frac{\delta}{\delta\rho_{kk'}}}$
  should be corrected to ${\displaystyle\frac{\delta}{\delta\rho_{k'k}}}$.
}.

Matrix elements of the two-body interaction can be computed as follows.
Applying the inverse transformation of Eq.~(\ref{eq:Four1}),
\be f_n^{(\mathrm{C})}(r_{ij}) = \frac{1}{(2\pi)^3}\int d^3q\,
  \tilde f_n^{(\mathrm{C})}(q)\,e^{i\vect{q}\cdot\vect{r}_{ij}}
  = \frac{1}{(2\pi)^3}\int d^3q\,\tilde f_n^{(\mathrm{C})}(q)\,
    e^{i\vect{q}\cdot\vect{r}_i} e^{-i\vect{q}\cdot\vect{r}_j}\,,
\label{eq:Four2}\ee
a non-antisymmetrized element of $v^{(\mathrm{C})}$ can be decomposed
via~\cite{Horie-Sasaki}
\be \bra \mu_i'\mu_j'|f_n^{(\mathrm{C})}(r_{ij})|\mu_i\mu_j\ket_{\mathrm{n.a.}}
=  \frac{1}{(2\pi)^3}\int d^3q\,\tilde f_n^{(\mathrm{C})}(q)\,
  \bra\mu_i'|e^{i\vect{q}\cdot\vect{r}_i}|\mu_i\ket\,
  \bra\mu_j'|e^{-i\vect{q}\cdot\vect{r}_j}|\mu_j\ket\,.
\label{eq:int-me-decomp}\ee
The analytic expression of the s.p. element
$\bra\mu'|e^{\pm i\vect{q}\cdot\vect{r}}|\mu\ket$,
which is independent of the function $f_n^{(\mathrm{C})}(r)$,
is obtained for the basis functions of Eq.~(\ref{eq:basis}).
The angular integration is also implemented analytically
for the right-hand side (r.h.s.) of Eq.~(\ref{eq:int-me-decomp})
via the Racah algebra.
Thus it is reduced to an integral of the single variable $q$,
which is the only part depending on the function $f_n^{(\mathrm{C})}(r)$.
Although analytic expression may also be derived
for the $q$-integration~\cite{Nakada-Sato_2002,Nakada-Sato_2002_erratum},
it is not numerically advantageous
because of the round-off errors~\cite{Nakada_2006}.
The decomposition of Eq.~(\ref{eq:int-me-decomp}) is made
at the expense of the additional variable $\vect{q}$.
This structure may be conceived in the context
of the tensor network~\cite{TensorNetwork_2014}.
Formulas for calculating the one- and two-body matrix elements are given
in Refs.~\refcite{Nakada-Sato_2002,Nakada-Sato_2002_erratum}
and \refcite{Nakada_2006}.

To apply the spherical bases of Eq.~(\ref{eq:basis})
to the HFB calculations,
as well as to the deformed MF calculations
discussed in the Secs.~\ref{subsec:defcal} and \ref{sec:deformed},
truncation for the orbital angular momentum $\ell$ is unavoidable.
Hereafter the cut-off value of $\ell$ is denoted by $\ell_\mathrm{cut}$.
It is convenient to consider $\ell_\mathrm{cut}$
in terms of the conventional HO functions,
since it provides a first approximation of the s.p. orbits.
Denote the number of quanta in a HO function
by $N_\mathrm{osc}\,(=0,1,2,\cdots)$.
It has been found that
$\ell_\mathrm{cut}\geq N_\mathrm{osc}^\mathrm{F}+2$ should be taken
to handle the pairing properties,
where $N_\mathrm{osc}^\mathrm{F}$ is defined to be
the highest $N_\mathrm{osc}$ for the s.p. levels
in the major shell near the Fermi level~\cite{Nakada_2006}.
The $\ell_\mathrm{cut}$ value for the deformed MF calculations
is discussed in Subsec.~\ref{subsec:defcal}.

The advantages of the GEM are listed.
\begin{romanlist}
\item\label{it:asymp} By superposing different ranges of Gaussians,
  exponential and even oscillatory asymptotics at large $r$,
  which depends on the s.p. (or q.p.) energies,
  can be described efficiently~\cite{Nakada-Sato_2002,Nakada_2006}.
\item\label{it:2body} It is relatively easy to compute
  matrix elements of various two-body interactions~\cite{Nakada-Sato_2002,
  Nakada-Sato_2002_erratum},
  including $v^{(\mathrm{LS})}$ and $v^{(\mathrm{TN})}$
  as well as $v^{(\mathrm{C})}$.
  Interactions can be switched through the replacement of single subroutine
  for the $q$-integration in Eq.~(\ref{eq:int-me-decomp}) by another,
  which helps to ensure the reliability of the code.
\item\label{it:param} The basis parameters are
  insensitive to nuclide~\cite{Nakada_2008}.
  It is even practical to apply a single set of bases
  to almost the whole range of the nuclear chart~\cite{Nakada-Sugiura_2014,
    Nakada-Sugiura_2014_erratum} [see Eq.~(\ref{eq:basis-param})],
  facilitating systematic calculations by storing the two-body matrix elements.
\item\label{it:Coul-cm} The Coulomb and the c.m. Hamiltonian
  $V_C$ and $H_\mathrm{c.m.}$ can fully be included,
  up to the exchange and the pairing terms~\cite{Nakada-Sato_2002}.
\end{romanlist}
Evidence for the point (\ref{it:asymp}) will be shown
in Sec.~\ref{subsec:def-halo}.
This point is linked to efficiency of this method
also for coupling to the continuum,
as tested in the HFB calculations~\cite{Nakada_2006}
and by extensive calculations in the random-phase approximation (RPA)~\cite{
  Nakada-Mizuyama-Yamagami-Matsuo}.
The point (\ref{it:2body}) helps to apply the GEM to the M3Y-type interactions.
Concerning the point (\ref{it:param}),
it should be recalled that full convergence for the parameters
is not easy to attain.
This difficulty holds for all the other SCMF calculations;
even when it looks convergent by enlarging the model space slightly,
it does not necessarily guarantee full convergence.
The results are compared from the variational viewpoints,
and it seems legitimate to assert that
the present method often gives good enough precision or even better
than the other existing calculations~\cite{Nakada-Sato_2002,Nakada_2008},
particularly when applied with finite-range interactions.

To handle $Z$ or $N=\mathrm{odd}$ nuclei
with keeping the time-reversal ($\mathcal{T}$) symmetry,
the equal-filling approximation (EFA) is employed,
which can be interpreted as an approximation assuming an ensemble
of the degenerate states~\cite{PerezMartin-Robledo_2008}.
The EFA is also used in the spherical HF calculations
for nuclei in which nucleons partially occupy a spherical s.p. orbit.

The s.p. energy for the level $k$ is defined by
$\epsilon(k)=\delta E/\delta n(k)$,
where $n(k)$ represents the occupation probability on $k$.
For the Hamiltonian of Eqs.~(\ref{eq:Hamil},\ref{eq:effint}),
\be\begin{split}
\epsilon(k) = \Big(1-\frac{1}{A}\Big)&\,\bra k|\frac{\vect{p}^2}{2M}|k\ket
+ \sum_{k'} n(k')\,\bra kk'|V|kk'\ket
+ \bra\Phi|\mathit{\Delta}V^\mathrm{rearr.}(k)|\Phi\ket\,;\\
V &= V_N + V_C
- \frac{1}{A}\sum_{i<j} \frac{\vect{p}_i\cdot\vect{p}_j}{M}\,,\\
\mathit{\Delta}V^\mathrm{rearr.}(k) &=
\sum_{i<j}\bigg[\Big\{\frac{\delta C_\mathrm{SE}}{\delta\rho}\,P_\mathrm{SE}
  + \frac{\delta C_\mathrm{TE}}{\delta\rho}\,P_\mathrm{TE}\Big\}\,
\delta(\vect{r}_{ij}) \\
&\qquad\quad + 2i\frac{\delta D}{\delta\rho}\,
 \vect{p}_{ij}\times\delta(\vect{r}_{ij})\,\vect{p}_{ij}\cdot
 (\vect{s}_i+\vect{s}_j)\bigg]\,
    \frac{\delta\rho(\vect{R}_{ij})}{\delta n(k)}\,.
\end{split}\label{eq:spe}\ee
In later sections contribution of a specific channel of the two-body interaction
$v_{ij}^{(\mathrm{X})}$ (\textit{e.g.} $\mathrm{X}=\mathrm{TN}$)
to the total energy $E$ and $\epsilon(k)$
will be investigated, through
\be\begin{split}
E^{(\mathrm{X})} &= \big\bra\Phi\big|\sum_{i<j}v_{ij}^{(\mathrm{X})}
\big|\Phi\big\ket\,,\\
\epsilon^{(\mathrm{X})}(k)
&= \sum_{k'} n(k')\,\big\bra kk'\big|\sum_{i<j}v_{ij}^{(\mathrm{X})}
\big|kk'\big\ket\,.
\end{split}\label{eq:E&spe-X}\ee

\subsection{Describing deformed nuclei}\label{subsec:defcal}

As mentioned above, a single set of spherical GEM bases is applicable
to a wide range of masses.
This is because the GEM basis-set of Eq.~(\ref{eq:basis-param})
is adaptable to the variable radial lengthening of the s.p. w.f.'s.
For the same reason,
the spherical GEM bases are applicable to deformed nuclei
without serious loss of precision, unless the deformation is too strong,
as have been shown for axially deformed cases~\cite{
  Nakada_2008,Suzuki-Nakada-Miyahara_2016,Miyahara-Nakada_2018,
  Nakada-Takayama_2018}.

Let us here assume the rotational symmetry
with respect to the intrinsic $z$-axis,
the parity conservation, the $\mathcal{T}$-symmetry,
and the symmetry with respect to the rotation
around the $y$-axis by the angle $\pi$
(\textit{i.e.} the $\mathcal{R}$-symmetry)~\cite{Bohr-Mottelson_2}.
The s.p. w.f. is represented
by expansion with the spherical GEM bases (\ref{eq:basis}),
\be \varphi_{k\pi m}(\vect{r}) = \sum_{\nu\ell j} c^{(k)}_{\nu\ell jm}\,
  \phi_{\nu\ell jm}(\vect{r})\,,
\label{eq:spwf}\ee
where the subscript $\pi$ on the left-hand side (l.h.s.) stands for the parity.
The sum of $\ell$ and $j$ on the r.h.s. runs over
all possible values satisfying $\pi=(-)^\ell$, $j=\ell\pm 1/2$ and $j\geq |m|$.
Owing to the $\mathcal{R}$ and the $\mathcal{T}$ symmetries,
the coefficient $c^{(k)}_{\nu\ell jm}$ is taken to be real.

The precision of the GEM for the deformed MF has been examined
via the axially symmetric HO potential,
\be h = \frac{\vect{p}^2}{2M} + \frac{M}{2}\left[
 \omega_\perp^2(x^2+y^2)+\omega_z^2 z^2)\right]
 = \frac{\vect{p}^2}{2M} + \frac{M}{2} \omega_0^2\,r^2
 \left[1-\frac{4}{3}\sqrt{\frac{4\pi}{5}}\delta_\mathrm{def}\,
  Y^{(2)}_0(\hat{\vect{r}})\right]\,,
\label{eq:AHO}\ee
where $\omega_0^2=(2\omega_\perp^2+\omega_z^2)/3$
and $\delta_\mathrm{def}=(\omega_\perp^2-\omega_z^2)/2\omega_0^2$.
By truncating the basis-set by $\ell\leq\ell_\mathrm{cut}$,
approximate eigenvalues of the s.p. Hamiltonian of Eq.~(\ref{eq:AHO})
are calculated with the spherical GEM bases
of Eqs.~(\ref{eq:basis},\ref{eq:basis-param}),
and compared with the exact ones
$\epsilon^\mathrm{exact}(n_\perp n_z)
=\omega_\perp(n_\perp+1)+\omega_z(n_z+\frac{1}{2})$ ($n_\perp, n_z= 0,1,\cdots$).

\begin{figure}
\centerline{\includegraphics[scale=0.5]{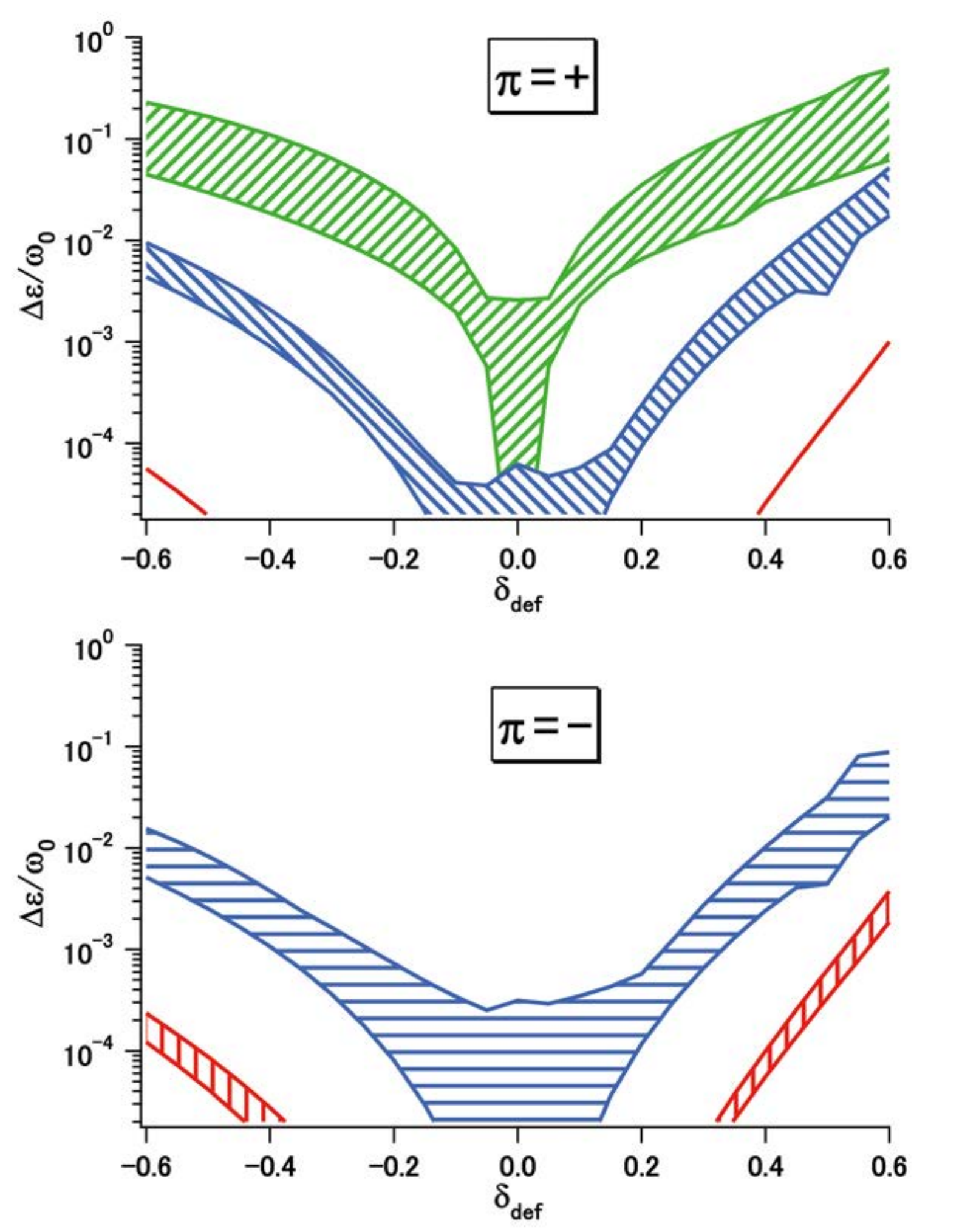}}
\vspace{1cm}
\caption{Errors of the s.p. energies of the anisotropic harmonic oscillator
  $\mathit{\Delta}\epsilon(n_\perp n_z)$ for $\ell_\mathrm{cut}=7$.
  Upper panel: Errors for the $\pi=+$ levels;
  red, blue and green areas show bounds of errors
  for the $N_\mathrm{osc}=0$, $2$ and $4$ levels.
  Lower panel: Errors for the $\pi=-$ levels;
  red and blue areas represent bounds of errors
  for the $N_\mathrm{osc}=1$ and $3$ levels.
  Quote from Ref.~\protect\refcite{Nakada_2008}.
\label{fig:def-ho}}
\end{figure}

In Fig.~\ref{fig:def-ho}, errors of the energy eigenvalues of the GEM,
$\mathit{\Delta}\epsilon(n_\perp n_z)=
\epsilon^\mathrm{GEM}(n_\perp n_z)-\epsilon^\mathrm{exact}(n_\perp n_z)$,
obtained with the basis-set of Eq.~(\ref{eq:basis-param})
and $\ell_\mathrm{cut}=7$, are plotted.
For $h$ of (\ref{eq:AHO}),
$(\omega_\perp^2 \omega_z)^{1/3}=41.2\,A^{-1/3}\,\mathrm{MeV}$
with $A=24$ is assumed at each $\delta_\mathrm{def}$,
whereas the relative errors are insensitive to $\omega_0$.
The parameters for the GEM bases are those of (\ref{eq:basis-param}).
At $\delta_\mathrm{def}=0$,
$\mathit{\Delta}\epsilon(n_\perp n_z)$ is irrelevant to the $\ell$-truncation,
coming only from the radial part in Eq.~(\ref{eq:basis-param}).
The small errors at $\delta_\mathrm{def}=0$ indicate that
the errors due to the $\ell$-truncation dominate
those at $\delta_\mathrm{def}\ne 0$.

If the criterion $\mathit{\Delta}\epsilon\lesssim 0.01\,\omega_0$ is imposed,
which yields $\mathit{\Delta}\epsilon\lesssim 0.1\,\mathrm{MeV}$
for $\omega_0\sim 10\,\mathrm{MeV}$,
the present GEM set satisfies it for all the $N_\mathrm{osc}\leq 3$ levels
at $-0.5\lesssim\delta_\mathrm{def}\lesssim +0.35$.
This bound correlates to the crossing point
with the higher $N_\mathrm{osc}$ levels,
beyond which the $\ell$-truncation is not justified.
In Table~\ref{tab:def-conv},
convergence for $\ell_\mathrm{cut}$ is presented
for the HFB calculations of $^{24}$Mg and $^{40}$Mg with M3Y-P6.
Definition of $\sqrt{\bra r^2\ket}$ and $q_0$ will be given
in Eqs.~(\ref{eq:r2-matter},\ref{eq:q0}), respectively.
Though not fully convergent particularly for $^{40}$Mg,
influence on physical quantities is weak already with $\ell_\mathrm{cut}=7$.
In the following calculations,
I adopt $\ell_\mathrm{cut}=\max(N_\mathrm{osc}^\mathrm{F}+2,7)$
when the spherical symmetry is assumed,
and $\ell_\mathrm{cut}=\max(N_\mathrm{osc}^\mathrm{F}+4,7)$ otherwise.
Normally-deformed nuclei can be described to reasonable precision
with this $\ell_\mathrm{cut}$.
For stronger deformation,
it would be more suitable to adopt deformed GEM bases.

\begin{table}[pt]
  \caption{Convergence of binding energies $-E$,
    r.m.s. matter radii $\sqrt{\bra r^2\ket}$
    and mass quadrupole moments $q_0$ for $\ell_\mathrm{cut}$.
    Shown are the HFB results of $^{24}$Mg and $^{40}$Mg with M3Y-P6.
 \label{tab:def-conv}}
\centerline
{\begin{tabular}{cccrrrr}
\toprule 
&&&~~$\ell_\mathrm{cut}=7$~~&~~$\ell_\mathrm{cut}=8$~~
&~~$\ell_\mathrm{cut}=9$~~&~$\ell_\mathrm{cut}=10$~~\\ \colrule 
$^{24}$Mg & $-E$ &(MeV)& $189.815$ & $189.817$ & $189.818$ & $189.818$ \\
& $\sqrt{\bra r^2\ket}$ &(fm)& $2.999$ & $2.999$ & $2.999$ & $2.999$ \\
& $q_0$ &(fm$^2$)& $118.3$ & $118.3$ & $118.3$ & $118.3$ \\
$^{40}$Mg & $-E$ &(MeV)& $257.796$ & $257.880$ & $257.949$ & $258.005$ \\
& $\sqrt{\bra r^2\ket}$ &(fm)& $3.661$ & $3.664$ & $3.666$ & $3.669$ \\
& $q_0$ &(fm$^2$)& $219.8$ & $218.3$ & $217.2$ & $216.6$ \\
\botrule 
\end{tabular}}
\end{table}

\section{Magic numbers off the $\beta$-stability}\label{sec:magic}

In arguing shell structure in nuclei, including magic numbers,
the s.p. orbits should be defined properly.
In the HF theory, the s.p. orbits are formed from scratch
so as to fulfill the variational principle
and the HF condition~\cite{Ring-Schuck}.
The s.p. orbitals are determined self-consistently,
without artificial postulate.
The s.p. (or q.p.) states under the presence of the pair correlation
can be investigated via the HFB approaches.
Thus the SCMF theory supplies a framework
desirable to investigate nuclear shell structure
and its evolution from stable to unstable nuclei,
up to effects of the nucleonic interaction.

In this section I shall show magic numbers predicted by the SCMF calculations
mainly with the M3Y-P6 semi-realistic interaction
and related topics,
paying particular attention to effects of 
$v^{(\mathrm{TN})}$ and $v^{(\mathrm{C})}_\mathrm{OPEP}$.
The M3Y-P6a interaction is applied
to a specific problem in Subsec.~\ref{subsec:r_c}.

\subsection{Performance for doubly magic nuclei and pairing properties}
\label{subsec:doubly&pairing}

Before arguing magic numbers off the $\beta$-stability,
the appropriateness of the effective interactions is tested
for known doubly magic nuclei.
It is reasonably expected that the spherical HF calculations give
a good approximation for the doubly magic nuclei.
In Table~\ref{tab:DMprop},
the spherical HF results with M3Y-P6 and D1S are tabulated
for the binding energies and the r.m.s. radii,
the charge radii as well as the matter radii,
in comparison with the experimental data~\cite{AtomicMass_2003,
  Khoa-Than-Grasso_1978,Ozawa-Suzuki-Tanihata_2001,
  Alkhazov-Belostotsky-Vorobyov_1978,Angeli-Marinova_2013}.
The matter radii are calculated by~\cite{Nakada-Sato_2002}
\be \bra r^2\ket = \frac{1}{A}
 \Big\bra\Phi\Big|\sum_{i=1}^A (\vect{r}_i-\vect{R})^2\Big|\Phi\Big\ket
 = \frac{1}{A}\Big\bra\Phi\Big| \sum_{i=1}^A \vect{r}_i^2\Big|\Phi\Big\ket
 - \bra\Phi|\vect{R}^2|\Phi\ket\,,
\label{eq:r2-matter}\ee
and the charge radii by~\cite{Friar-Negele_1975}
\be\begin{split} \bra r^2\ket_c =& \frac{1}{Z}\bigg[
 \Big\bra\Phi\Big|\sum_{i\in p}(\vect{r}_i-\vect{R})^2\Big|\Phi\Big\ket
 + Z\,\bra r^2_p\ket_c + N\,\bra r^2_n\ket_c \\
 &\quad + \frac{1}{M^2}\sum_{\tau=p,n}(2\mu_\tau-e_\tau)\,
 \Big\bra\Phi\Big|\sum_{i\in\tau}\vect{\ell}_i\cdot\vect{s}_i\Big|\Phi\Big\ket
 \bigg]\,,
\end{split}\label{eq:r2-charge}\ee
where $\bra r^2_\tau\ket_c$
is the mean-square (m.s.) charge radius of a single nucleon $\tau\,(=p,n)$,
$\mu_\tau$ is the magnetic moment of $\tau$
in the unit of $\mu_N$~\cite{PDG_2018},
and $e_\tau\,(=1,0)$ is the electric charge of $\tau$.
Note that,
while the data on the binding energies and the charge radii are very accurate,
the matter radii are extracted from hadronic scatterings
through some reaction models and are not always accurate.
Some of the parameters in the effective interactions for the SCMF calculations
were fitted to these energies and radii.
For instance, binding energies and radii of $^{16}$O and $^{208}$Pb were used
when fixing the parameters in M3Y-P6.
It is still worth pointing out
that the SCMF calculations are capable of describing these properties
in a vast range of $A$,
covering $A\approx 10$ to $A\gtrsim 200$ at narrowest.

\begin{table}[pt]
 \caption{Binding energies ($-E$), r.m.s. matter ($\sqrt{\bra r^2\ket}$)
   and charge ($\sqrt{\bra r^2\ket_c}$) radii of several doubly magic nuclei.
 Experimental data are taken
 from Refs.~\protect\refcite{AtomicMass_2003,Khoa-Than-Grasso_1978,Ozawa-Suzuki-Tanihata_2001,Alkhazov-Belostotsky-Vorobyov_1978,Angeli-Marinova_2013}.
 \label{tab:DMprop}}
\centerline
{\begin{tabular}{cccrrrr}
\toprule 
&&&~~~Exp.~~&~~M3Y-P6~&~~~~~D1S~~\\ \colrule 
$^{16}$O & $-E$ &(MeV)& $127.6$ & $126.3$ & $129.5$ \\
& $\sqrt{\bra r^2\ket}$ &(fm)& $2.61$ & $2.59$ & $2.61$ \\
& $\sqrt{\bra r^2\ket_c}$ &(fm)& $2.70$ & $2.71$ & $2.73$ \\
$^{24}$O & $-E$ &(MeV)& $168.5$ & $166.2$ & $168.6$ \\
& $\sqrt{\bra r^2\ket}$ &(fm)& $3.19$ & $3.05$ & $3.01$ \\
& $\sqrt{\bra r^2\ket_c}$ &(fm)& --- & $2.76$ & $2.77$ \\
$^{40}$Ca & $-E$ &(MeV)& $342.1$ & $335.9$ & $344.6$ \\
& $\sqrt{\bra r^2\ket}$ &(fm)& $3.47$ & $3.37$ & $3.37$ \\
& $\sqrt{\bra r^2\ket_c}$ &(fm)& $3.48$ & $3.47$ & $3.48$ \\
$^{48}$Ca & $-E$ &(MeV)& $416.0$ & $413.8$ & $416.8$ \\
& $\sqrt{\bra r^2\ket}$ &(fm)& $3.57$ & $3.51$ & $3.51$ \\
& $\sqrt{\bra r^2\ket_c}$ &(fm)& $3.48$ & $3.48$ & $3.49$ \\
$^{56}$Ni & $-E$ &(MeV)& $484.0$ & $473.7$ & $483.8$ \\
& $\sqrt{\bra r^2\ket}$ &(fm)& --- & $3.65$ & $3.64$ \\
& $\sqrt{\bra r^2\ket_c}$ &(fm)& --- & $3.76$ & $3.75$ \\
$^{90}$Zr & $-E$ &(MeV)& $783.9$ & $781.1$ & $785.9$ \\
& $\sqrt{\bra r^2\ket}$ &(fm)& $4.32$ & $4.23$ & $4.24$ \\
& $\sqrt{\bra r^2\ket_c}$ &(fm)& $4.27$ & $4.25$ & $4.26$ \\
$^{100}$Sn & $-E$ &(MeV)& $824.8$ & $822.5$ & $831.6$ \\
& $\sqrt{\bra r^2\ket}$ &(fm)& --- & $4.36$ & $4.36$ \\
& $\sqrt{\bra r^2\ket_c}$ &(fm)& --- & $4.47$ & $4.47$ \\
$^{132}$Sn & $-E$ &(MeV)& $1102.9$ & $1097.8$ & $1104.1$ \\
& $\sqrt{\bra r^2\ket}$ &(fm)& --- & $4.78$ & $4.77$ \\
& $\sqrt{\bra r^2\ket_c}$ &(fm)& $4.71$ & $4.70$ & $4.70$ \\
$^{208}$Pb & $-E$ &(MeV)& $1636.4$ & $1634.5$ & $1639.0$ \\
& $\sqrt{\bra r^2\ket}$ &(fm)& $5.49$ & $5.53$ & $5.51$ \\
& $\sqrt{\bra r^2\ket_c}$ &(fm)& $5.50$ & $5.48$ & $5.48$ \\
\botrule 
\end{tabular}}
\end{table}

\begin{figure}
\centerline{\includegraphics[scale=0.7]{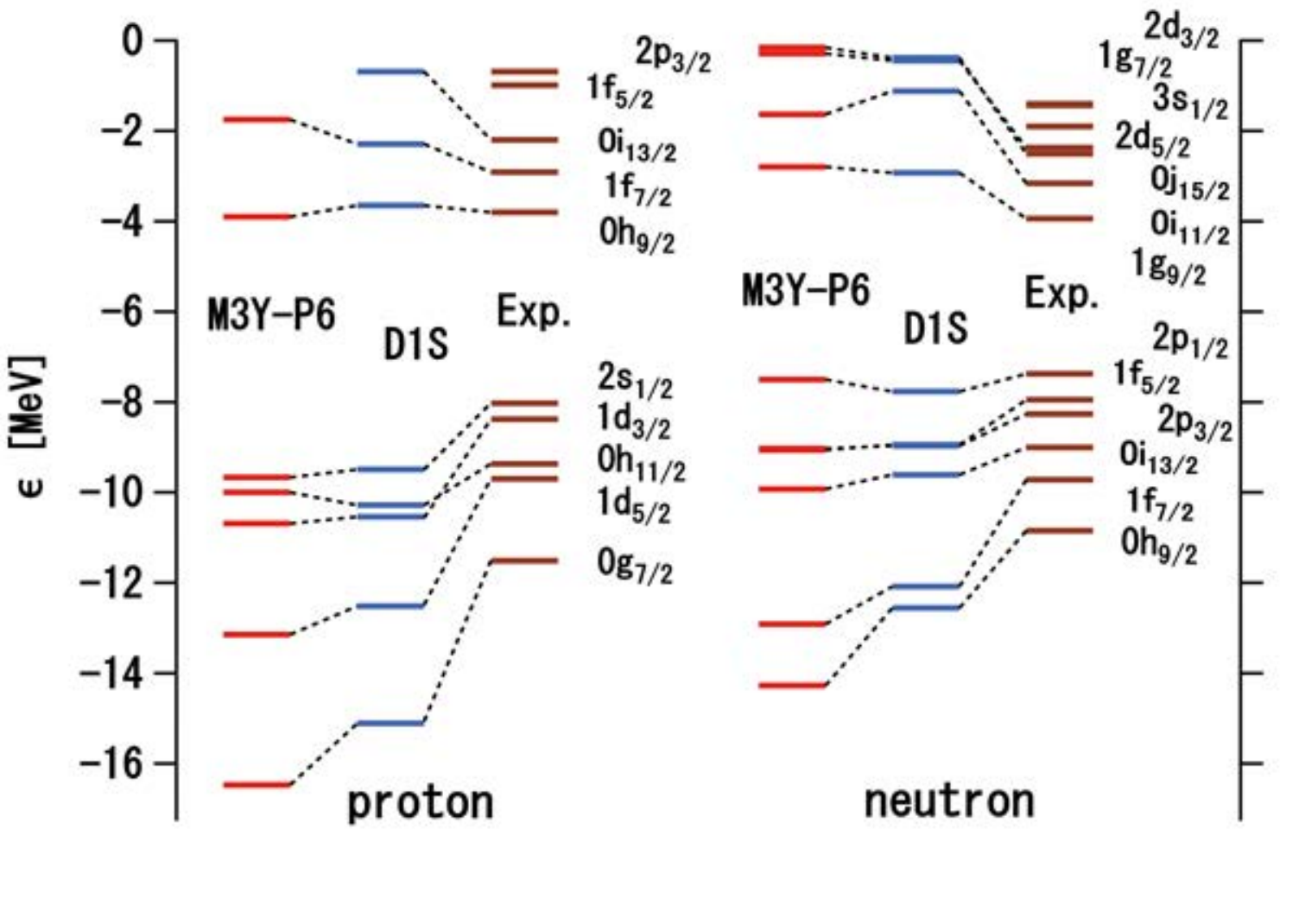}}
\caption{Single-particle levels for $^{208}$Pb.
  Those obtained with D1S and M3Y-P6 are compared with the data,
  which are extracted from the lowest-lying levels~\protect\cite{
    AtomicMass_2003,TableOfIsotopes}.
\label{fig:Pb_spe}}
\end{figure}

The s.p. levels around the Fermi level in $^{208}$Pb are depicted
in Fig.~\ref{fig:Pb_spe}.
The SCMF results are compared with the experimental levels.
The experimental levels are extracted from the difference
of the binding energies between $^{208}$Pb
and its neighboring nuclei $^{207}$Tl, $^{209}$Bi,
$^{207,209}$Pb~\cite{AtomicMass_2003},
and the excitation energies of these neighbors~\cite{TableOfIsotopes}.
I here used only the lowest states of individual spin-parity,
which carry dominant portions of the s.p. strengths.
However, to be more precise,
the s.p. strengths are fragmented due to correlations,
in connection to the effect of the so-called $\omega$-mass~\cite{
  Mahaux_1985,Migdal_1967}.
Wider level spacing in the HF results than in the data
is attributed to this effect.
It is confirmed that the SCMF calculations well describe
the sequence of the s.p. levels.

\begin{figure}
\hspace*{1cm}\includegraphics[scale=0.5]{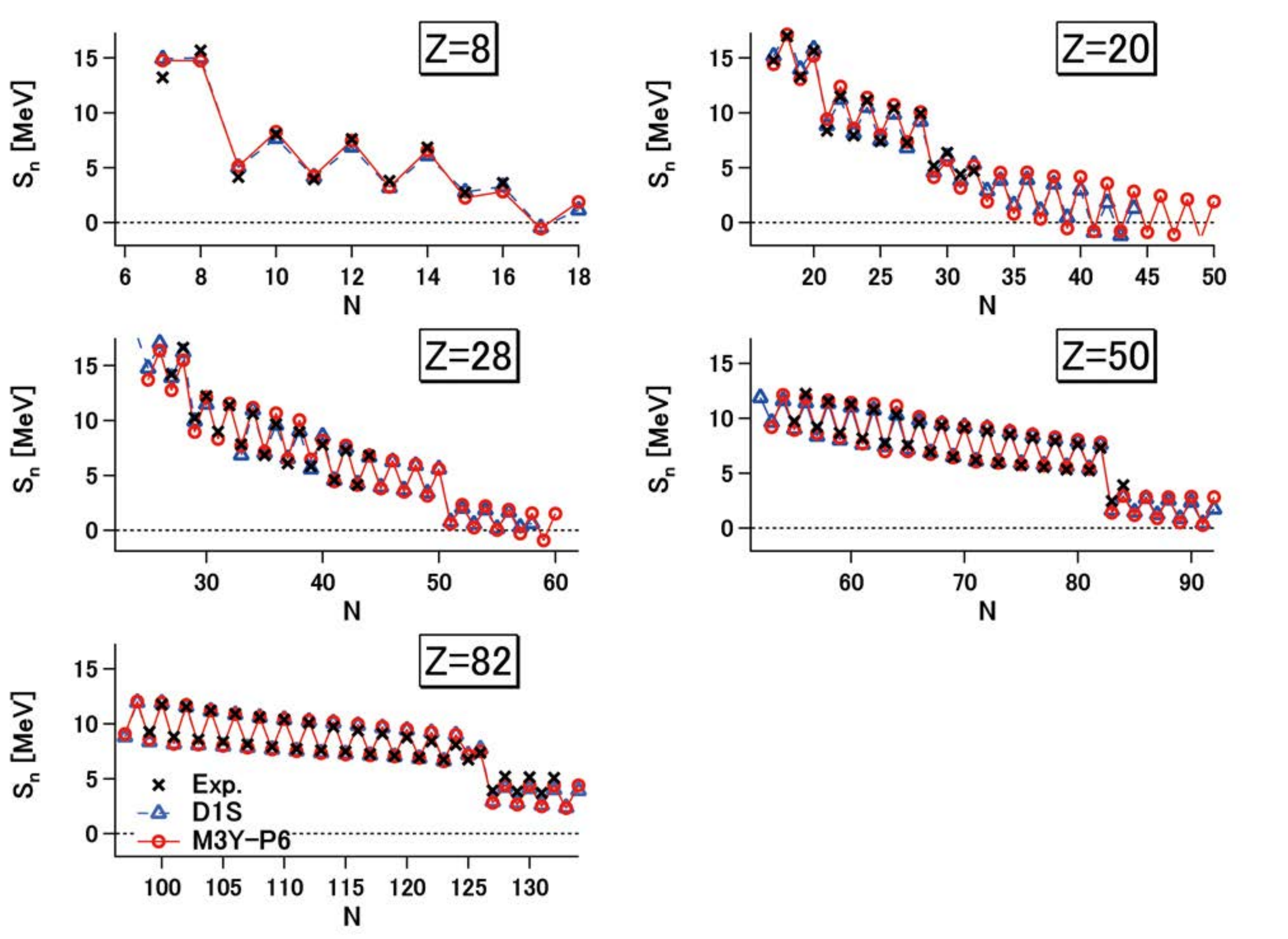}\vspace*{1cm}
\caption{Neutron separation energies
for $Z=8$, $20$,$28$, $50$ and $82$ nuclei,
calculated with D1S (blue triangles) and M3Y-P6 (red circles).
Lines are drawn to guide eyes.
Experimental values are taken from Ref.~\protect\refcite{AtomicMass_2003}
and presented by the crosses.
\label{fig:Sn}}
\end{figure}

\begin{figure}
\hspace*{1cm}\includegraphics[scale=0.5]{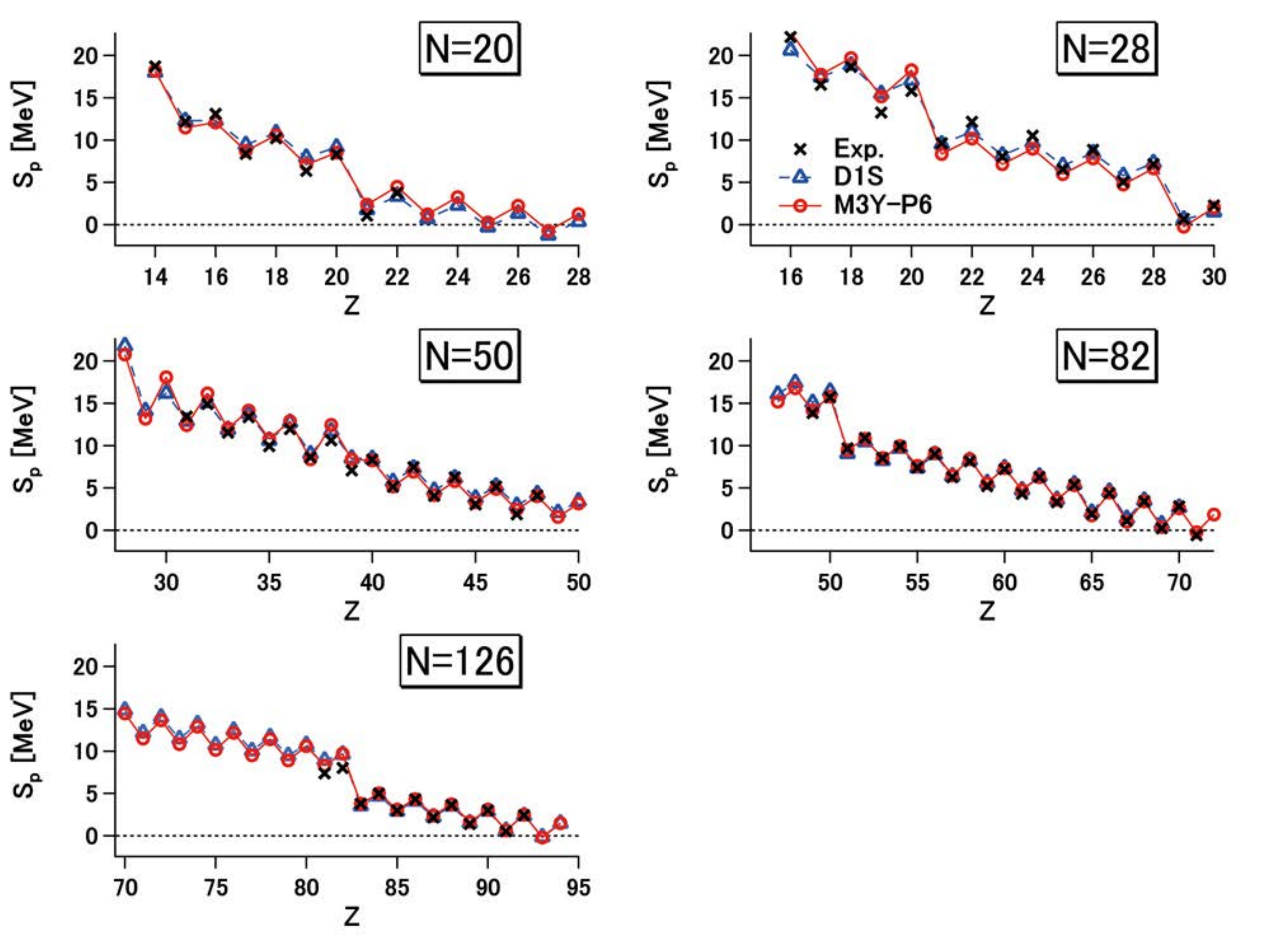}\vspace*{1cm}
\caption{Proton separation energies
for $N=20$, $28$, $50$, $82$ and $126$ nuclei.
See Fig.~\protect\ref{fig:Sn} for the conventions.
\label{fig:Sp}}
\end{figure}

Pair correlations enter the g.s. of nuclei with non-magic $Z$ or $N$.
Nuclei having magic $Z$ ($N$) and non-magic $N$ ($Z$)
usually keep the spherical shape,
but gains pair correlation among neutrons (protons).
They serve as a good testing ground for the pairing channel
of the effective interaction.
Pairing properties of such nuclei obtained by the spherical HFB calculations
are shown in Figs.~\ref{fig:Sn} and \ref{fig:Sp}.
The pairing gaps are frequently used to check the pairing properties,
for which experimental information is obtained
by differentiating the binding energies of the adjacent nuclei.
However, since it is not easy to remove ambiguities
in extracting the pairing gaps both experimentally and theoretically,
the one-neutron (one-proton) separation energies $S_n$ ($S_p$)
are displayed here,
whose staggering depending on even or odd $N$ ($Z$)
represents the degree of the pair correlation.
The so-called three-point formula for the pairing gap~\cite{Bohr-Mottelson_1}
corresponds to the difference of $S_n$ or $S_p$ between the neighboring nuclei.
It is confirmed from Figs.~\ref{fig:Sn} and \ref{fig:Sp}
that the pairing channel of M3Y-P6 is comparably good to that of D1S.

\subsection{Tensor-force effects on isotopic variation
  of proton-hole energies in Ca}
\label{subsec:tensor}

Effects of the tensor force are among the topics of this review.
I first summarize what has been pointed out as tensor-force effects
on nucleons occupying spherical s.p. orbitals~\cite{Otsuka-etal_2005}.
These points are vital to understanding the arguments below.
\begin{romanlist}
\item\label{it:TNS-pn}
  The tensor force acts primarily between protons and neutrons.
\item\label{it:TNS-close}
  The tensor-force effects vanish when all the $\ell s$ partners,
  \textit{i.e.} pairs of the $j=\ell\pm 1/2$ orbitals, are filled.
\item\label{it:TNS:ls}
  When protons occupy a $j=\ell+1/2$ orbit but not its $\ell s$ partner,
  the tensor force acts attractively (repulsively)
  on neutrons occupying a $j'=\ell'-1/2$ ( $j'=\ell'+1/2$) orbit,
  and \textit{vice versa}.
\end{romanlist}
The point (\ref{it:TNS-close}),
which was given in Ref.~\refcite{Otsuka-etal_2005},
has been proven in more general cases~\cite{Suzuki-Nakada-Miyahara_2016}
that the tensor force effects are canceled
in the spin-saturated systems.

As shown in Fig.~\ref{fig:Pb_spe} for $^{208}$Pb,
the s.p. states obtained in the HF calculations do not directly correspond
to the individual observed states
even in the vicinity of doubly magic nuclei.
However, there are some cases
in which the spectroscopic factors have been measured for fragmented states
and their sum reaches unity to good precision.
The proton-hole states near $^{40}$Ca and $^{48}$Ca are
noteworthy examples~\cite{Doll-Wagner-Knopfle-Mairle_1976,Ogilvie-etal_1987}.
By averaging the energies weighted by the spectroscopic factors,
experimental s.p. energies $\epsilon(p1s_{1/2})$ and $\epsilon(p0d_{3/2})$
are evaluated,
which may be compared to the s.p. energies calculated in the HF approaches.
Interestingly, $p1s_{1/2}$ and $p0d_{3/2}$ are inverted
from $^{40}$Ca to $^{48}$Ca.

\begin{figure}
\centerline{\includegraphics[scale=0.5]{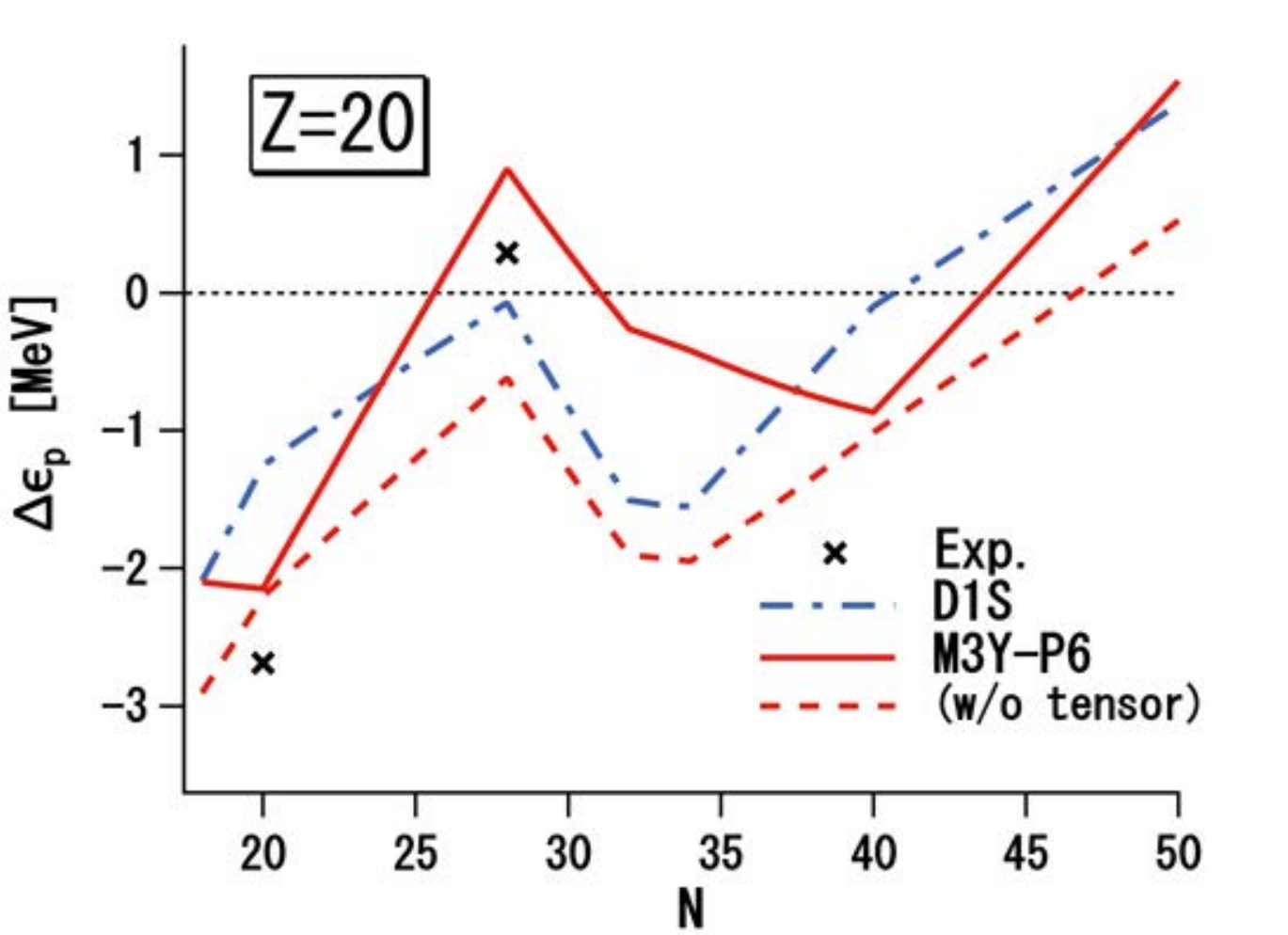}}
\caption{Difference of the s.p. energies between $p1s_{1/2}$ and $p0d_{3/2}$
  ($\mathit{\Delta}\epsilon_p$) in the Ca isotopes.
  Blue and red lines represent the results
  of the spherical HF calculations with D1S and M3Y-P6, respectively.
  Red dashed line is obtained from M3Y-P6
  but by removing the contribution of the tensor force;
  \textit{i.e.}
  $\mathit{\Delta}\epsilon_p-\mathit{\Delta}\epsilon_p^{(\mathrm{TN})}$.
  Crosses are experimental values at $^{40}$Ca and $^{48}$Ca,
  which are averages weighted by the spectroscopic factors~\cite{
    Doll-Wagner-Knopfle-Mairle_1976,Ogilvie-etal_1987}.
\label{fig:de13}}
\end{figure}

The s.p. energy difference under interest is denoted simply by
$\mathit{\Delta}\epsilon_p=\epsilon(p1s_{1/2})-\epsilon(p0d_{3/2})$.
Figure~\ref{fig:de13} shows $N$-dependence of $\mathit{\Delta}\epsilon_p$
in the Ca isotopes
obtained by the spherical HF calculations,
in comparison with the experimental values in $^{40}$Ca and $^{48}$Ca.
$N$-dependence of $\mathit{\Delta}\epsilon_p$ with other interactions
is found in Refs.~\refcite{Grasso-etal_2007}
and \refcite{Wang-Gu-Zhang-Dong_2011}.
Although $\mathit{\Delta}\epsilon_p$ rapidly changes
between $^{40}$Ca and $^{48}$Ca,
most interactions without explicit tensor force, including D1S,
fail to reproduce the slope of $\mathit{\Delta}\epsilon_p$.
With D1S, the inversion of $p1s_{1/2}$ and $p0d_{3/2}$ is not described.
If D1M is used, $\mathit{\Delta}\epsilon_p$ slightly shifts upward
and the inversion is obtained,
but at the expense of discrepancy worse at $^{40}$Ca.
In contrast, M3Y-P6 gives the slope of $\mathit{\Delta}\epsilon_p$
quite close to the experimental one,
reproducing the inversion of $p1s_{1/2}$ and $p0d_{3/2}$.
Figure~\ref{fig:de13} also shows the s.p. energy difference
after removing the contribution of the tensor force from the M3Y-P6 result,
$\mathit{\Delta}\epsilon_p-\mathit{\Delta}\epsilon_p^{(\mathrm{TN})}$.
This $\mathit{\Delta}\epsilon_p-\mathit{\Delta}\epsilon_p^{(\mathrm{TN})}$
varies almost in parallel to $\mathit{\Delta}\epsilon_p$ with D1S,
confirming that $v^{(\mathrm{TN})}$ plays a crucial role
in the $N$-dependence of $\mathit{\Delta}\epsilon_p$.
The variation of $\mathit{\Delta}\epsilon_p$ from $N=20$ to $28$
is a result of the occupation of the $n0f_{7/2}$ orbit.
Recall the point (\ref{it:TNS:ls}) above.
As $n0f_{7/2}$ is occupied,
the tensor force lowers $p0d_{3/2}$ but not $p1s_{1/2}$,
resultantly raising $\mathit{\Delta}\epsilon_p$.
It is emphasized that the realistic tensor force
is consistent with the observed variation of $\mathit{\Delta}\epsilon_p$
from $^{40}$Ca to $^{48}$Ca.
Similar results have been obtained with the other parameter-sets
of the M3Y-type interaction~\cite{Nakada-Sugiura-Margueron},
and with the SLy5+$T_w$ interaction~\cite{Wang-Gu-Zhang-Dong_2011,Bai-etal_2010}
which contains the zero-range tensor force
determined from the $G$-matrix~\cite{Stancu-Brink-Flocard_1977}.

It should be commented that the experimental s.p. energies could be influenced
if tiny missing spectroscopic strengths lie at high energies.
Still, $\mathit{\Delta}\epsilon_p$ can be relatively insensitive
to such strengths,
as they tend to be canceled
between $\epsilon(p1s_{1/2})$ and $\epsilon(p0d_{3/2})$
after taking the difference.


Figure~\ref{fig:de13} exhibits $\mathit{\Delta}\epsilon_p$
below $N=20$ and above $N=28$, as well.
The tensor force greatly affects
the $N$-dependence of $\mathit{\Delta}\epsilon_p$ in $N<20$ and $32<N<40$.
The former is accounted for by the occupation of $n0d_{3/2}$,
and the latter by the occupation of $n0f_{5/2}$.
Although another inversion of $p1s_{1/2}$ and $p0d_{3/2}$ is predicted
on the way to $^{70}$Ca,
it is delayed due to the tensor force.

\subsection{Chart of magic numbers}\label{subsec:chart}

It is of interest whether and how well the SCMF frame
with a specific effective interaction describe
the magic numbers indicated by experiments,
and what numbers are predicted to be magic in the region
where experiments have not reached.

Magic numbers have been experimentally identified
by irregularities in energies indicating relative stability.
Theoretically, a magic number is a result
of the quenching of the many-body correlations,
which takes place owing to the large shell gap.
Without a clear definition,
it is reasonable to identify magic numbers in theoretical studies
when the spherical HF solution gives the absolute energy minimum,
as realized in the doubly magic nuclei.
Nuclei near magic numbers usually keep the spherical or near-spherical shape.
For spherical nuclei,
the pairing among like nucleons provides a dominant correlation beyond the HF.
Therefore it should be a good first step to investigate magic numbers
through the stability of the spherical HF state against the pair correlation.

\begin{figure}[!hp]
\centerline{\includegraphics[scale=0.65,angle=90]{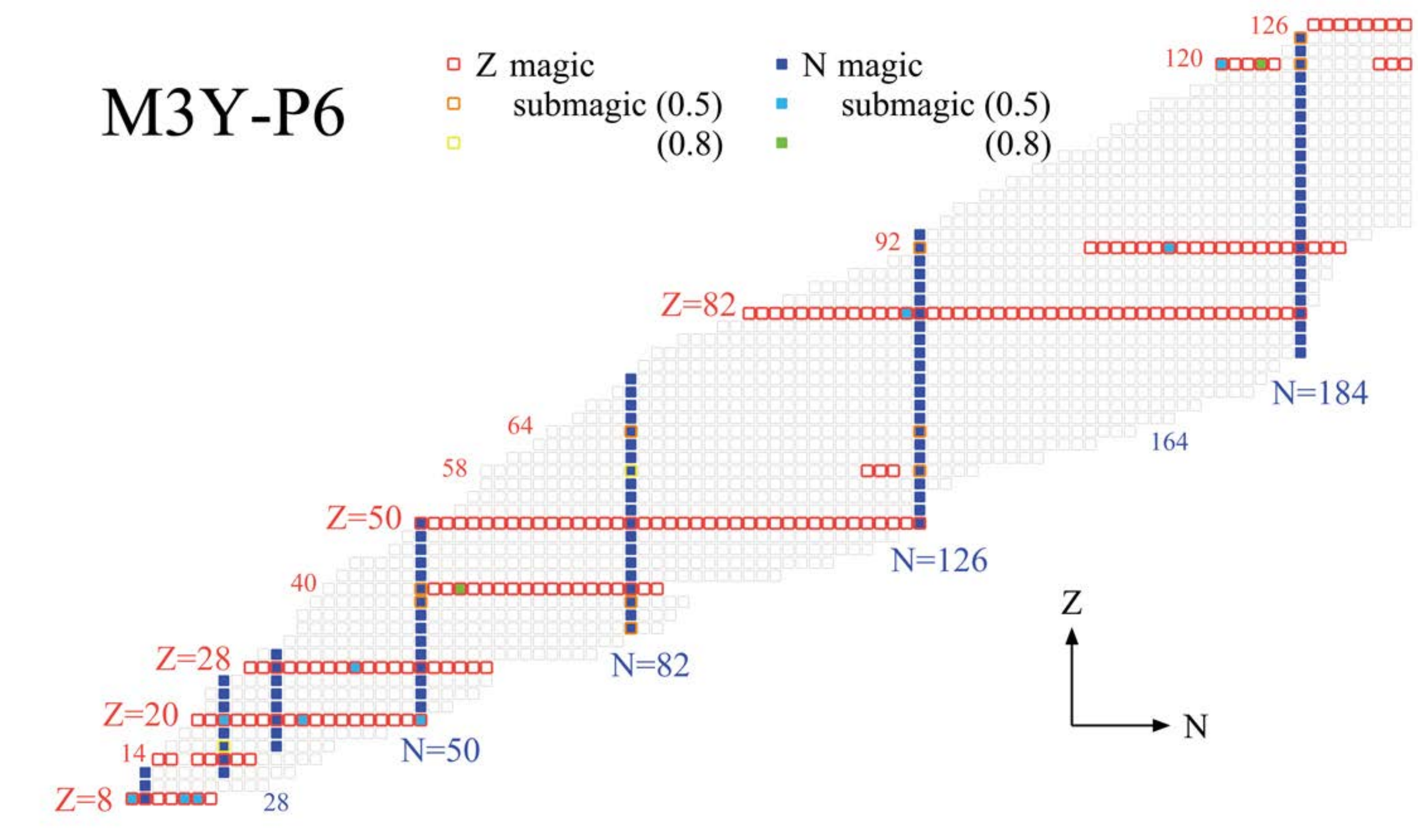}}
\caption{Chart showing magic numbers predicted with the M3Y-P6 interaction.
Individual boxes correspond to even-even nuclei.
Magic (submagic) $Z$'s are represented by the red-colored
(orange- or yellow-colored) frame,
and magic (submagic) $N$'s by filling the box with the blue
(skyblue or green) color.
The $\lambda_\mathrm{sub}$ values for the submagic numbers (in MeV)
are as parenthesized.
Quote from Ref.~\protect\refcite{Nakada-Sugiura_2014}.}
\label{fig:magic_M3Yp6}
\end{figure}

In Fig.~\ref{fig:magic_M3Yp6},
a chart of magic numbers is drawn in the region $8\leq Z\leq 126,~ N\leq 200$,
by comparing the spherical HF and the HFB results with M3Y-P6.
Each box stands for an even-even nucleus.
The proton magicity is displayed by the colored frames,
and the neutron magicity by the filled boxes.
Since deformation is not taken into account,
the boundaries in Fig.~\ref{fig:magic_M3Yp6} are somewhat arbitrary,
not precisely representing the drip lines.
$Z$ ($N$) is identified to be magic
when the proton (neutron) pair correlation vanishes
in the spherical HFB calculation.
In addition, the number $Z$ ($N$) is regarded as submagic,
when the HF and HFB energies, $E^\mathrm{HF}$ and $E^\mathrm{HFB}$,
are close in the magic $N$ ($Z$) nucleus,
even if the proton (neutron) pair correlation does not entirely vanish.
Correlation is suppressed in these nuclei ,
quite possibly resulting in \textit{e.g.} high excitation energy,
as at $^{68}$Ni~\cite{Nakada_2010b}
and $^{146}$Gd~\cite{Matsuzawa-Nakada-Ogawa-Momoki,
  Matsuzawa-Nakada-Ogawa-Momoki_erratum}.
In practice, submagic numbers are identified for a nucleus
in which $E^\mathrm{HF}-E^\mathrm{HFB}$,
\textit{i.e.} energy gain due to the pair correlation,
is narrower than a certain value $\lambda_\mathrm{sub}$.
In Fig.~\ref{fig:magic_M3Yp6}
results with $\lambda_\mathrm{sub}=0.5\,\mathrm{MeV}$ and $0.8\,\mathrm{MeV}$
are presented.
For further detail, see Ref.~\refcite{Nakada-Sugiura_2014}.

It is found that $N=14,16$ ($N=32$) are picked up as submagic numbers
in the O (Ca) isotopes,
while the $N=20$ ($28$) magicity is lost at $^{30}$Ne ($^{40}$Mg and $^{42}$Si).
$N=40$ is indicated to be submagic at $^{68}$Ni, but not at $^{60}$Ca.
$Z=38$ at $^{88}$Sr, $Z=40$ at $^{90}$Zr, and $Z=64$ at $^{146}$Gd
are categorized as submagic numbers,
as have been established experimentally.
$N=56$ at $^{96}$Zr and $Z=58$ at $^{140}$Ce are also indicated
to be submagic,
which are in harmony with their high first excitation energies.
In the superheavy region, no magicity is found at $Z=114$,
while $Z=120$ can be magic, depending on $N$.
The conventional magic numbers $N=50,82,126$ seem to hold
even in the proton-deficient region,
indicating no drastic influence on our understanding of the path
of the $r$-process nucleosynthesis~\cite{Iliadis_2007}.
$N=184$ is predicted to be a stable magic number as well.

It is remarked that the distribution of magic numbers predicted with M3Y-P6
is compatible with most experimental data.
Although similar calculations were carried out
also with M3Y-P7, D1M and D1S~\cite{Nakada-Sugiura_2014},
none of them coincide with the experimental information
at the comparable level to M3Y-P6.
Besides the difference between M3Y-P6 and P7,
a part of the reason is attributable to the tensor force. 
There are two regions in which the results of Fig.~\ref{fig:magic_M3Yp6}
contradict to the experiments.
One is a single nucleus $^{32}$Mg,
at which the $N=20$ magicity is known to be break down~\cite{
  GuillemaudMueller-etal_1984}
while given as magic in Fig.~\ref{fig:magic_M3Yp6}.
The other is the $60\leq N\leq 70$ region of Zr,
in which $Z=40$ is predicted be magic
but deformation has been established experimentally~\cite{
  Cheifetz-Jared-Thompson-Wilhelmy_1970,Hotchkis-etal_1991,
  Sumikama-etal_2011,Paul-etal_2017}.
In both cases, the origin of the discrepancy is ascribed
to the quadrupole deformation,
which is a source of breaking of magicity independent of the pairing.
Investigation on these nuclei with taking account of the quadrupole deformation
will be shown in Sec.~\ref{subsec:def-TNS}.
It is also noted that by other SCMF studies deformation has been predicted
in neutron-rich Sn and Pb nuclei~\cite{D1S-results,Geng-Toki-Meng_2004},
for which no data are available at present.
Deformed SCMF calculations on them with M3Y-P6 are of interest.

\begin{figure}\begin{center}
\includegraphics[scale=0.6]{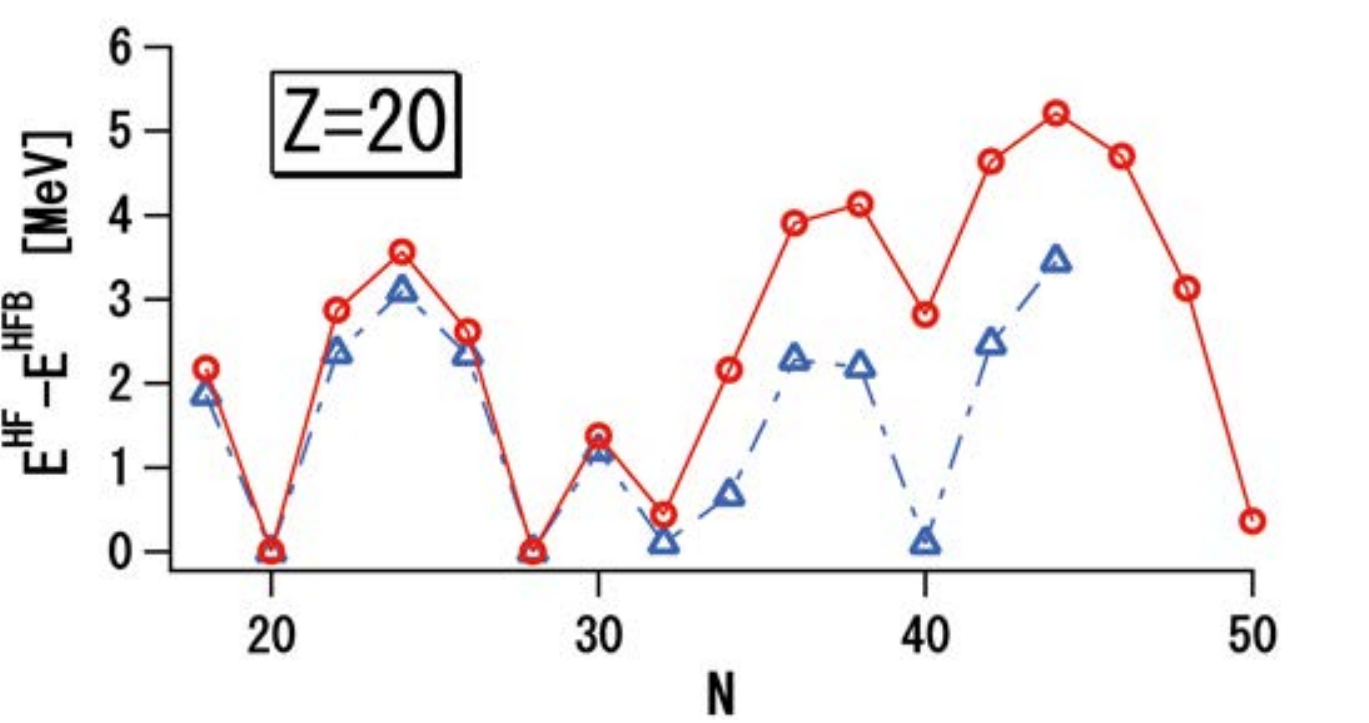}\\
\includegraphics[scale=0.6]{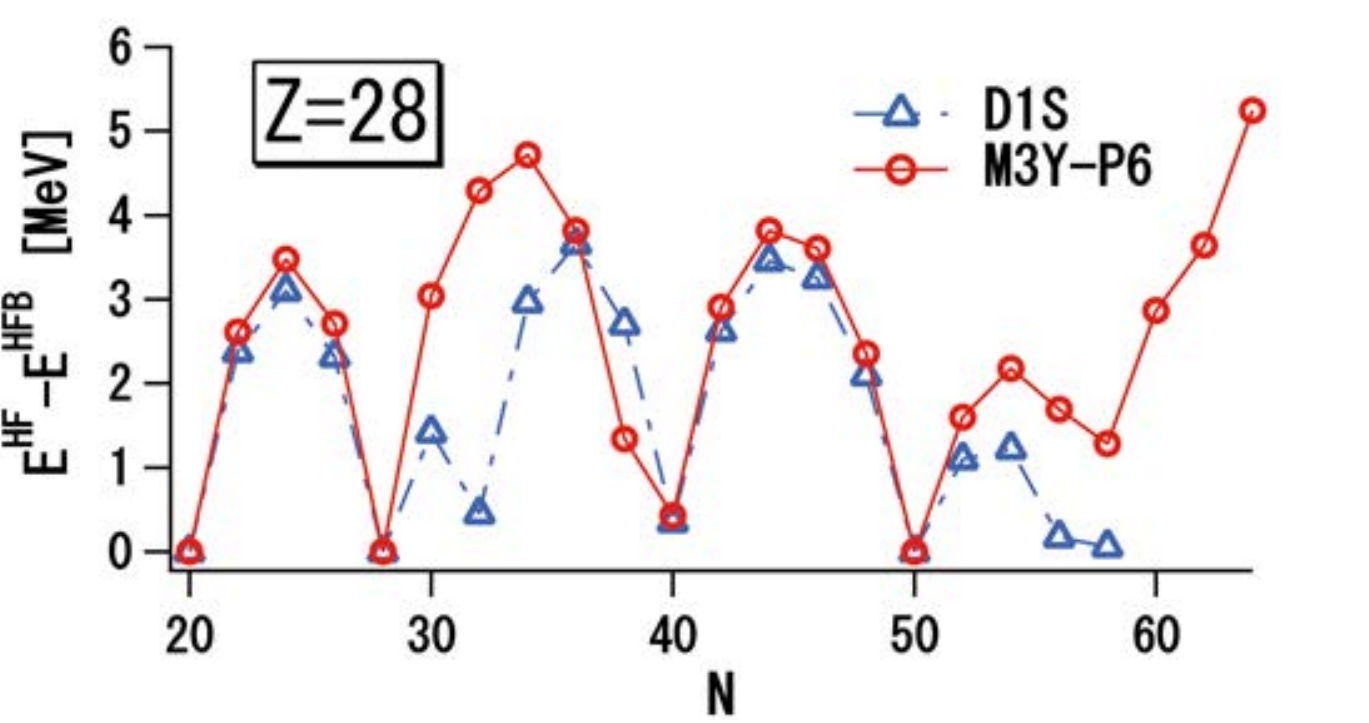}
\end{center}
  \caption{$E^\mathrm{HF}-E^\mathrm{HFB}$ for the Ca and the Ni nuclei.
    Red circles (blue triangles) are the results with M3Y-P6 (D1S).
  Lines are drawn to guide eyes.}
\label{fig:dE_HF-HFB}
\end{figure}

To discuss the magic and submagic numbers in the Ca and Ni region,
$E^\mathrm{HF}-E^\mathrm{HFB}$ is displayed in Fig.~\ref{fig:dE_HF-HFB}.
Results with M3Y-P6 are presented and compared to those with D1S.
In both Ca and Ni, the energy difference vanishes at $N=20$ and $28$
with both of the interactions.
M3Y-P6 predicts that $N=32$ and $40$ behave distinctively between Ca and Ni,
in contrast to D1S.
Whereas the pair correlation at $N=32$ is quenched for Ca,
it is sizable for Ni with M3Y-P6 though small with D1S.
While D1S predicts that $N=40$ is nearly magic both for Ca and Ni,
M3Y-P6 indicates that it is submagic for Ni but not for Ca.
For the neutron-rich Ni region,
the vanishing pair energy with D1S suggests that $N=56$ and $58$
could be magic or submagic.
The $N=58$ magicity is indicated with M3Y-P5$'$~\cite{Nakada_2010},
an older parameter-set of the M3Y-type interaction,
and with some of the Skyrme interactions~\cite{Terasaki-Engel_2006}.
With M3Y-P6, $E^\mathrm{HF}-E^\mathrm{HFB}$ takes
a distinctive local minimum at $N=58$,
though not vanishing.

\begin{figure}
\centerline{\includegraphics[scale=0.5]{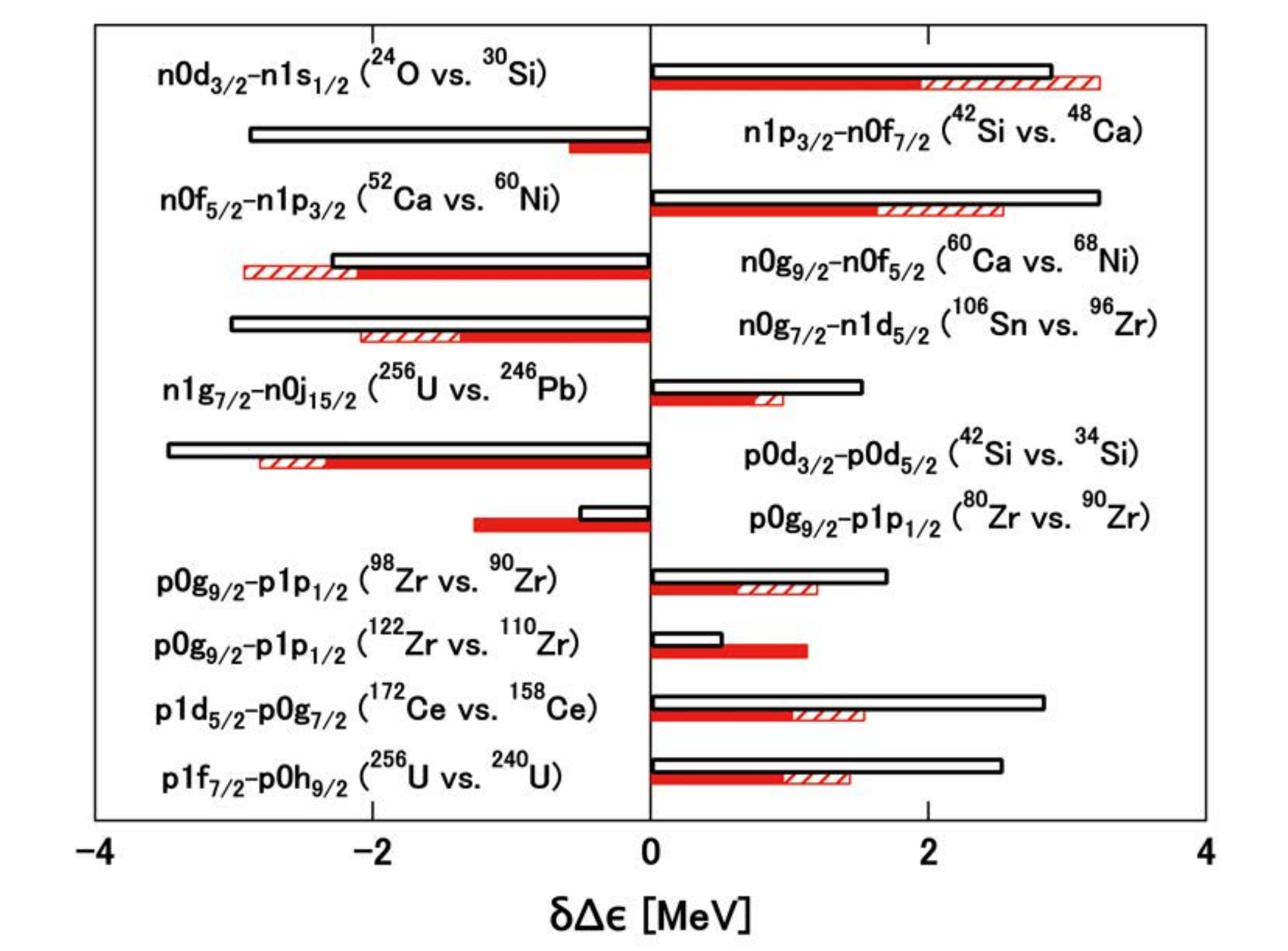}}
\caption{Difference of the shell gaps
 between two members of isotopes or isotones
 $\delta\mathit{\Delta}\epsilon(j_2\,\mbox{-}\,j_1)$ (open bars)
 obtained from the HF results with M3Y-P6.
 Contributions of $v^{(\mathrm{TN})}$ (red filled bars)
 and $v_\mathrm{OPEP}^{(\mathrm{C})}$ (red hatched bars) to it,
 $\delta\mathit{\Delta}\epsilon^{(\mathrm{TN})}(j_2\,\mbox{-}\,j_1)$ and
 $\delta\mathit{\Delta}\epsilon^{(\mathrm{OPEP})}(j_2\,\mbox{-}\,j_1)$,
 are also displayed.
Quote from Ref.~\protect\refcite{Nakada-Sugiura_2014_erratum}.}
\label{fig:ddspe_M3Yp6}
\end{figure}

To examine origin of the $Z$- or $N$-dependence of the magic numbers,
double difference of the s.p. energies is considered,
denoted by $\delta\mathit{\Delta}\epsilon(j_2\,\mbox{-}\,j_1)$;
$\epsilon(j_2)-\epsilon(j_1)$ at a certain nuclide $(Z_b,N_b)$
relative to that at a reference nuclide $(Z_a,N_a)$.
This quantity $\delta\mathit{\Delta}\epsilon(j_2\,\mbox{-}\,j_1)$
typifies $Z$- or $N$-dependence of the shell gap,
and the corresponding quantities
$\delta\mathit{\Delta}\epsilon^{(\mathrm{TN})}(j_2\,\mbox{-}\,j_1)$
and $\delta\mathit{\Delta}\epsilon^{(\mathrm{OPEP})}(j_2\,\mbox{-}\,j_1)$
represent contributions of $v^{(\mathrm{TN})}$ and $v_\mathrm{OPEP}^{(\mathrm{C})}$
to the shell gap.
Figure~\ref{fig:ddspe_M3Yp6} summarizes
$\delta\mathit{\Delta}\epsilon(j_2\,\mbox{-}\,j_1)$ for several regions
in the HF results with the M3Y-P6 interaction.

The top row of Fig.~\ref{fig:ddspe_M3Yp6}
exhibits $\delta\mathit{\Delta}\epsilon(n0d_{3/2}\,\mbox{-}\,n1s_{1/2})$,
the $N=16$ shell gap at $^{24}$O relative to that at $^{30}$Si.
The shell gap is larger at $^{24}$O than at $^{30}$Si
by $2.9\,\mathrm{MeV}$,
accounting for the emergence of the $N=16$ magicity at $^{24}$O.
It is difficult to obtain this enhancement
without either $v^{(\mathrm{TN})}$ or $v_\mathrm{OPEP}^{(\mathrm{C})}$;
$\delta\mathit{\Delta}\epsilon^{(\mathrm{TN})}(n0d_{3/2}\,\mbox{-}\,
n1s_{1/2})=2.0\,\mathrm{MeV}$
and $\delta\mathit{\Delta}\epsilon^{(\mathrm{OPEP})}(n0d_{3/2}\,\mbox{-}\,
n1s_{1/2})=1.3\,\mathrm{MeV}$.
In the second top row of Fig.~\ref{fig:ddspe_M3Yp6},
the $N=28$ shell gap at $^{42}$Si relative to that at $^{48}$Ca,
$\delta\mathit{\Delta}\epsilon(n1p_{3/2}\,\mbox{-}\,n0f_{7/2})$,
is presented.
The shell gap is quenched at $^{42}$Si
with $\delta\mathit{\Delta}\epsilon(n1p_{3/2}\,\mbox{-}\,n0f_{7/2})
=-2.9\,\mathrm{MeV}$,
while $v^{(\mathrm{TN})}$ gives only $-0.6\,\mathrm{MeV}$
and $v_\mathrm{OPEP}^{(\mathrm{C})}$ contribution is small but positive
(not visible in Fig.~\ref{fig:ddspe_M3Yp6}).
However, $v^{(\mathrm{TN})}$ plays a significant role
in the deformation of $^{42}$Si, as discussed in Sec.~\ref{subsec:def-TNS}.
The third and fourth rows show that
the contributions of $v^{(\mathrm{TN})}$ and $v_\mathrm{OPEP}^{(\mathrm{C})}$
are crucial in the difference of the $N=32$ and $40$ magicities
between Ca and Ni viewed in Fig.~\ref{fig:dE_HF-HFB}.
For $\delta\mathit{\Delta}\epsilon(p0g_{9/2}\,\mbox{-}\,p1p_{1/2})$
evaluated for $^{80}$Zr relative to $^{90}$Zr,
the full M3Y-P6 result is $-0.5\,\mathrm{MeV}$,
while $\delta\mathit{\Delta}\epsilon^{(\mathrm{TN})}
(p0g_{9/2}\,\mbox{-}\,p1p_{1/2})=-1.3\,\mathrm{MeV}$.
Namely,
the sign of $\delta\mathit{\Delta}\epsilon(p0g_{9/2}\,\mbox{-}\,p1p_{1/2})$
is inverted owing to $v^{(\mathrm{TN})}$,
which could be crucial in the erosion of the $Z=40$ magicity at $N\approx 40$.
Analogous sign inversion due to $v^{(\mathrm{TN})}$ contributes
to the persistence of the $N=82$ magicity around $^{122}$Zr.

Several effects of the tensor-force and OPEP are confirmed:
\begin{itemlist}
\item The tensor force $v^{(\mathrm{TN})}$
 often (though not always) plays a significant role
 in the $Z$- or $N$-dependence of the shell gap,
 accounting for appearance and disappearance of magicity.
\item The central spin-isospin channel from the OPEP,
 $v^{(\mathrm{C})}_\mathrm{OPEP}$,
 tends to enhance the tensor-force effect.
 Strong $Z$- or $N$-dependence of the shell gap often coincides
 with their cooperative contribution.
\item These effects strongly appear when an orbit having high $\ell$
 is occupied.
\end{itemlist}

There could be some cases in which additional correlations,
even those beyond MF, destroy the magicity.
However, the results shown here suggest
that the semi-realistic interactions often give a simple SCMF picture
for the appearance and the disappearance of the magicity,
resonating the spirit of the DFT.

\subsection{Observability of bubble structure in nuclear density distributions}

Although there has been no experimental evidence,
in some nuclei there could be a spatial region
in which density is distinctively depleted, usually around the nuclear center.
This exotic structure is called nuclear `bubble'.
I here discuss whether and in what nucleus bubble can hopefully be observed.

Nuclear density distribution in a nucleus is accessible
by scattering experiments.
Charge densities are unambiguously measured
by the elastic electron scattering~\cite{Friar-Negele_1975,deForest-Walecka},
which primarily reflect proton distribution.
Whereas the target nuclei had been restricted to stable ones until recently,
the new SCRIT technology makes it possible
to handle short-lived nuclei~\cite{Suda-etal_2009}.
Neutron or matter densities are extracted via hadronic probes,
for which it is difficult to get rid of the model-dependence.
To establish the depletion of density,
model-independent analysis is highly desired.
For this reason I focus on proton bubbles.
Two possibilities have been pointed out for proton bubbles.
One is a bubble created by the Coulomb repulsion~\cite{
  Decharge-Berger-Girod-Dietrich_2003}.
This mechanism requires the Coulomb force to be strong
and therefore is limited to extremely heavy nuclei,
quite probably beyond the reach of current and near-future
density measurements.
The other is a bubble produced by holes of an $s$-orbit,
because only protons occupying $s$-orbits can contribute
to the density at the center.
For a bubble of this type to be actualized,
the hole state should be a pure $s$-state to a good precision~\cite{
  Saxena-etal_2019}.
Any correlations mixing $\ell$'s,
\textit{e.g.} pairing and deformation, act against forming the bubble.
Hence $Z$ should be magic
where the lowest unoccupied proton s.p. state is an $s$-state.

There are no more than a few candidates
which have an $s$-hole proton bubble with detectable size.
Whereas one might consider from the s.p. level sequence
displayed in Fig.~\ref{fig:de13} that $^{46}$Ar is one of them~\cite{
  Khan-Grasso-Margueron-VanGiai_2008,ToddRutel-Piekarewicz-Cottle_2004}
since $p1s_{1/2}$ is the highest occupied proton level at $^{48}$Ca,
the small $\mathit{\Delta}\epsilon_p$ leads to sizable pair correlation
among protons at $^{46}$Ar,
preventing the bubble~\cite{Nakada-Sugiura-Margueron}.
Despite the inversion of $p1s_{1/2}$ and $p0d_{3/2}$ in Fig.~\ref{fig:de13},
the Ar nuclei become unbound at $N>42$
in the spherical HFB calculation with M3Y-P6,
while Ca is bound up to $N=50$.
Although deformation may extend the drip line for Ar,
it prohibits the bubble.
Thus it is unlikely for any Ar nuclei to have proton bubble structure.

\begin{figure}
\centerline{\includegraphics[scale=0.5]{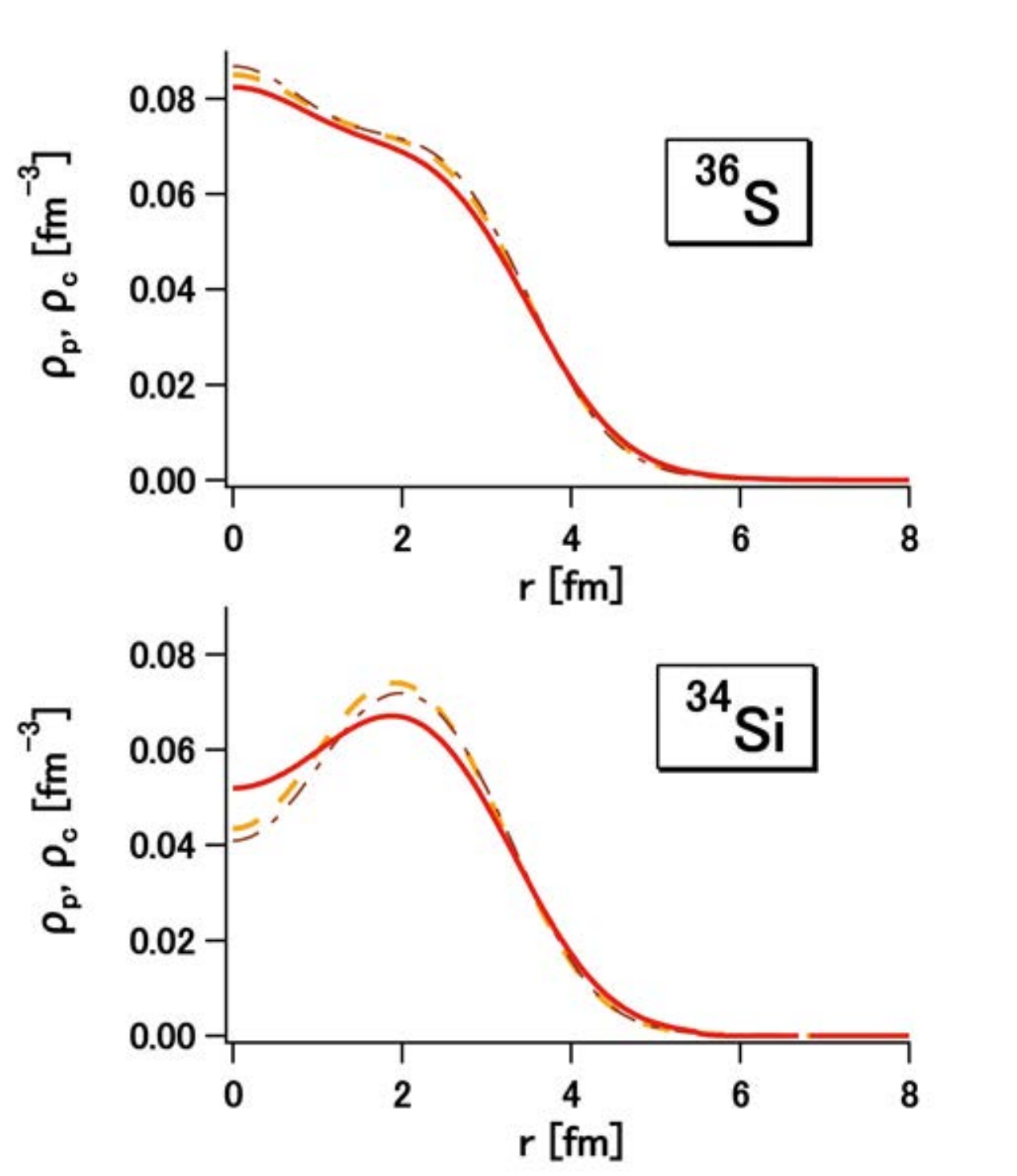}}\vspace*{1cm}
\caption{Proton and charge density distributions
  in $^{36}$S and $^{34}$Si
  obtained from the HFB calculations with M3Y-P6.
  Point-proton densities without (with) c.m. correction
  are displayed by orange dashed (thin brown dot-dashed) lines,
  and charge density by red solid lines.
\label{fig:rho_Si34}}
\end{figure}

The possibility of proton bubble structure has been pointed out
also for $^{34}$Si~\cite{Grasso-etal_2009}.
The energy difference $\epsilon(p1s_{1/2})-\epsilon(p0d_{5/2})$
exceeds $5\,\mathrm{MeV}$ for $^{36}$S in the spherical HF calculations.
We note that $\epsilon(p0d_{3/2})-\epsilon(p1s_{1/2})$
is less than $2\,\mathrm{MeV}$,
giving rise to sizable pair excitation at $^{36}$S.
However, owing to the large gap $\epsilon(p1s_{1/2})-\epsilon(p0d_{5/2})$,
the g.s. of $^{34}$Si is expected to have
almost pure $(p1s_{1/2})^{-2}$ configuration.
Proton and charge density distributions of $^{34}$Si
predicted by the HFB calculation with M3Y-P6
are depicted in Fig.~\ref{fig:rho_Si34},
in comparison with those of $^{36}$S.
See \ref{app:cmcorr} for the c.m. and nucleon finite-size corrections.
Since $\epsilon(p1s_{1/2})-\epsilon(p0d_{5/2})$ is sufficiently large,
a prominent proton bubble structure is predicted for $^{34}$Si.
The bubble remains in the charge density,
although it is somewhat smeared due to the nucleon finite-size effects.
The large $\epsilon(p1s_{1/2})-\epsilon(p0d_{5/2})$ makes $^{34}$Si doubly magic,
giving identical density distribution between HF and HFB.
The result for $^{34}$Si is not sensitive to the effective interaction.
It has been argued~\cite{Yao-Baroni-Bender-Heenen}
that correlations beyond the MF regime
could wash out the central depletion of the proton density in $^{34}$Si.
However, the results in Ref.~\refcite{Yao-Baroni-Bender-Heenen}
seem to overestimate correlation effects.
The emptiness of the $p1s_{1/2}$ has been reported
experimentally~\cite{Mutschler-etal_2017},
supporting the possibility of the proton bubble. 
Future experiments on the charge density of this nucleus are awaited.

The next possibility of a proton bubble
could be given by a $p2s_{1/2}$ hole state.
Although there have been arguments
in Hg nuclei~\cite{ToddRutel-Piekarewicz-Cottle_2004},
the pair correlation stays sizable in the spherical HFB calculations
with M3Y-P6,
prohibiting bubbles.

\subsection{$3N$-force effect on charge radii}\label{subsec:r_c}

Important information about nuclear structure has been supplied
by atomic experiments.
Frequencies of electromagnetic waves in atomic deexcitations
slightly vary among isotopes,
depending on the charge distribution of the nuclei
as well as on the reduced mass.
The difference in m.s. charge radii among the isotopes
is extracted from accurate measurement on the shifts of the frequencies,
\textit{i.e.} the isotope shifts.
It has been known that there are kinks at magic $N$
in the charge radii in many isotopes~\cite{Angeli-Marinova_2013}.
This is partly accounted for as an effect of deformation,
since the nuclear radius becomes greater in the deformed shape
than in the spherical shape [see Eq.~(\ref{eq:r2-def})]~\cite{Bohr-Mottelson_1}.
However, kinks have also been observed in the isotopes with magic $Z$,
which quite likely stay spherical,
as typified by a kink at $N=126$
in the Pb isotopes~\cite{Aufmuth-Heilig-Steudel_1987}.

In the Pb nuclei,
the proton configuration is unlikely to change significantly;
the protons occupy the s.p. levels up to the $Z=82$ shell gap.
Therefore the evolution of the neutron distribution
should be responsible for the differential charge radii among isotopes
through the attraction between protons and neutrons.
The broad distribution of neutrons swells individual proton s.p. w.f.'s
and increases the charge radius.
Whereas the kink at $N=126$ in Pb was hardly described
with the conventional Skyrme interaction~\cite{Tajima-etal_1993},
a RMF calculation yields a kink analogous to the observed one~\cite{
  Sharma-Lalazissis-Ring_1994}.
It has been recognized that this model-dependence originates
primarily from the $\ell s$ potential.
The spatial distributions of the s.p. orbits
are slightly displaced between the $\ell s$ partners.
For a nucleon occupying a $j=\ell-1/2$ ($j=\ell+1/2$) orbital,
the $\ell s$ potential acts repulsively (attractively)
and makes its distribution wider (narrower).
Moreover, this effect is the stronger for the higher $\ell$.
The observed level sequence indicates
that the lowest neutron s.p. orbit above the $N=126$ shell gap
is $1g_{9/2}$, and $0i_{11/2}$ lies above it~\cite{TableOfIsotopes}.
The $n0i_{11/2}$ orbit can be partially occupied at $N>126$
owing to the pair correlation.
Because of the larger radius of $n0i_{11/2}$
than those of the surrounding orbits,
occupancy on the $n0i_{11/2}$ orbit plays an important role
in the kink~\cite{Reinhard-Flocard_1995}.
Degree of the occupation depends on the s.p. energy difference
$\epsilon(n0i_{11/2})-\epsilon(n1g_{9/2})$.
With the conventional Skyrme interaction,
the energy difference is too large
for $n0i_{11/2}$ to be sizably occupied.
In contrast, the RMF gives a small energy difference,
which results from its isospin content of the $\ell s$ potential~\cite{
  Sharma-Lalazissis-Konig-Ring_1995}
and is connected to the pseudo-spin symmetry~\cite{Liang-Meng-Zhou_2015}.
This observation leads to the extension
of the Skyrme EDF~\cite{Reinhard-Flocard_1995}.
However, in the results of the RMF and the extended Skyrme EDF,
the kink is obtained at the expense of too good pseudo-spin symmetry.
It was shown that the kink at $N=126$ is difficult to be reproduced
unless $n1g_{9/2}$ and $n0i_{11/2}$ are
nearly degenerate or even inverted~\cite{Goddard-Stevenson-Rios_2013},
incompatible with the observed energy levels~\cite{TableOfIsotopes}.

While the $\ell s$ potential yields the $\ell s$ splitting
and size of the $\ell s$ splitting is known experimentally,
its origin based on the nucleonic interaction has not been well understood.
In addition to the bare $2N$ LS interaction,
the tensor force contributes to the $\ell s$ splitting
at its 2nd order~\cite{Terasawa_1960,Arima-Terasawa_1960}.
Still, they are not enough to account for
the observed $\ell s$ splitting~\cite{Ando-Bando_1981}.
There have been suggestions that
correlation effects may cure this problem~\cite{Suzuki-Okamoto-Kumagai_1987},
and that the $3N$ force affects significantly~\cite{Pieper-Pandharipande_1993}.
Relatively recently,
it has been indicated that the $3N$ interaction derived from the $\chi$EFT,
which effectively gives significant density-dependence in the LS channel,
substantially enhances the $\ell s$ potential~\cite{Kohno_2012,Kohno_2013,
  Kohno_2013_erratum}.
Inspired by this work,
the density-dependent LS channel $v^{(\mathrm{LS}\rho)}$
has been introduced in Ref.~\refcite{Nakada-Inakura_2015},
which yields a variant of the M3Y-P6 interaction called M3Y-P6a.
In M3Y-P6, $v^{(\mathrm{LS})}$ was enhanced from the $G$-matrix result
so as to reproduce the s.p. level sequence.
Instead of enhancing $v^{(\mathrm{LS})}$,
$v^{(\mathrm{LS}\rho)}$ is added in M3Y-P6a
while $v^{(\mathrm{LS})}$ is returned to the original one
determined by the $G$-matrix.
Because the $\chi$EFT is not yet well convergent at present,
the functional form of $D[\rho]$ is taken as Eq.~(\ref{eq:DinLS})
to be compatible with the $\chi$EFT suggestion,
but its strength is fixed in a phenomenological manner.
The $d_1$ term of the denominator of $D[\rho]$
is employed only to avoid instability for high $\rho$
and has been assumed to be $1.0\,\mathrm{fm}^3$.
To keep most of the M3Y-P6 results of the shell structure,
the parameter $w_1$ has been fitted
to the splitting of the $n0i$ orbits obtained with M3Y-P6
at $^{208}$Pb~\cite{Nakada-Inakura_2015}.
Other $\ell s$ splitting is hardly influenced as well.

\begin{figure}
\centerline{\includegraphics[scale=0.6]{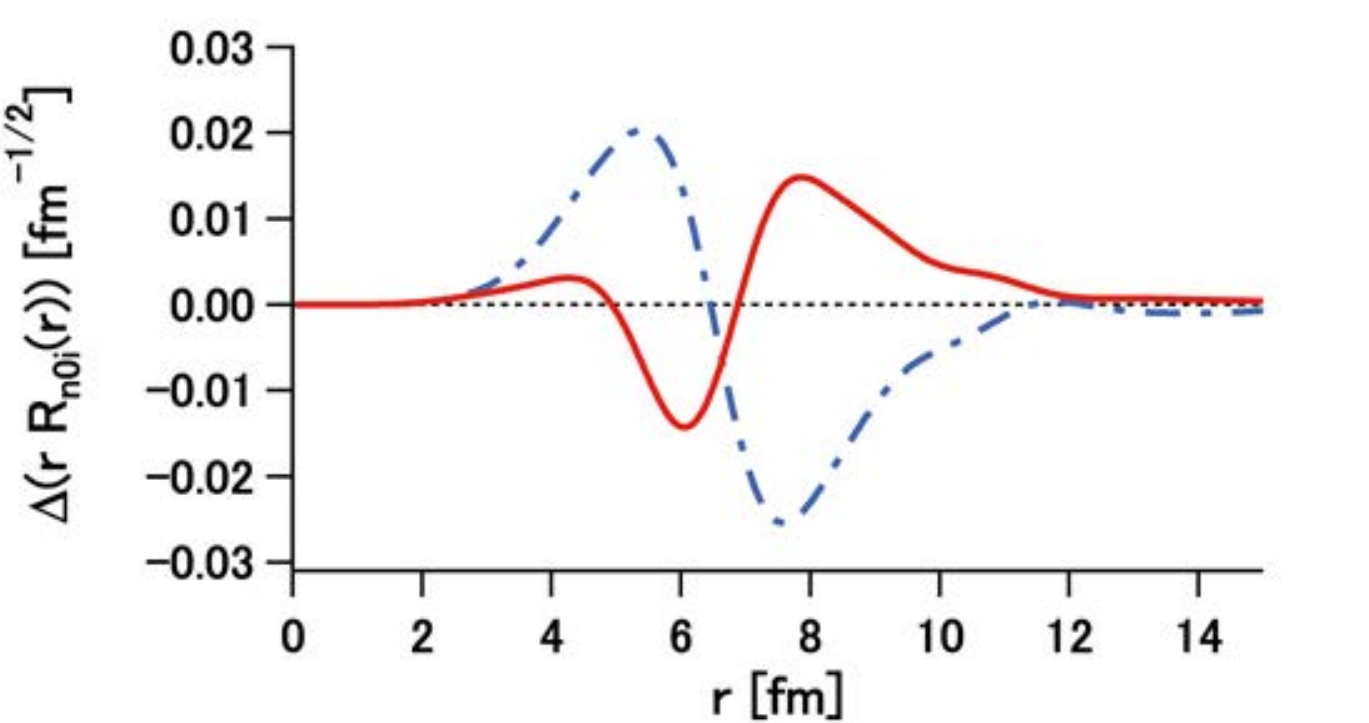}}
\caption{Difference of the radial part of the s.p. functions
for $n0i_{13/2}$ (blue dot-dashed line) and $n0i_{11/2}$ (red solid line);
$r\,R_{n\ell j}(r)$ obtained with M3Y-P6a relative to that with M3Y-P6,
by the HF calculations at $^{208}$Pb.
Quote from Ref.~\protect\refcite{Nakada-Inakura_2015}.
\label{fig:dspwf_n0i}}
\end{figure}

For understanding the effects of $v^{(\mathrm{LS}\rho)}$,
it is useful to consider its contribution to the $\ell s$ potential.
The density-dependence gives an additional term to the $\ell s$ potential.
Under the spherical symmetry,
$v^{(\mathrm{LS}\rho)}$ yields the $\ell s$ potential as follows,
\begin{equation}
\frac{1}{r}\bigg[\,D[\rho(r)]\,
 \frac{d}{dr}\Big(\rho(r)+\rho_\tau(r)\Big)
+ \frac{1}{2}\,\frac{\delta D}{\delta\rho}[\rho(r)]\,
 \Big(\rho(r)+\rho_\tau(r)\Big)\,\frac{d\rho(r)}{dr}\bigg]\,
 \mbox{\boldmath$\ell$}\cdot\vect{s}\,.\quad(\tau=p,n) \label{eq:lspot}
\end{equation}
Because of $D[\rho]$ of Eq.~(\ref{eq:DinLS}),
$j=\ell-1/2$ ($j=\ell+1/2$) orbits tend to shift outward (inward)
so as for the $\ell s$ potential to act more weakly (strongly).
Note that the density rearrangement term
containing $\delta D/\delta\rho$ enhances this tendency.
To confirm this effect,
the difference of the radial functions $R_j(r)$ between M3Y-P6 and M3Y-P6a
is depicted in Fig.~\ref{fig:dspwf_n0i},
for $j=n0i_{13/2}$ and $n0i_{11/2}$ at $^{208}$Pb.
Concerning the phase, $R_{n0i}(r)\geq 0$ is assumed as usual.
When the interaction is switched from M3Y-P6 to M3Y-P6a,
the m.s. radius of $R_{n0i_{11/2}}(r)$ increases by $0.49\,\mathrm{fm}^2$.

\begin{figure}
\includegraphics[scale=0.4]{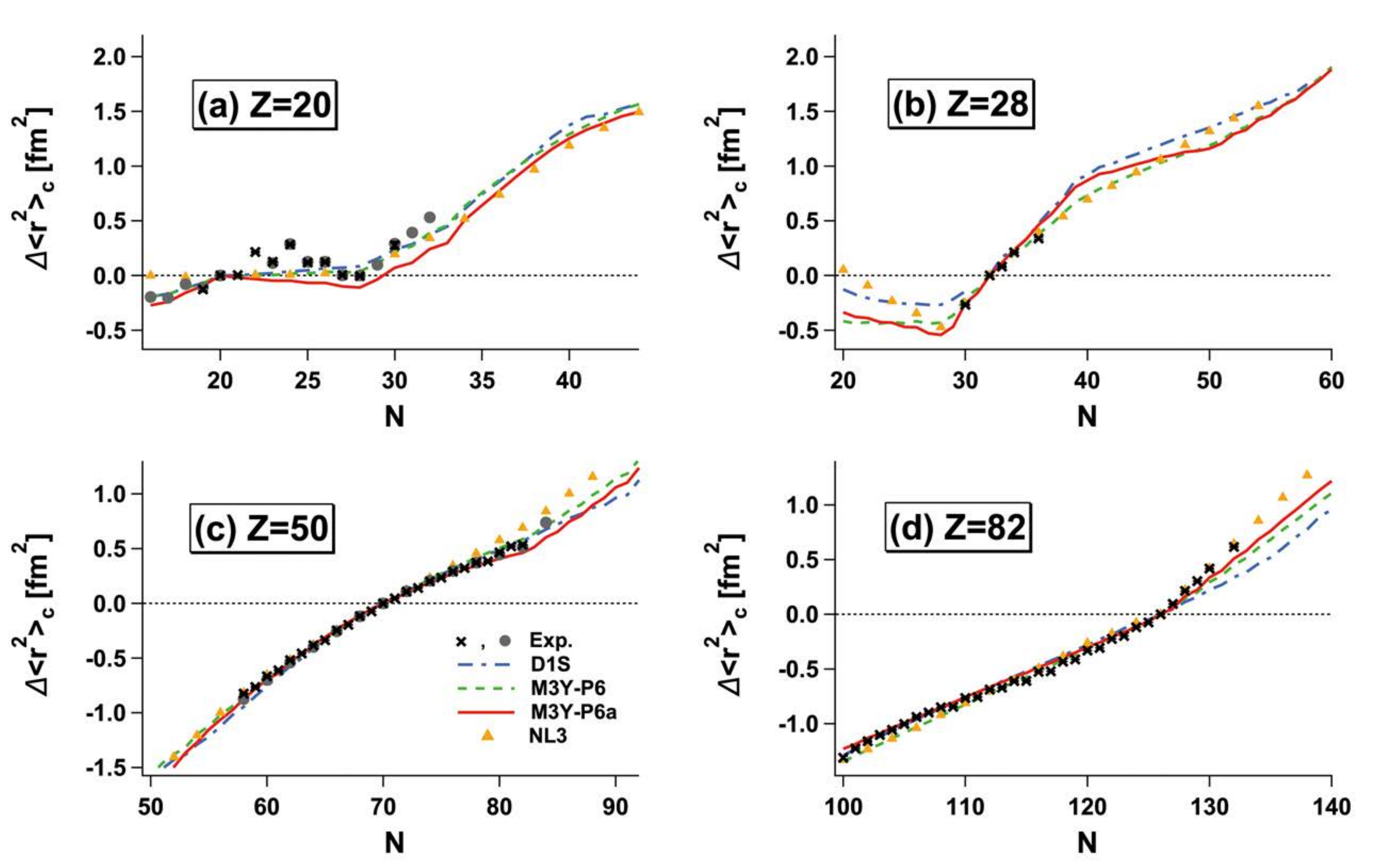}
\caption{$\mathit{\Delta}\bra r^2\ket_c$
  of (a) $Z=20$, (b) $Z=28$, (c) $Z=50$ and (d) $Z=82$ nuclei.
  Spherical HFB results with D1S (blue dot-dashed line),
  M3Y-P6 (green dashed line) and M3Y-P6a (red solid line) are presented.
  RMF results for even-$N$ nuclei are taken
  from Ref.~\cite{Lalazissis-Ring_1999} (orange triangles),
  in which the neutron finite-size effects are ignored.
  Experimental data are taken
  from Refs.~\protect\refcite{Angeli-Marinova_2013} (crosses) for all panels,
  \protect\refcite{Miller-etal_2019} (for $N\leq 20$)
  and \protect\refcite{GarciaRuiz-etal_2016} (for $N\geq 23$)
  in (a) (gray circles),
  \protect\refcite{Gorges-etal_2019} in (c) (gray circles).
  Quote from Ref.~\protect\refcite{Nakada_2019}.
\label{fig:drc_Z-magic}}
\end{figure}

By the spherical HFB calculations,
the isotopic variation of the charge radii is investigated
for the $Z=\mathrm{magic}$ nuclei, Ca, Ni, Sn and Pb.
The m.s. charge radius of the $^A Z$ nucleus
is calculated via Eq.~(\ref{eq:r2-charge}).
Thanks to the atomic experiments,
accurate and abundant data are available on the differential m.s. charge radii.
The differential m.s. charge radius of $^A Z$
is given by $\mathit{\Delta}\bra r^2\ket_c(\mbox{$^A Z$})
=\bra r^2\ket_c(\mbox{$^A Z$})-\bra r^2\ket_c(\mbox{$^{A_0} Z$})$,
where $^{A_0} Z$ is the reference nuclide.
The results are presented in Fig.~\ref{fig:drc_Z-magic}.
Comparison of the M3Y-P6a results to the M3Y-P6 ones
reveals effects of $v^{(\mathrm{LS}\rho)}$,
when keeping size of the $\ell s$ splittings.
Results with the D1S interaction~\cite{Gogny_D1S}
and those of the RMF~\cite{Lalazissis-Ring_1999}
are also shown.
Detailed discussion is given below for individual isotopes.

\subsubsection{Pb isotopes}

I start from the Pb isotopes shown in Fig.~\ref{fig:drc_Z-magic}(d).
For the Pb nuclei, $^{208}$Pb is taken as the reference,
$\mathit{\Delta}\bra r^2\ket_c(\mbox{$^A$Pb})
=\bra r^2\ket_c(\mbox{$^A$Pb})-\bra r^2\ket_c(\mbox{$^{208}$Pb})$.

The LS channel of D1S has the same form
as that of the original Skyrme interaction.
As a result, $n0i_{11/2}$ is hardly occupied just beyond $N=126$
and thereby no apparent kink emerges at $N=126$,
although $\mathit{\Delta}\bra r^2\ket_c$ gradually increases
beyond $N\gtrsim 135$.
Having both the TE and TO channels in $v^{(\mathrm{LS})}$,
M3Y-P6 gives a kink at $N=126$,
since $n0i_{11/2}$ lies closer to $n1g_{9/2}$ than in the D1S case.
However, the kink is much weaker than the experimental one.
In contrast, an obvious kink is obtained with M3Y-P6a.
Owing to $v^{(\mathrm{LS}\rho)}$,
the s.p. w.f. of $n0i_{11/2}$ shifts outward further,
which makes qualitative difference
in the $N$-dependence of $\mathit{\Delta}\bra r^2\ket_c$.
The s.p. w.f. of $n0i_{13/2}$ shifts inward,
improving the slope of $\mathit{\Delta}\bra r^2\ket_c$ in $N<126$.
The pseudo-spin symmetry is moderately broken
with $\epsilon(n0i_{11/2})-\epsilon(n1g_{9/2})=0.72\,\mathrm{MeV}$ at $^{208}$Pb,
which is comparable to the measured energy difference
between $(11/2)^+_1$ and $(9/2)^+_1$ at $^{209}$Pb~\cite{TableOfIsotopes}.
Although the kink at $N=126$ is yet weaker than the data,
it seems to provide a good indication
what physics is key to this phenomenon.
Whereas the RMF results look to reproduce $\mathit{\Delta}\bra r^2\ket_c$
quite well in Fig.~\ref{fig:drc_Z-magic}(d),
they do not describe relevant quantities as mentioned above,
yielding too good pseudo-spin symmetry.

The observed even-odd staggering
of $\mathit{\Delta}\bra r^2\ket_c(\mbox{$^A$Pb})$ at $N>126$
seems to rule out inversion of $n1g_{9/2}$ and $n0i_{11/2}$,
consistent with the present results.
It is also commented that $\mathit{\Delta}\bra r^2\ket_c(\mbox{$^{209}$Pb})$,
which also participates in the kink~\cite{Barzakh-etal_2018},
cannot yet be accounted for.
This problem might implicate the weak breaking of the $Z=82$ core
at this nucleus.

\subsubsection{Ca isotopes}

For the Ca nuclei displayed in Fig.~\ref{fig:drc_Z-magic}(a),
$^{40}$Ca is taken as the reference nuclide,
$\mathit{\Delta}\bra r^2\ket_c(\mbox{$^A$Ca})
=\bra r^2\ket_c(\mbox{$^A$Ca})-\bra r^2\ket_c(\mbox{$^{40}$Ca})$.
The inversion of $p0d_{3/2}$ and $p1s_{1/2}$ from $^{40}$Ca to $^{48}$Ca
discussed in Sec.~\ref{subsec:tensor}
is also reproduced with M3Y-P6a,
owing significantly to $v^{(\mathrm{TN})}$.

Recent experiments in $\mathit{\Delta}\bra r^2\ket_c(\mbox{$^A$Ca})$
have shown a kink at $N=28$~\cite{GarciaRuiz-etal_2016}
and an inverted kink at $N=20$~\cite{Miller-etal_2019},
which is called `anti-kink' in Ref.~\refcite{Nakada_2019}.
While both are qualitatively described
by the spherical HFB calculations also with D1S and M3Y-P6,
$v^{(\mathrm{LS}\rho)}$ in M3Y-P6a makes the kink and the anti-kink stronger.
The anti-kink at $N=20$ is not apparent in the RMF results.
The kink and the anti-kink in the calculations
are related in part to the flat $\mathit{\Delta}\bra r^2\ket_c(\mbox{$^A$Ca})$
in $20\leq N\leq 28$,
where $n0f_{7/2}$ is being filled.
The relatively small radius of $n0f_{7/2}$ suppresses
an increase of the charge radii.

The fluctuation observed in $^{42-46}$Ca
cannot be reproduced by the calculations presented here.
There is a suggestion from the shell model~\cite{Caurier-etal_2001}
that this fluctuation is linked to excitations out of the $1s0d$-shell.
Fayans' EDF~\cite{Fayans_1998} provides the fluctuation,
but without breaking the $Z=20$ core~\cite{Fayans-etal_2000}.
In that result, the generalized pairing term
that couples to the density gradient is found
to give rise to the fluctuation~\cite{Reinhard-Nazarewicz_2017}.
Fayans' EDF has been claimed to successfully explain
$\mathit{\Delta}\bra r^2\ket_c(\mbox{$^A$Ca})$ also
in the neutron-deficient region~\cite{Miller-etal_2019}.

\subsubsection{Ni isotopes}

Figure~\ref{fig:drc_Z-magic}(b) shows
$\mathit{\Delta}\bra r^2\ket_c(\mbox{$^A$Ni})
=\bra r^2\ket_c(\mbox{$^A$Ni})-\bra r^2\ket_c(\mbox{$^{60}$Ni})$.
The calculations with M3Y-P6a predict prominent kinks at $N=28,50$,
and an anti-kink at $N=40$.
It is indicated that kinks and anti-kinks well correspond to the magicity.
Recall that $^{68}$Ni is close to doubly magic~\cite{
  Broda-etal_1995,Nakada-Sugiura_2014}.
The kink at $N=50$ and the anti-kink at $N=40$ are not conspicuous
in the other results.
The kink at $N=28$ is also obtained with D1S and M3Y-P6,
though weaker than with M3Y-P6a and in the RMF.
With the significant interaction-dependence,
measurements around $^{56,68,78}$Ni should provide
interesting and important information of nucleonic interaction,
particularly of the $3N$ force what affects the $\ell s$ potential.

\subsubsection{Sn isotopes}

For the Sn isotopes,
Fig.~\ref{fig:drc_Z-magic}(c) is drawn by adopting $^{120}$Sn
as the reference nuclide,
$\mathit{\Delta}\bra r^2\ket_c(\mbox{$^A$Sn})
=\bra r^2\ket_c(\mbox{$^A$Sn})-\bra r^2\ket_c(\mbox{$^{120}$Sn})$.
It is found that M3Y-P6a describes
$\mathit{\Delta}\bra r^2\ket_c(\mbox{$^A$Sn})$
remarkably well, in a long chain of the Sn isotopes.
In particular, it predicted a kink at $N=82$,
which is discovered in a recent experiment~\cite{Gorges-etal_2019}.
In the prediction of the kink with M3Y-P6a,
$v^{(\mathrm{LS}\rho)}$ plays an essential role as in the case of Pb,
shifting the $n0h_{9/2}$ ($n0h_{11/2}$) orbit outward (inward)
and making $\mathit{\Delta}\bra r^2\ket_c(\mbox{$^A$Sn})$ steeper above $N=82$
(less steep below $N=82$).
No other interactions and EDFs except Fayans'~\cite{Fayans-etal_2000}
predicted the kink.

\subsubsection{Further discussions}

The kinks and the anti-kinks argued above are connected to the magic numbers.
The mechanism originating from the $\ell s$ potential,
including the effects of $v^{(\mathrm{LS}\rho)}$,
implies the following general rule~\cite{Nakada_2019}.

There are two types of nuclear magic numbers:
the $\ell s$-closed magic numbers and the $jj$-closed ones.
At a $jj$-closed magic number
a high-$j$ orbit with $j=\ell+1/2$ is filled,
and its $\ell s$ partner with $j=\ell-1/2$ starts occupied
above the magic number.
Even if the $j=\ell-1/2$ orbit does not lie lowest above the magic number,
the approximate symmetry with respect to the pseudo-spin~\cite{
  Hecht-Adler_1969,Arima-Harvey-Shimizu_1969}
ensures that it is not far from the lowest orbit,
and it has sizable occupancy owing to the pair correlation.
Since the s.p. w.f. of $j=\ell-1/2$ distributes relatively widely
and that of $j=\ell+1/2$ narrowly,
it should be generic for kinks to come out in $\mathit{\Delta}\bra r^2\ket_c$
at $jj$-closed $N$'s.
On the contrary, an $\ell s$-closed magic number usually occurs
after a $j=\ell-1/2$ orbit is filled,
and then a $j=\ell+1/2$ orbit having higher $\ell$ starts occupied above it.
Thereby anti-kinks are expected at $\ell s$-closed $N$'s.
Although the influence of deformation could obscure this effect,
it is expected for anti-kinks to be observed
by selecting isotopes keeping sphericity.
Anti-kinks could be good evidence for the $3N$-force effect
on the $\ell s$ potential.
As well as in the charge radii,
kinks and anti-kinks are predicted in the matter radii~\cite{Nakada_2019}.

The recently discovered kink at $^{132}$Sn
was predicted only by M3Y-P6a and Fayans' EDF.
These two models account for the kink through different physics mechanisms.
Fayans' EDF may be advantageous in describing the fluctuation
of $\mathit{\Delta}\bra r^2\ket_c(\mbox{$^A$Ca})$ in $^{42-46}$Ca
within a single model,
and have been successfully applied to the Fe and the Cd nuclei~\cite{
  Minamisono-etal_2016,Hammen-etal_2018}.
However, it is so far done with $Z$-dependent parameters,
lacking microscopic justification,
and tends to give too strong even-odd staggering
in $\mathit{\Delta}\bra r^2\ket_c$.
Apart from these advantages and disadvantages,
it is of interest to discriminate by other data
which mechanism is dominant.
The anti-kink, which is predicted \textit{e.g.} at $^{68}$Ni,
may supply possibility.

\section{Several topics in deformed nuclei}\label{sec:deformed}

As topics concerning deformed nuclei,
I shall argue tensor-force effects and deformed halos in this section.
To investigate them,
the axial MF calculations assuming the parity conservation,
the $\mathcal{R}$ and the $\mathcal{T}$ symmetries are applied. 

\subsection{Tensor-force effects on deformation}\label{subsec:def-TNS}

The HF frame, rather than HFB, is suitable
for investigating tensor-force effects on deformation.
The tensor force does not influence the w.f.'s significantly~\cite{
  Suzuki-Nakada-Miyahara_2016},
and its effects are represented perturbatively
by $E^{(\mathrm{TN})}$ defined in Eq.~(\ref{eq:E&spe-X}).
Moreover, M3Y-P6 is appropriate
because of the realistic nature of the tensor force in it.
I shall show results of the axial HF calculations with M3Y-P6
for the proton-deficient $N=20,28$ nuclei,
which lie at the `shore' of the `island of inversion',
and for the Zr isotopes whose shapes alter several times depending on $N$.
For reference, axial HF results with D1M are also shown.
Though D1M was designated for calculations beyond MF,
comparison at the HF level would be useful
whether and how tensor-force effects are incorporated
in phenomenological interactions without explicit tensor force.
It is recalled that $N=20$ at $^{32}$Mg and $Z=40$ in $60\leq N\leq 70$
are erroneously picked up as candidates of magic numbers
by the spherical MF calculations in Sec.~\ref{subsec:chart}.
It is also investigated whether calculations taking account of deformation
can resolve this problem.

Bender \textit{et al.} applied a class of the Skyrme interactions
including the tensor channels~\cite{Lesinski-etal_2007} to deformed nuclei,
and analyzed their influence~\cite{Bender-etal_2007}.
Their study had already disclosed several important effects
of the tensor force on the nuclear deformation.
Since the realistic tensor force based on the $G$-matrix is applied,
the SCMF study with the semi-realistic interaction
is of value in confirming and further elucidating
the real effects of the tensor force.

\subsubsection{$N=20$ and $28$}\label{subsubsec:N20&28}

The structure of neutron-rich nuclei in the $20\lesssim N\lesssim 28$ region
has attracted great interest.
It has been discovered that the $N=20$ and $28$ magicities are broken
in some nuclei~\cite{Thibault-etal_1975,GuillemaudMueller-etal_1984,
  Bastin-etal_2007,Takeuchi-etal_2012,Crawford-etal_2019},
which form the `island of inversion'~\cite{Warburton-Becker-Brown_1990}.
For the $N=20$ nuclei $^{30}$Ne and $^{32}$Mg,
neutron excitation out of the $1s0d$-shell is suggested
by shell-model calculations~\cite{Poves-Retamosa_1987,
  Utsuno-Otsuka-Mizusaki-Honma_1999},
meaning deformation.
However, it has not been easy to capture deformation
in the SCMF calculations~\cite{Terasaki-Flocard-Heenen-Bonche_1997,
  Peru-Girod-Berger_2000},
although deformation could be realized
via correlations beyond the MF framework~\cite{Peru-Girod-Berger_2000,
  RodriguezGuzman-Egido-Robledo_2002,RodriguezGuzman-Egido-Robledo_2003,
  Kimura-Horiuchi_2002,Shimada-etal_2016}
and interpretation other than deformation
has not fully been ruled out~\cite{Yamagami-VanGiai_2004}.
For $N=28$,
many SCMF calculations predicted the breakdown of the $N=28$ magicity
due to quadrupole deformation at $^{40}$Mg and $^{42}$Si~\cite{
  Terasaki-Flocard-Heenen-Bonche_1997,Peru-Girod-Berger_2000,
  RodriguezGuzman-Egido-Robledo_2002,Shimada-etal_2016,Li-etal_2011}.

\begin{figure}\begin{center}
  \includegraphics[scale=0.6]{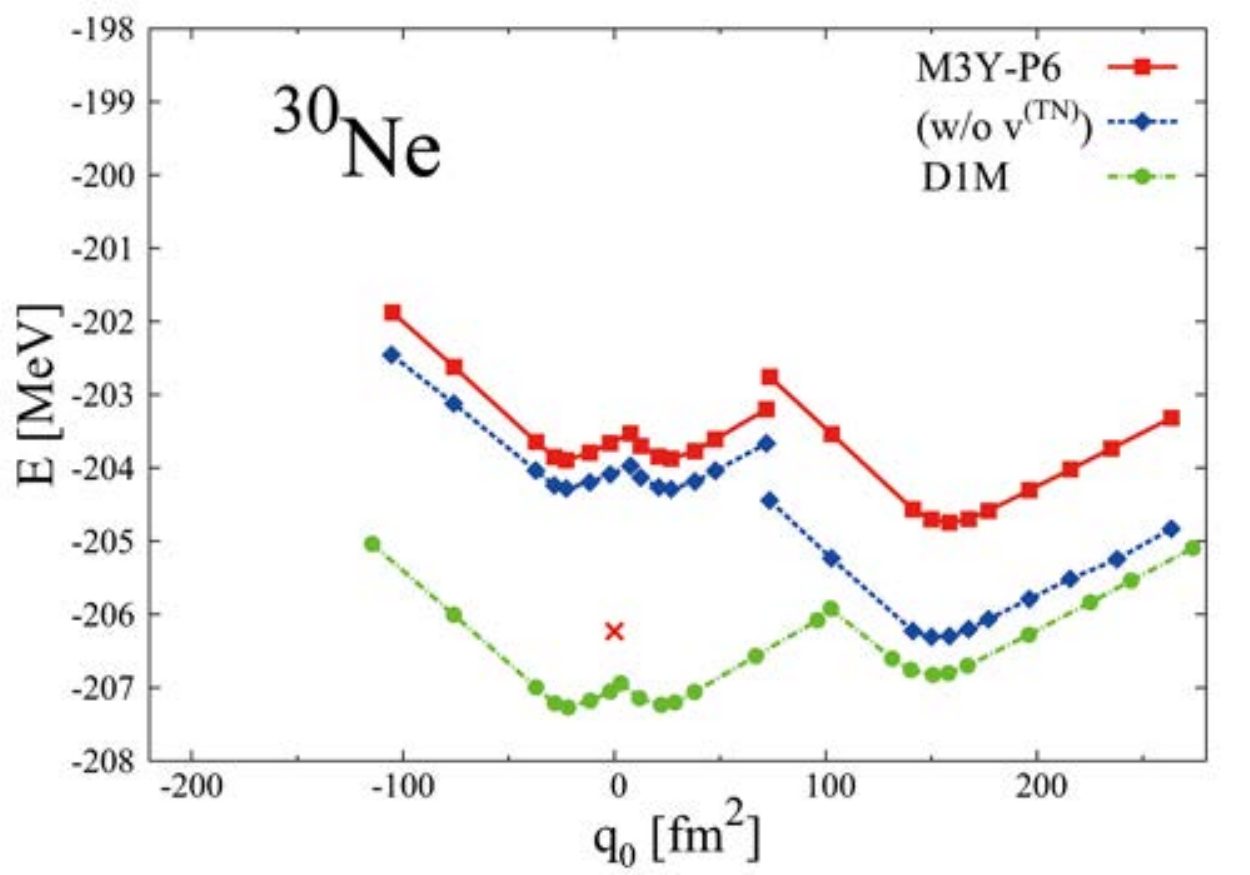}\\
  \includegraphics[scale=0.6]{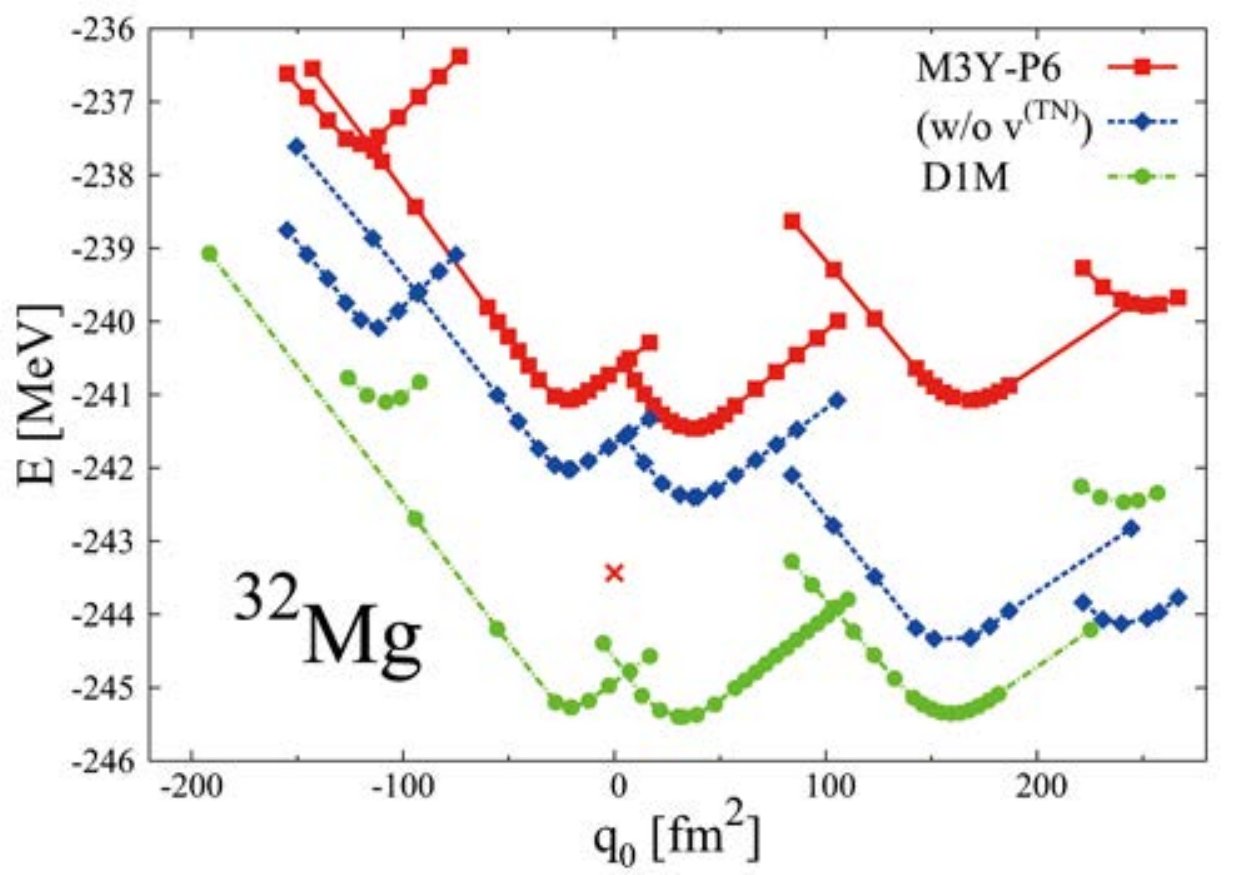}\\
  \includegraphics[scale=0.6]{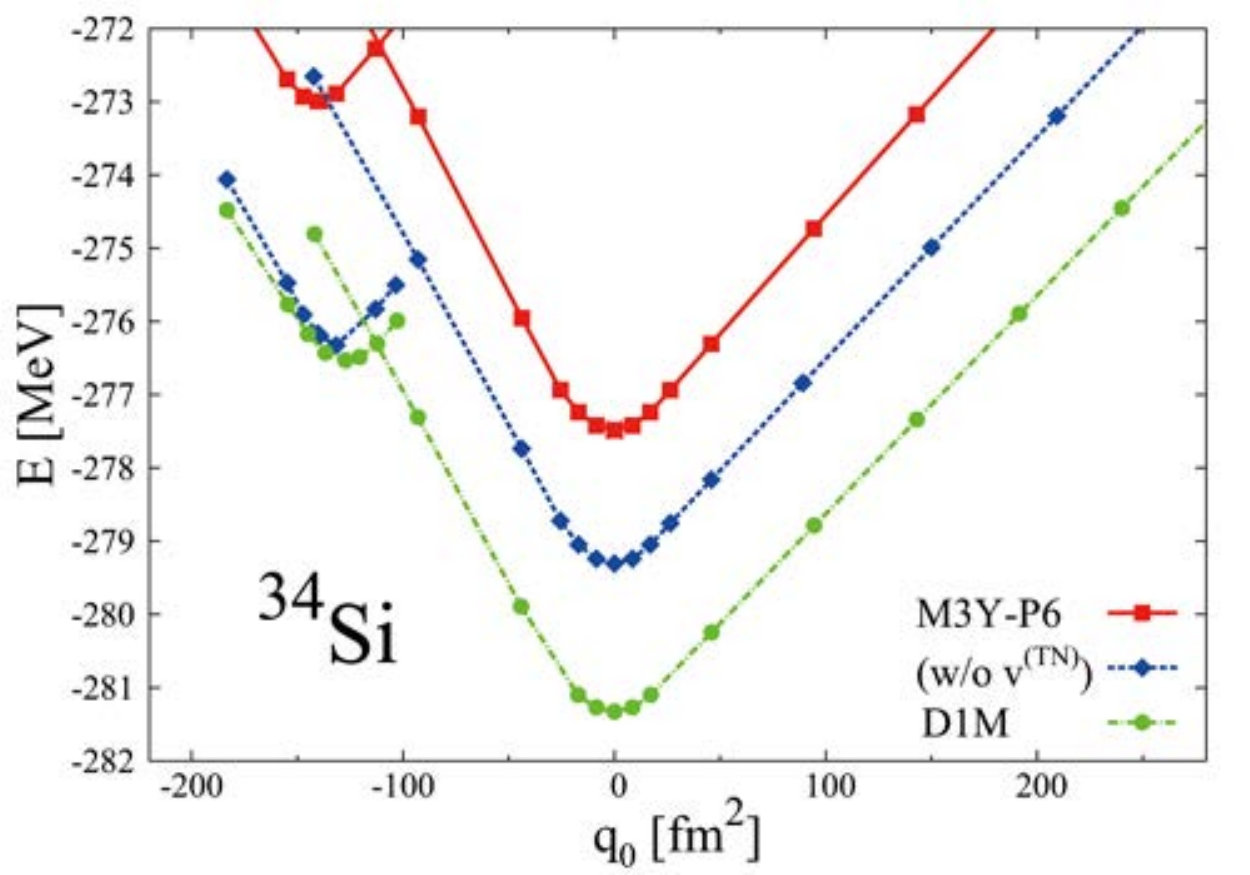}\\
  \end{center}
  \caption{CHF results of $E(q_0)$ (red squares)
    and $\big[E-E^{(\mathrm{TN})}\big](q_0)$ (blue diamonds)
    for $^{30}$Ne, $^{32}$Mg and $^{34}$Si, which are obtained with M3Y-P6.
    For comparison,
    the energy obtained from the spherical HFB calculation (red cross)
    and $E(q_0)$ with D1M (green circles) are also plotted.
    Lines are drawn to guide the eyes.
    Quote from Ref.~\protect\refcite{Suzuki-Nakada-Miyahara_2016}.
    \label{fig:N20_E-q0}}
\end{figure}

\begin{figure}\begin{center}
  \includegraphics[scale=0.6]{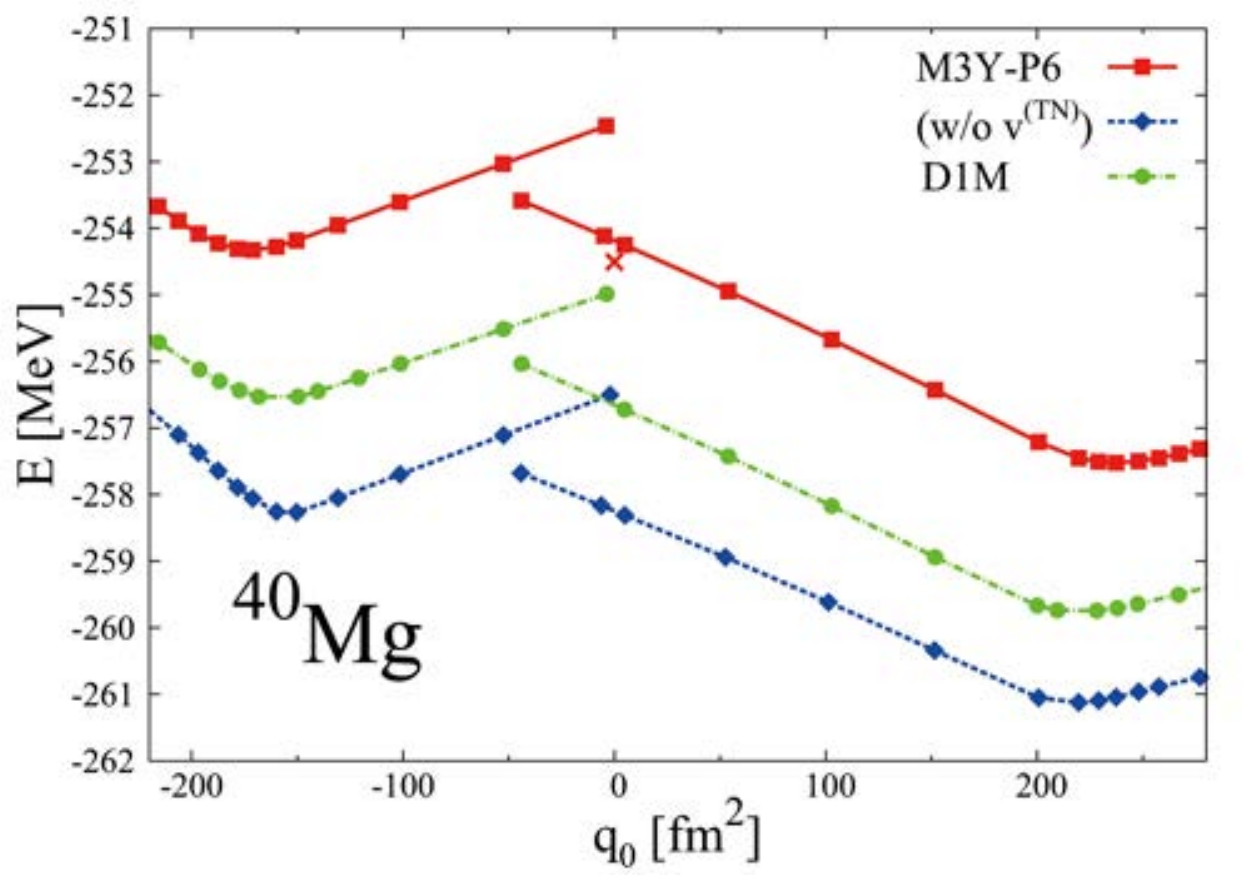}\\
  \includegraphics[scale=0.6]{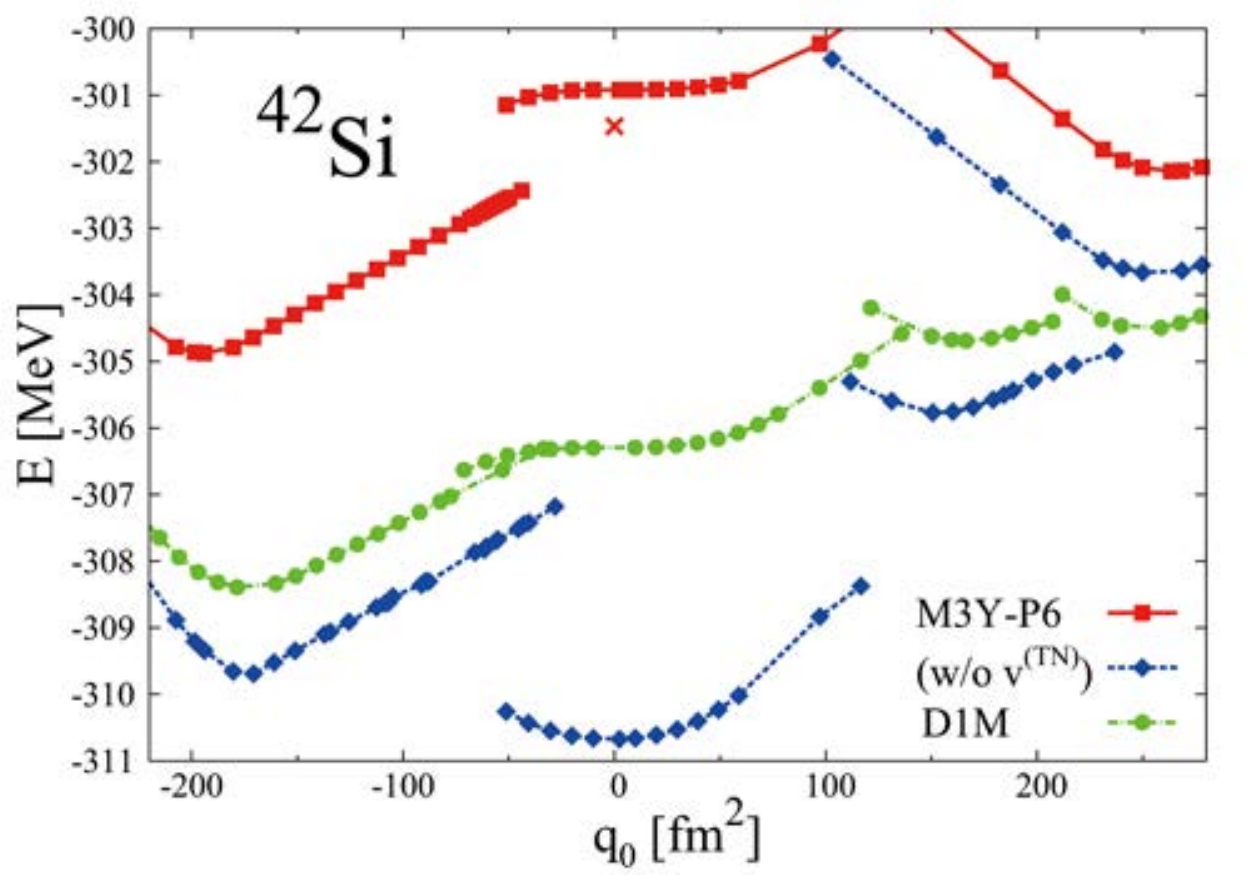}\\
  \includegraphics[scale=0.6]{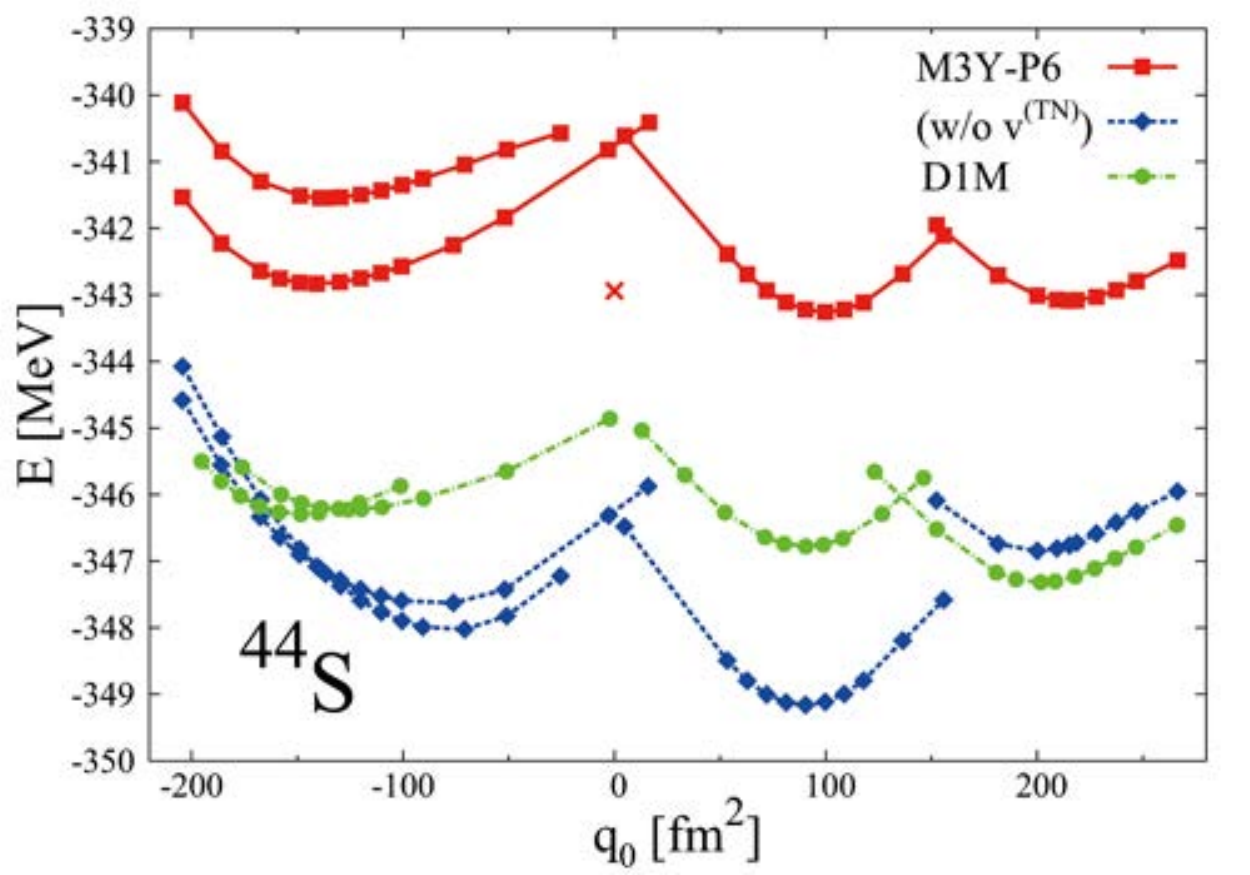}\\
  \end{center}
  \caption{CHF results for $^{40}$Mg, $^{42}$Si and $^{44}$S.
    See caption to Fig.~\protect\ref{fig:N20_E-q0} for legend.
    Quote from Ref.~\protect\refcite{Suzuki-Nakada-Miyahara_2016}.
    \label{fig:N28_E-q0}}
\end{figure}

The semi-realistic interaction M3Y-P6 has been applied
to the axial HF and the constrained HF (CHF) calculations.
In the latter,
a term constraining the mass quadrupole moment $q_0$ is added to $H$.
For the CHF state $|\Phi(q_0)\ket$ at each $q_0$,
where
\be\begin{split} q_0 &= \sqrt{\frac{16\pi}{5}}\Big\bra\Phi\Big|
\sum_i (\vect{r}_i-\vect{R})^2\,Y^{(2)}_0(\widehat{\vect{r}_i-\vect{R}})
\Big|\Phi\Big\ket \\
&= \sqrt{\frac{16\pi}{5}}\bigg[\Big\bra\Phi\Big|
  \sum_i r_i^2\,Y^{(2)}_0(\hat{\vect{r}}_i)\Big|\Phi\Big\ket
 - \bra\Phi|R^2\,Y^{(2)}_0(\hat{\vect{R}})|\Phi\ket\bigg]\,,
\end{split}\label{eq:q0}\ee
$E(q_0)=\bra\Phi(q_0)|H|\Phi(q_0)\ket$ and $E^{(\mathrm{TN})}(q_0)$ is evaluated.
In Fig.~\ref{fig:N20_E-q0},
the energy curves $E(q_0)$'s and $\big[E-E^{(\mathrm{TN})}\big](q_0)$'s
are depicted for the $N=20$ isotones $^{30}$Ne, $^{32}$Mg and $^{34}$Si.
The axial HF results of $E(q_0)$'s with D1M are also displayed for reference.
The energy curves for the $N=28$ isotones $^{40}$Mg, $^{42}$Si and $^{44}$S
are presented in Fig.~\ref{fig:N28_E-q0}.

It is found that the $q_0$ values at the local minima
are determined primarily by the configurations,
\textit{i.e.} which s.p. levels are occupied.
They hardly depend on the interactions,
D1M or M3Y-P6, with or without $v^{(\mathrm{TN})}$.
The slope of $E(q_0)$ is also insensitive to the interactions.
Namely, $E(q_0)$ for individual configuration shifts almost by a constant
among different interactions.
Therefore, the energy curves of different interactions are well speculated
once the energy shifts are evaluated for individual configurations
at a certain $q_0$.
However, the values of the energy shift significantly
depend on the interactions and the configurations.

From the comparison between $E(q_0)$ and $\big[E-E^{(\mathrm{TN})}\big](q_0)$,
the following conclusions are addressed as tensor-force effects~\cite{
  Suzuki-Nakada-Miyahara_2016},
in addition to those listed in Sec.~\ref{subsec:tensor}.
\begin{romanlist}
\setcounter{enumi}{3}
 \item\label{it:repel} The tensor force acts repulsively (at the MF level).
 \item\label{it:ls-jj} The tensor force tends to lower the spherical state
   relative to the deformed ones at the $\ell s$-closed magic numbers
   (\textit{e.g.} $N=20$),
   while the opposite holds at the $jj$-closed magic numbers
   (\textit{e.g.} $N=28$).
\end{romanlist}
These are consistent with those of Ref.~\refcite{Bender-etal_2007}.
Recall the point (\ref{it:TNS:ls}) in Sec.~\ref{subsec:tensor}.
Since $j=\ell+1/2$ orbits lie lower and have higher occupation probability
than its $\ell s$ partners both for protons and neutrons,
the tensor force is necessarily repulsive at energy minima within the MF frame.
For the point (\ref{it:ls-jj}),
it is a key that the tensor force feels the spin d.o.f.,
in combination with the point (\ref{it:TNS-close}) in Sec.~\ref{subsec:tensor}.
At an $\ell s$-closed magic number,
the repulsive effect of the tensor force becomes minimal
at the spherical configuration.
On the contrary, at a $jj$-closed magic number
the spin d.o.f. are active at the sphericity
because a $j=\ell+1/2$ orbit is filled but its $\ell s$ partner is empty.
Deformation drives the system toward spin saturation
and the repulsion is weakened,
accounting for the point (\ref{it:ls-jj}).
These effects have been confirmed in terms of the s.p. energies~\cite{
  Suzuki-Nakada-Miyahara_2016}.

$E(q_0)$ with D1M is closer to $E(q_0)$ with M3Y-P6
than to $\big[E-E^{(\mathrm{TN})}\big](q_0)$,
after shifting the energies by an appropriate constant for each nuclide.
In this respect, it may be said that D1M includes some tensor-force effects
in an effective manner.
However, different between D1M and M3Y-P6 is apparent
in the $Z$-dependence.
For instance, $E(q_0\approx 150\,\mathrm{fm}^2)-E(q_0\approx 0)$
becomes larger at $^{30}$Ne than at $^{32}$Mg in the D1M results,
but the opposite is correct for $E(q_0)$'s with M3Y-P6.
Because of the close energy,
the prolate and the spherical minima at $^{32}$Mg can be inverted
when additional correlations,
\textit{e.g.} the rotational correlations, are taken into account,
possibly resolving the problem of the magicity in this nucleus.
The close energies of these states seem consistent
with the shape coexistence indicated by experiments~\cite{Wimmer-etal_2010}.

Among the $N=28$ nuclei,
it is remarked that the tensor force shifts the lowest state
from the spherical configuration to the oblate one at $^{42}$Si.
The tensor-force effect is partly incorporated in D1M
in an effective manner again.

\subsubsection{$Z=40$}

The proton number $Z=40$ is $\ell s$-closed magic at the sphericity.
However, it is not so stiff,
and the structure of Zr strongly depends on the neutron number $N$.
Experimentally, it is established that
the Zr nuclei are deformed at $N=40$~\cite{Lister-etal_1987},
spherical at $N=50$~\cite{NuDat},
and become deformed again
at $N=60$~\cite{NuDat,Cheifetz-Jared-Thompson-Wilhelmy_1970}.
Whether and how well such shape evolution is described
can be a good and interesting test of theoretical models and their inputs.

\begin{figure}
\centerline{\includegraphics[scale=0.8]{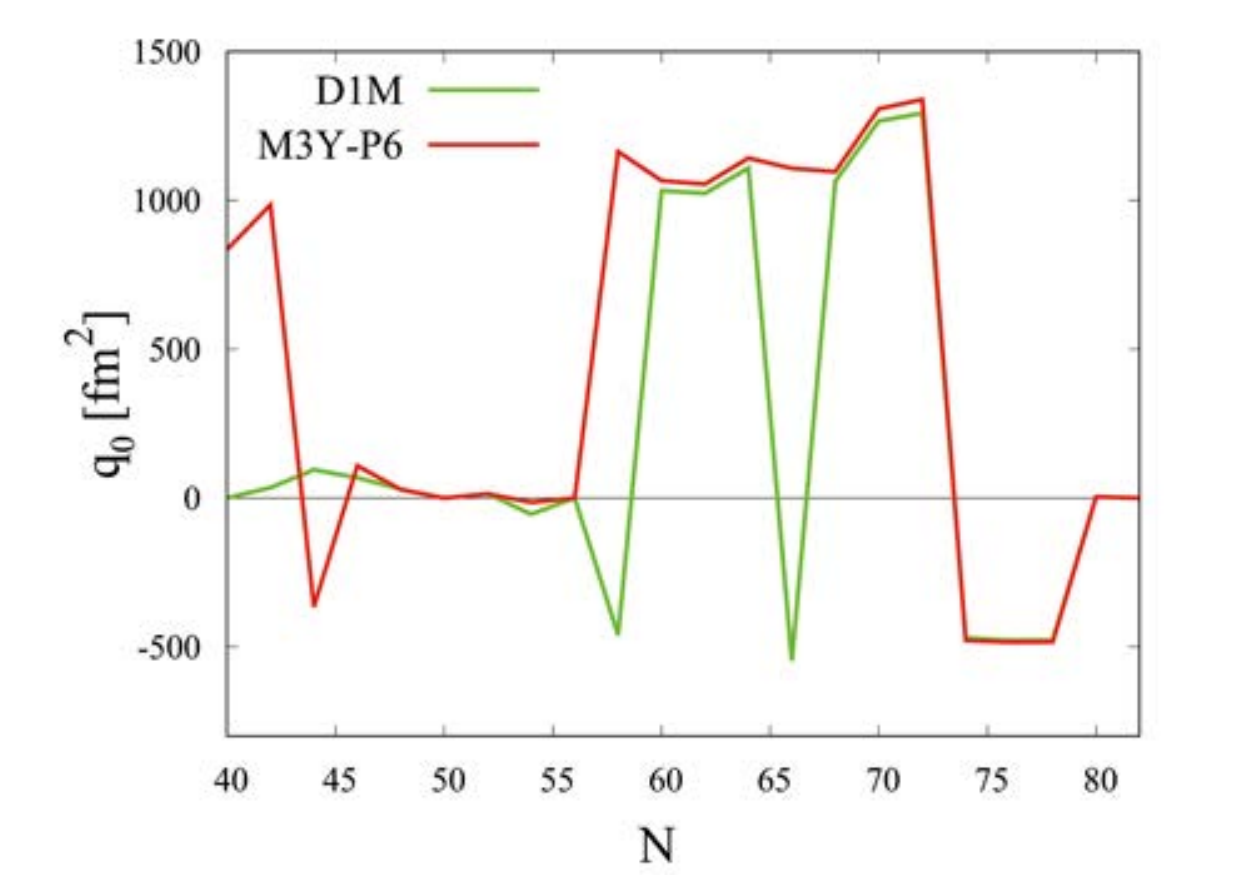}}\vspace*{1cm}
\caption{Values of $q_0$ that give the lowest energy
  for the individual nucleus in the axial HF calculations
  with M3Y-P6 (red line) and D1M (green line).
Quote from Ref.~\protect\refcite{Miyahara-Nakada_2018}.
\label{fig:q0-N}}
\end{figure}

To investigate shape evolution,
the energy minimum is searched for each Zr nucleus
by the axial HF calculations.
Figure~\ref{fig:q0-N} shows $q_0$ values
that give the lowest energy at individual $N$.
Results of M3Y-P6 are presented and compared to those of D1M.

In the present calculations with M3Y-P6,
the Zr nuclei are deformed with the prolate shape at $N\approx 40$,
are spherical at $N\approx 50$,
and become deformed again at $N\approx 60$,
almost consistent with the experimental data.
These results nearly resolve the problem of the magic-number prediction
in the Zr region in Sec.~\ref{subsec:chart}.
Notably, deformation around $^{80}$Zr is naturally obtained.
In contrast, D1M gives spherical shape at $N=40$ at the HF level,
as D1S in the HFB~\cite{D1S-results}.
$N=56$ is indicated to be submagic at $^{96}$Zr
in Fig.~\ref{fig:magic_M3Yp6},
and the spherical shape obtained in the axial HF calculation
supports this indication,
consistently with the high $E_x(2^+_1)$
in measurement~\cite{NuDat,Sadler-etal_1975}.
The spherical-to-prolate shape change occurs at $N=58$ in the present result,
earlier than the experimental indication
of the sudden change from $^{98}$Zr to $^{100}$Zr~\cite{NuDat}.
It was argued that this shape change may be interpreted
as a quantum phase transition~\cite{Togashi-Tsunoda-Otsuka-Shimizu_2016}.
The shape of the Zr nuclei is predicted to stay prolate
with stable values of $q_0$ in $58\leq N\leq 72$,
reminiscent that the measured $E_x(2^+_1)$'s in $60\leq N\leq 70$
are low and close to one another~\cite{Sumikama-etal_2011,Paul-etal_2017}.
In $64\leq N\leq 72$,
the second minimum is obtained on the oblate side.
An oblate minimum becomes lowest in $74\leq N\leq 78$,
and the shape returns to spherical at $N=80$.
The shape evolution predicted in the axial HF with D1M is not quite different.
The difference is found around $N=40$ mentioned above,
at $N=58$ where the lowest minimum is oblate with D1M,
and at $N=66$ where the prolate and oblate minima lie close in energy.

\begin{figure}
\centerline{\includegraphics[scale=0.7]{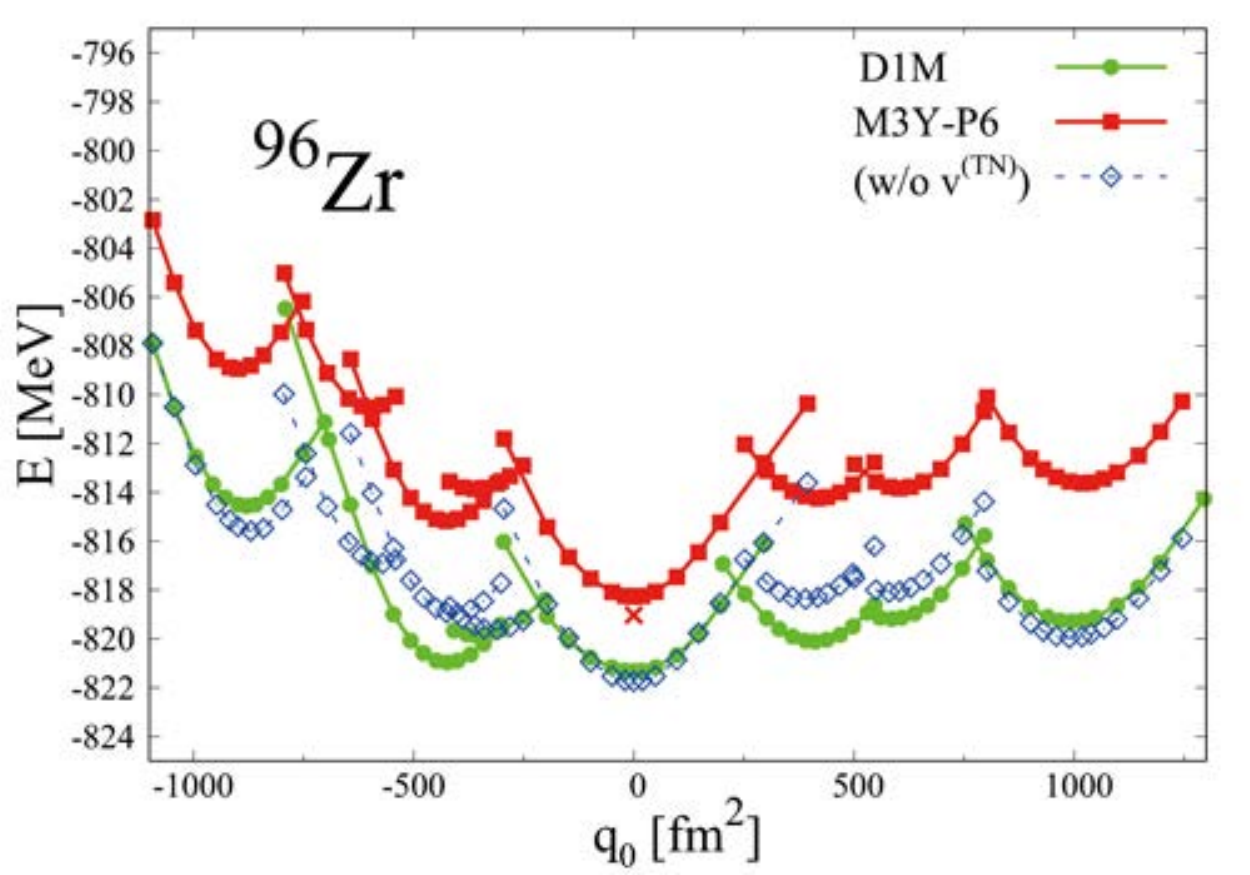}}\vspace*{1cm}
\caption{$E(q_0)$ (red squares)
and $[E-E^{(\mathrm{TN})}](q_0)$ (blue open diamonds)
for $^{96}$Zr, which are obtained by the axial HF and CHF calculations
with M3Y-P6.
$E(q_0)$ with D1M (green circles) are also plotted.
Lines are drawn to guide the eyes.
For reference, the energy obtained by the spherical HFB calculation~\cite{
  Nakada-Sugiura_2014}
is shown by the red cross (at $q_0=0$).
Quote from Ref.~\protect\refcite{Miyahara-Nakada_2018}.
\label{fig:Zr96_E-q0}}
\end{figure}

The effects of the tensor force on the shape evolution of the Zr nuclei
have been investigated in Ref.~\refcite{Miyahara-Nakada_2018}.
I here pick up $^{96}$Zr as an example, which is useful to show an effect
additional to those discussed in Subsec.~\ref{subsubsec:N20&28}.
The energy curve $E(q_0)$ for $^{96}$Zr is depicted in Fig.~\ref{fig:Zr96_E-q0}.
On account of the $\ell s$-closed nature of $Z=40$ at the sphericity,
the repulsive tensor-force effect is suppressed around $q_0=0$.
Although $E(q_0)$ is lowest at $q_0=0$ in all the results here,
energy difference between the spherical and the deformed minima
is smaller in $\big[E-E^{(\mathrm{TN})}\big](q_0)$ and in the D1M results
than in $E(q_0)$ with M3Y-P6.
A shallow minimum around $q_0=0$ was reported
in the HFB result with D1S~\cite{D1S-results}.
Thus, the tensor force makes the spherical minimum more stable,
enhancing the magicity in this nucleus.
Several minima on the prolate and oblate sides likely lead to shape coexistence
as indicated by experiments~\cite{Kremer-etal_2016}.

\begin{figure}
\centerline{\includegraphics[scale=1.0]{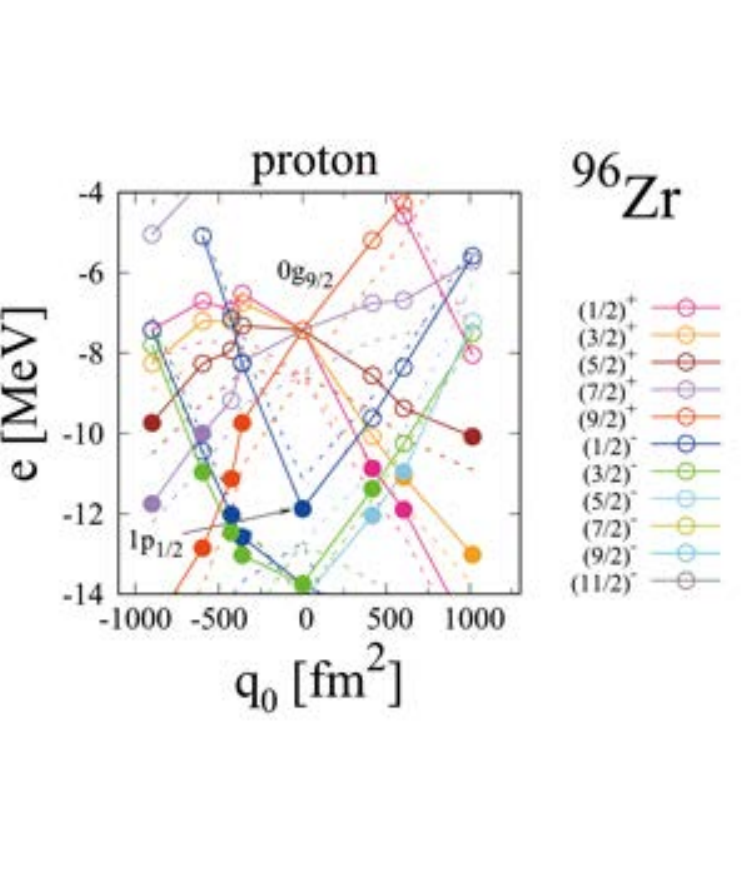}
\hspace*{-0.7cm}\includegraphics[scale=1.0]{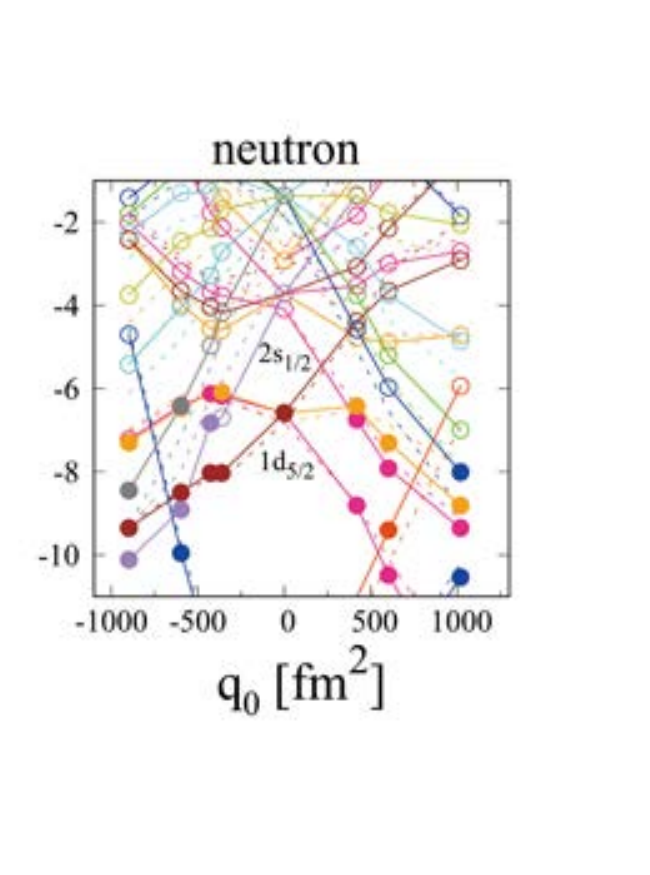}}
\caption{Proton and neutron s.p. energies $\epsilon(k)$
in $^{96}$Zr obtained by the axial HF calculations with M3Y-P6,
at the minima shown in Fig.~\ref{fig:Zr96_E-q0}.
Occupied (unoccupied) levels are represented by the filled (open) circles
connected by the solid lines.
The quantum number of each level $\Omega^\pi$ is distinguished by colors
and is indicated in the middle.
Dashed lines show $\big[\epsilon-\epsilon^{(\mathrm{TN})}\big](k)$.
Labels for several spherical orbits are given for reference.
Quote from Ref.~\protect\refcite{Miyahara-Nakada_2018}.
\label{fig:Zr96_spe}}
\end{figure}

The roles of the tensor force are investigated in more detail
via the s.p. levels.
In Fig.~\ref{fig:Zr96_spe}
the s.p. levels $\epsilon(k)$ near the Fermi energy are displayed
and compared with $\big[\epsilon-\epsilon^{(\mathrm{TN})}\big](k)$.
As $n0g_{9/2}$ and $n1d_{5/2}$ are occupied,
the tensor force enhances the shell gap between $p1p_{1/2}$ and $p0g_{9/2}$
at the spherical minimum.
For neutrons, the gap between $n1d_{5/2}$ and $n0g_{7/2}$ is relatively large,
and $n2s_{1/2}$ becomes the lowest unoccupied level.
It was shown in Figs. 6 and 9 of Ref.~\refcite{Nakada-Sugiura_2014}
that D1M gives smaller shell gaps than M3Y-P6 both for protons and neutrons,
which make the energies of the deformed minima
close to that of the spherical minimum.

Under the presence of a unique-parity orbit (\textit{e.g.} $n0h_{11/2}$),
the tensor force has an additional effect favoring sphericity.
If the unique-parity orbit is occupied,
the spin d.o.f. are active
and thereby the repulsive effects of the tensor force become strong.
Figure~\ref{fig:Zr96_spe} shows that a s.p. level dominated by $n0h_{11/2}$
is occupied at the $q_0\approx 1000\,\mathrm{fm}^2$ minimum,
raising its energy in Fig.~\ref{fig:Zr96_E-q0}.
This effect should be stronger in the lower part of the major shell
than in the upper part of the major shell.
In the lower part of the major shell,
the unique-parity orbit is almost empty at the spherical limit,
while gains higher occupancy as the deformation grows.
For nuclei in which the unique-parity orbit is located
above but not distant from the Fermi energy,
energies of the deformed states are raised by the tensor force.
As a result, the tensor force further enhances the magicity of $N=56$
at this nucleus.
On the other hand, in the upper part of the major shell
the unique-parity orbit is partially occupied at the spherical limit,
and deformation does not strengthen the repulsion due to the tensor force
so much.

\subsection{Deformed halo}\label{subsec:def-halo}

As argued in Sec.~\ref{subsec:halo},
nuclear halos have been observed in light nuclei.
Since beams of heavier unstable nuclei come available,
experimental evidence for halos has been reported up to $^{37}$Mg.
Because the $s$ or the $p$-wave should be dominant in the halos,
deformation significantly affects halos,
which mixes components having different $\ell$ values.
I here argue halos in neutron-rich Mg nuclei,
for which the pairing among neutrons works crucially,
using the axial HFB results with M3Y-P6.
Many of the Mg nuclei have been known to be well-deformed.
The neutron-rich Mg isotopes have been indicated to be deformed
as well~\cite{Doornenbal-etal_2013,Crawford-etal_2019}.
In these nuclei,
the pairing, the quadrupole deformation and the w.f. asymptotics
act cooperatively or competitively.
The SCMF framework, which relies on the variational principle,
is quite suitable to take account of them all that may be intertwined.

The heaviest halo nucleus ever observed is $^{37}$Mg,
for which enhancement of the reaction cross-section
has been discovered~\cite{Takechi-etal_2014}
and $p$-wave dominance of the last neutron
has been indicated~\cite{Kobayashi-etal_2014}.
The r.m.s. matter radii $\sqrt{\bra r^2\ket}$ of the Mg isotopes
including $^{37}$Mg have been extracted in Ref.~\refcite{Watanabe-etal_2014}.
As well as the enhancement at $^{37}$Mg,
irregular behavior was also found at $^{35}$Mg,
with $\sqrt{\bra r^2\ket}$ smaller than the average
of the neighboring even-$N$ nuclei $^{34}$Mg and $^{36}$Mg.

\begin{figure}\begin{center}
  \includegraphics[scale=0.45]{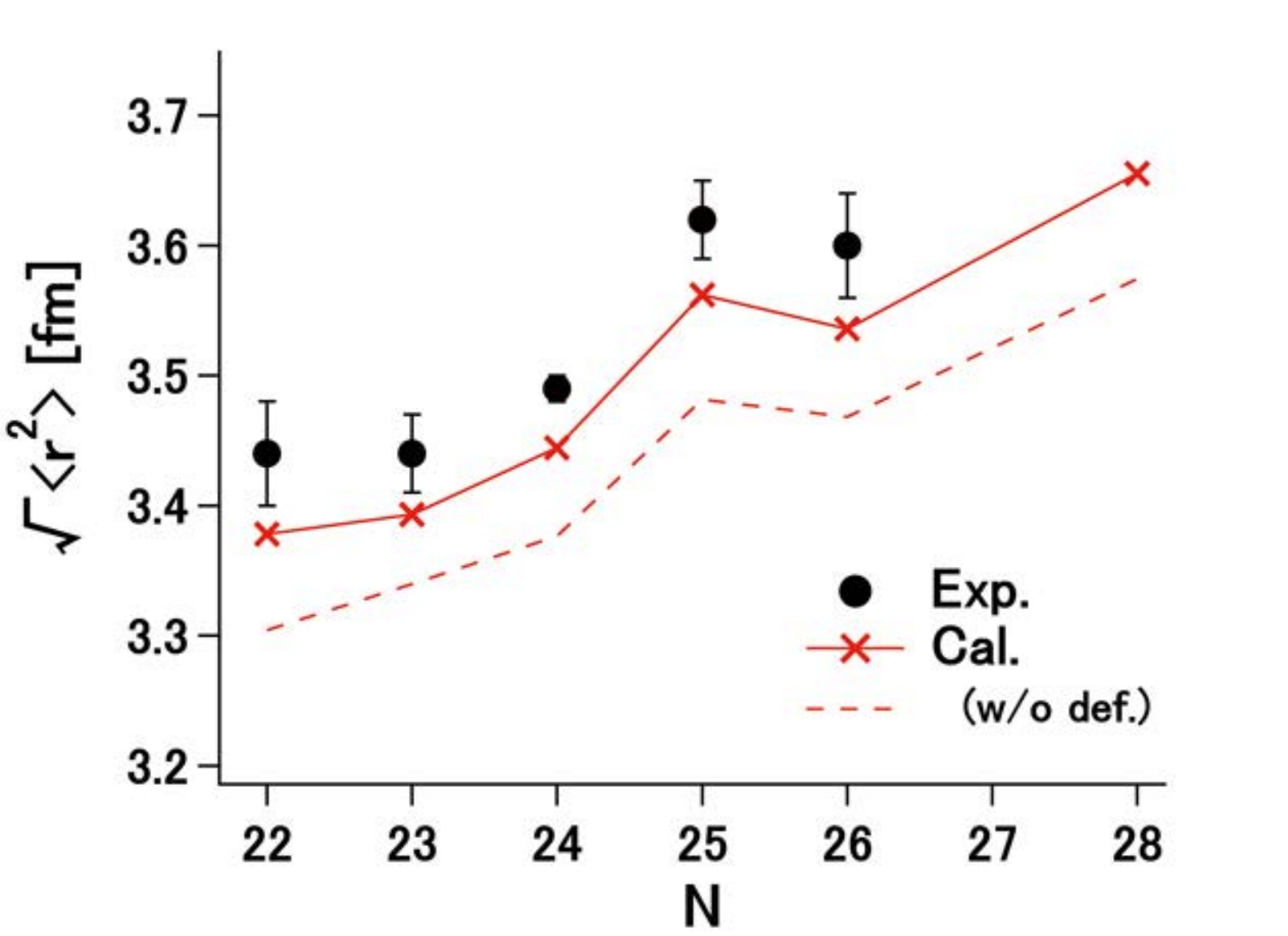}\\
  \includegraphics[scale=0.52]{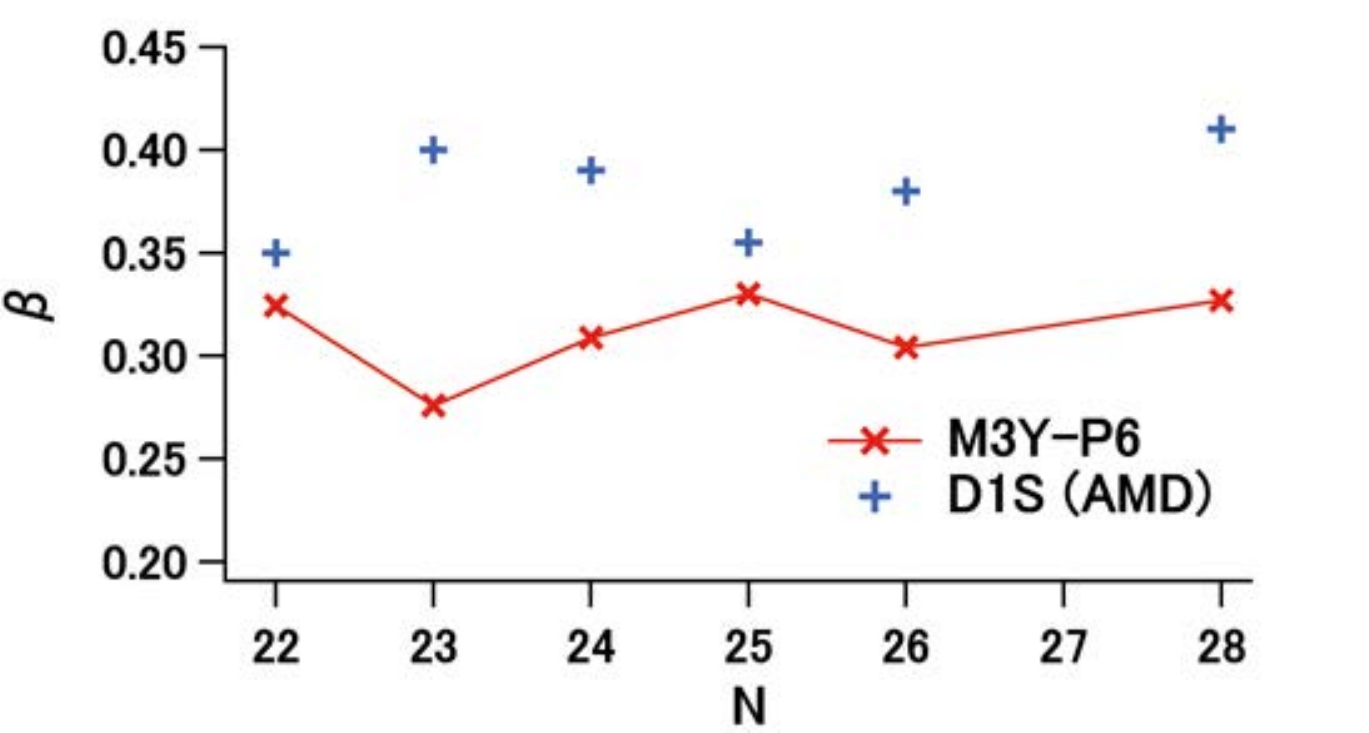}\vspace*{3mm}
  \caption{
    Upper panel: R.m.s. matter radii $\sqrt{\bra r^2\ket}$ in $^{34-40}$Mg.
    The crosses connected by the solid line
    represent the HFB results with M3Y-P6,
    and the dots with error bars are experimental values
    extracted from reaction cross-sections~\cite{Watanabe-etal_2014}.
    For reference, $\bar{r}_0$ values [see Eq.~(\ref{eq:rbar-beta})]
    are plotted by the dashed line.
    Lower panel: Deformation parameter $\beta$.
    The crosses are obtained from the present HFB results with M3Y-P6
    via Eq.~(\ref{eq:rbar-beta}).
    The pluses are the AMD results quoted
    from Ref.~\protect\refcite{Watanabe-etal_2014}.
    Quote from Ref.~\protect\refcite{Nakada-Takayama_2018}.
\label{fig:Mg_rad}}\end{center}
\end{figure}

The axial HFB results of $\sqrt{\bra r^2\ket}$ in $^{34-38}$Mg and at $^{40}$Mg
are presented in the upper panel of Fig.~\ref{fig:Mg_rad},
in comparison with the experimental values~\cite{Watanabe-etal_2014}.
The $^{39}$Mg nucleus is predicted to be unbound,
for which no bound state has been observed in experiments so far.
$^{40}$Mg has been produced~\cite{Baumann-etal_2007},
but no data of $\sqrt{\bra r^2\ket}$ are available.
Although the absolute values are slightly underestimated,
the present calculations reproduce the $N$-dependence
of $\sqrt{\bra r^2\ket}$ in $^{34-38}$Mg remarkably well.

Nuclear deformation influences nuclear radii,
as expressed as
\be \bra r^2\ket \approx \bar{r}_0^2\Big(1+\frac{5}{4\pi}\beta^2\Big)\,,
\label{eq:r2-def}\ee
for small deformation parameter $\beta$~\cite{Bohr-Mottelson_1}.
To distinguish the effects of halos from those of deformation,
the following relations~\cite{Watanabe-etal_2014} are assumed,
\be\begin{split}
    \bra r^2\ket &= \frac{\bar{r}_0^2}{3}
    \bigg[\exp\Big(2\sqrt{\frac{5}{4\pi}}\beta\Big)
      +2\exp\Big(-\sqrt{\frac{5}{4\pi}}\beta\Big)\bigg]\,,\\
    \frac{q_0}{A} &= \frac{\bar{r}_0^2}{3}
    \bigg[2\exp\Big(2\sqrt{\frac{5}{4\pi}}\beta\Big)
      -2\exp\Big(-\sqrt{\frac{5}{4\pi}}\beta\Big)\bigg]\,.
\end{split}\label{eq:rbar-beta}\ee
The parameter $\bar{r}_0$ on the r.h.s. corresponds to
the r.m.s. matter radius at the spherical limit,
which does not include effects of deformation
and therefore represents the effects of halos.
From the HFB results of $\bra r^2\ket$ and $q_0$,
$\bar{r}_0$ and $\beta$ are calculated via Eq.~(\ref{eq:rbar-beta})
for individual nuclei,
and depicted in Fig.~\ref{fig:Mg_rad}.

Halos are more clearly seen in the density distribution $\rho(\vect{r})$,
whereas in most cases $\rho(\vect{r})$ is not quick to be accessed
experimentally.
The asymptotic form of the q.p. w.f.'s at large $r$
has been given by Eq.~(\ref{qpwf-asymp}) in Sec.~\ref{subsec:asymp}.
For even-even nuclei, the density distribution is given by~\footnote{
  The c.m. correction discussed in \ref{app:cmcorr} is neglected here,
  because it hardly influences the asymptotics and halos.
  Namely, $\rho(\vect{r})$ here is nothing but $\rho^{(0)}(\vect{r})$
  in \ref{app:cmcorr}.}
\be \rho(\vect{r}) = \sum_k \big|V_k (\vect{r})\big|^2\,.
\label{eq:Rmat3}\ee
The asymptotic form of $\rho(\vect{r})$ is then obtained
as~\cite{Nakada_2006,Skyrme-P}
\be r^2\rho(\vect{r})\approx \exp(-2\eta^\mathrm{min}_+ r)\,,
\label{eq:dns-asymp}\ee
where $\eta^\mathrm{min}_\pm=\sqrt{2M(|\lambda|\pm\varepsilon^\mathrm{min})}$
with $\varepsilon^\mathrm{min}$ denoting the lowest q.p. energy.

Within the HFB frame, g.s. of an odd-$N$ nucleus
should have one q.p. on top of the HFB vacuum.
There are blocking effects due to the q.p.,
which can be handled by the interchange $(U,V)\,\leftrightarrow\,(V^\ast,U^\ast)$
for the q.p. state~\cite{Ring-Schuck}.
By denoting the q.p. state by $k_1$,
the density distribution of an odd-$N$ nucleus is obtained by
\be \rho(\vect{r}) = \sum_{k\,(\ne k_1)} \big|V_k (\vect{r})\big|^2
  + \big|U_{k_1} (\vect{r})\big|^2\,,
\label{eq:Rmat4}\ee
deriving the asymptotics as
\be r^2\rho(\vect{r})\approx \exp(-2\eta^\mathrm{min}_- r)\,,
\label{eq:dns-asymp2}\ee
instead of Eq.~(\ref{eq:dns-asymp}),
where $\varepsilon^\mathrm{min}=\varepsilon(k_1)$.
Since $\varepsilon(k)\geq 0$,
$\eta_{k-}\leq \sqrt{2M|\lambda|}\leq \eta_{k+}$ for any $k$ and therefore
$\eta^\mathrm{min}_{-}\leq \sqrt{2M|\lambda|}\leq \eta^\mathrm{min}_{+}$
are satisfied.
This inequality implies the tendency that densities of odd-$N$ nuclei
distribute more broadly than those of neighboring even-$N$ nuclei
near the drip line,
as long as they are bound.

To elucidate the roles of the pairing,
the q.p. energies in the HFB are approximated
by those of the HF+BCS scheme.~\footnote{
  Although the HF+BCS scheme does not give correct asymptotics
  as typically known as the neutron-gas problem~\cite{
    Dobaczewski-Nazarewicz-Werner_1996},
  it is here used only for assessing the q.p. energy $\varepsilon(k)$.
  As in Ref.~\refcite{Bennaceur-Dobaczewski-Ploszajczak},
  a similar argument applies
  to the canonical-basis representation of the HFB
  under an appropriate approximation.}
In the HF+BCS the q.p. energy is expressed
in terms of $\epsilon^\mathrm{HF}(k)$, the s.p. energy in the HF,
and the pairing gap $\Delta_k$,
by $\varepsilon(k)=\sqrt{[\epsilon^\mathrm{HF}(k)-\lambda]^2+\Delta_k^2}$.
Suppose that the s.p. level $k_1$ has $\epsilon^\mathrm{HF}(k_1)\approx\lambda$
and that $|\Delta_k|$ does not strongly depend on $k$ near the Fermi energy.
Then $\varepsilon^\mathrm{min}\approx|\Delta_{k_1}|$
and $\eta^\mathrm{min}_\pm\approx\sqrt{2M(|\lambda|\pm|\Delta_{k_1}|)}$ follow.
In the absence of the pair correlation,
$\Delta_{k_1}=0$ and hence $\eta^\mathrm{min}_\pm\approx\sqrt{2M|\lambda|}$
is obtained.
Compared to this normal-fluid case,
for even-$N$ nuclei $\rho(\vect{r})$ decays more rapidly
by the onset of the pair correlation.
This effect was known
as the pairing anti-halo effect~\cite{Bennaceur-Dobaczewski-Ploszajczak}.
In sharp contrast, for odd-$N$ nuclei
the pairing makes $\rho(\vect{r})$ damp more slowly,
likely enhancing a halo.
This new mechanism, called `unpaired-particle haloing'
in Ref.~\refcite{Nakada-Takayama_2018},
works strongly if $|\lambda|\approx|\Delta_{k_1}|$,
even when neither $\lambda$ nor $\Delta_{k_1}$ does not vanish.
Although the pair correlation could diminish
for small $|\lambda|$~\cite{Hamamoto_2004,Hamamoto-Mottelson_2004,
  Hamamoto-Sagawa_2004},
the unpaired-particle haloing starts bringing into action earlier,
at sizable $|\lambda|$.
Similar broadening mechanism for excited states
was pointed out in Ref.~\refcite{Yamagami_2005}.
There is also an argument
that the pairing could induce coupling to the continuum in some cases,
tending to enhance halos~\cite{Chen-Ring-Meng_2014,
  Zhang-Chen-Meng-Ring_2017}.

\begin{figure}
  \includegraphics[scale=0.48]{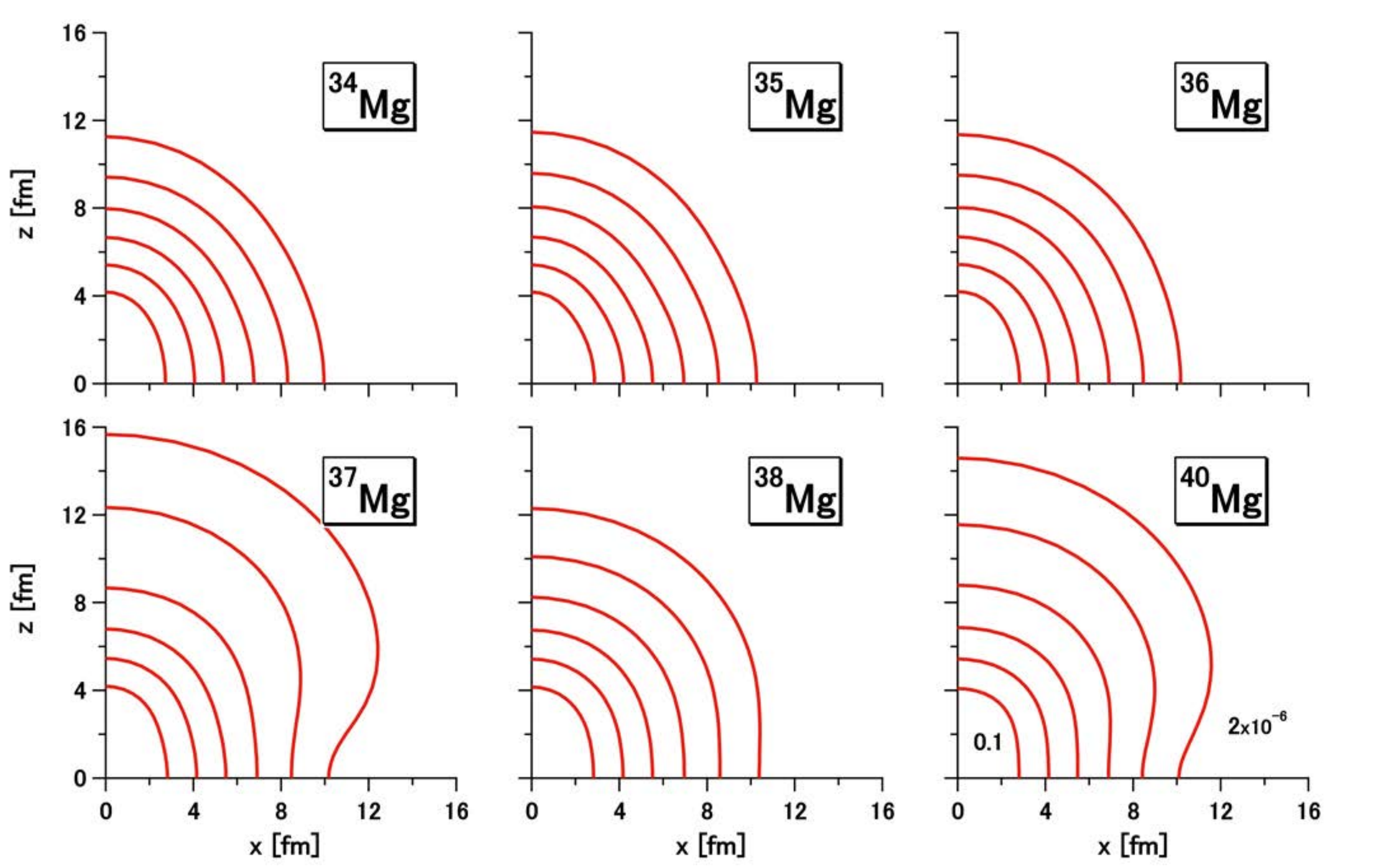}\vspace*{3mm}
  \caption{Contour plot of $\rho(\vect{r})$ on the $zx$-plane for $^{34-40}$Mg,
    obtained by the HFB calculations with the M3Y-P6 interaction.
    Positions of $\rho(\vect{r})=0.1$, $2\times 10^{-2}$, $2\times 10^{-3}$,
    $2\times 10^{-4}$, $2\times 10^{-5}$ and $2\times 10^{-6}\,\mathrm{fm}^{-3}$
    are presented.
    Quote from Ref.~\protect\refcite{Nakada-Takayama_2018}.
\label{fig:Mg_rhocnt}}
\end{figure}

The calculated density distributions of $^{34-40}$Mg are depicted
in Fig.~\ref{fig:Mg_rhocnt},
in terms of the equi-density lines on the $zx$-plane,
where the $z$-axis is the symmetry axis
and the $x$ coordinate represents the distance from the $z$-axis.
The $\mathcal{R}$-symmetry yields the reflection symmetry
with respect to the $xy$-plane.
As the equi-density lines are drawn for exponentially decreasing values
of $\rho(\vect{r})$,
the almost constant interval of the lines for large $r\,(=\sqrt{x^2+z^2})$
implies that the present numerical method discussed in Sec.~\ref{subsec:GEM}
well describes the exponential asymptotics
of Eqs.~(\ref{eq:dns-asymp},\ref{eq:dns-asymp2}).
Figure~\ref{fig:Mg_rhocnt} clearly shows halos in $^{37}$Mg
and $^{40}$Mg.
The peanut shape in the intrinsic states of these halos
are a result of the $p$-wave contribution.
In practice, the unpaired particle in $^{37}$Mg occupies
a level mainly comprised of the $p_{3/2}$ component,
having $\Omega^\pi=(1/2)^-$;
\textit{i.e.} $[N\,n_3\,\Lambda\,\Omega]=[3\,1\,0\,\frac{1}{2}]$
in terms of the Nilsson asymptotic quantum number.
The same orbital is responsible also for the halo in $^{40}$Mg.

In the present result, $^{37}$Mg gains a sizable pair correlation.
The neutron chemical potential is $-2.25\mathrm{MeV}$,
and the halo structure cannot be accounted for 
if the pair correlation is ignored.
However, the pairing leads to loose binding of q.p. states and a halo,
giving $\varepsilon^\mathrm{min}=2.02\,\mathrm{MeV}$
and therefore $|\lambda|-\varepsilon^\mathrm{min}\approx 0.2\,\mathrm{MeV}$.
This $|\lambda|-\varepsilon^\mathrm{min}$ value
gives rise to the long-tailed asymptotics of $\rho(\vect{r})$,
a manifestation of the unpaired-particle haloing.
For $^{40}$Mg, the pair correlation is quenched,
and this nucleus is free from the pairing anti-halo effect.
An analogous result was reported from the HFB calculation
with the D1S interaction in Ref.~\refcite{Nakada_2008}.

In addition to the pairing,
the deformation has an important effect
on the halos and the $N$-dependence of the radii in this region.
The role of the deformation at $^{35}$Mg is obvious,
when the present results are compared to
those of the antisymmetrized molecular dynamics (AMD)
in Ref.~\refcite{Watanabe-etal_2014},
which are successful in describing overall $N$-dependence
of radii in the Mg isotopes but are not so good in $^{34-38}$Mg.
The reduction of $\sqrt{\bra r^2\ket}(\mbox{$^{35}$Mg})$
is attributed to the smaller $\beta$ in $^{35}$Mg than in $^{34,36}$Mg,
as shown in the lower panel of Fig.~\ref{fig:Mg_rad}.
The value of $\beta$ is larger in $^{37}$Mg
than in the neighboring isotopes.
Although the larger $\bar{r}_0(\mbox{$^{37}$Mg})$ seems to account for
most of the enhancement of $\sqrt{\bra r^2\ket}$ in Fig.~\ref{fig:Mg_rad},
$\bar{r}_0$ and $\beta$ contribute cooperatively to the enhancement.
Deformation affects the ordering of the s.p. orbits~\cite{
  Misu-Nazarewicz-Aberg,Zhou-Meng-Ring-Zhao_2010}.
The occupation probability on the $\Omega^\pi=(1/2)^-$ q.p. level
(\textit{i.e.} $[3\,1\,0\,\frac{1}{2}]$)
is lower than that on the $\Omega^\pi=(5/2)^-$ level
(\textit{i.e.} $[3\,1\,2\,\frac{5}{2}]$) in $^{36,38}$Mg.
If the deformation were weak on the prolate side,
the last neutron should occupy the $\Omega^\pi=(5/2)^-$ level
dominated by the $0f_{7/2}$ component.
The larger deformation in $^{37}$Mg
is crucial for the q.p. state with $\Omega^\pi=(1/2)^-$ to be its g.s.,
which can be dominated by the $p$-wave and forms the halo.
Thus the deformation assists the unpaired-particle haloing
to operate,
and the halo drives the larger deformation to gain energy.

In Ref.~\refcite{Urata-Hagino-Sagawa_2017},
HFB calculations were performed in $^{36-38}$Mg
on top of the phenomenological deformed Woods-Saxon potential,
by taking the deformation $\beta$ as an $N$-independent parameter.
The staggering in the matter radii was attributed
to the pairing anti-halo effect in $^{38}$Mg,
whereas the g.s. of $^{37}$Mg stayed in the normal fluid phase.
In the present SCMF study,
the pair correlation survives at $^{37}$Mg,
which makes the staggering in $^{36-38}$Mg stronger
via the unpaired-particle haloing mechanism.

In Refs.~\refcite{Chen-Ring-Meng_2014,Zhou-Meng-Ring-Zhao_2010,Meng-etal_2006,Li-Meng-Ring-Zhao-Zhou_2012},
neutron halos up to more neutron-rich Mg isotopes ($^{42-46}$Mg)
have been argued
via the relativistic Hartree-Bogolyubov calculations,
though restricted to even-$N$.
However, the Mg nuclei beyond $N=28$ are not bound
in the present calculation using the M3Y-P6 interaction,
as in the HFB calculations with the Gogny-D1S interactions~\cite{D1S-results}.

\section{Summary and outlook}\label{sec:summary}

Ground-state properties of exotic nuclei
and their linkage to the nucleonic interaction have been reviewed,
based on the self-consistent mean-field (SCMF) calculations
with the M3Y-type semi-realistic interactions,
M3Y-P6 and its variant M3Y-P6a to be precise.

Plenty of striking phenomena have been disclosed
in nuclei far off the $\beta$-stability,
since the invention of the secondary beams.
They provide an opportunity to investigate
nuclear structure in close linkage to the nucleonic interaction.
The SCMF theories are useful for this purpose,
which give nuclear wave functions from scratch:
the input of the SCMF approaches is only the effective Hamiltonian,
and no artificial truncation of model space is required
in the SCMF calculations.
They are applicable to nuclei all over the nuclear chart, in principle.
However, while great efforts and certain progress have been made for many years,
there is no effective interaction that can supply reliable results
for all nuclei.
To overcome this situation,
it is important to examine effective interaction
in light of the microscopic nucleonic interaction.
The semi-realistic interactions are among such attempts,
which originate from the $G$-matrix
but with phenomenological modification in several respects.
It is noted that in M3Y-P6 and M3Y-P6a
the tensor force is kept unchanged from the one determined from the $G$-matrix,
and the longest-range channel of the central force
is kept equal to that of the one-pion exchange.
Some of the results are compared with those of the Gogny D1S or D1M interaction,
both of which are among widely applied effective interactions.
Though tensor force is not contained explicitly in D1S and D1M,
its effects are likely incorporated in part,
via adjustment of the parameters to experimental data.
It is a useful information which of the tensor-force effects
can be imitated by the other channels and which cannot.

If systematic data on finite nuclei are available,
they can be extrapolated to the infinite nuclear matter.
They enable us to examine the central channels of the effective interaction,
separated from the non-central channels.
Microscopic calculations on the neutron matter
seem helpful to constrain the isospin-dependence of the central force,
up to the density-dependence,
though further experimental confirmation is awaited.
The spin-dependent channels are not well constrained
from purely phenomenological standpoints.
Referencing the microscopic basis,
\textit{e.g.} keeping the channel of the one-pion exchange,
helps to avoid instability in $\rho\lesssim 4\rho_0$.
An appropriate combination of microscopic results and phenomenology
will be useful and vital in studying the structure of neutron stars,
consistently with properties of nuclei on earth.

The non-central channels of the effective interaction,
the LS and the tensor channels,
are significant in the structure of finite nuclei.
They are often key ingredients of the shell structure.
The tensor force has been demonstrated to affect $Z$- or $N$-dependence
of the shell structure,
typified by the proton-hole states of the Ca nuclei.
With the M3Y-P6 interaction,
magic numbers are predicted in a wide range of the nuclear chart,
consistently with almost all the available data.
The tensor force plays significant roles
in the appearance and disappearance of magic numbers,
and the spin-isospin channel in the central force often affects cooperatively.
At a specific nucleus, the magicity could lead to a proton bubble,
\textit{i.e.} depletion of the proton density at the nuclear center.
It has been argued that $^{34}$Si is good and probably the only candidate
that can be observed in the near future.
Effects of the tensor force on the nuclear deformation
have also been investigated
for $N=20$, $28$ and $Z=40$ nuclei.
Via the Hartree-Fock (HF) calculations,
it is elucidated that the tensor-force effects depend on the configuration
but are insensitive to the mass quadrupole moment for a fixed configuration.
The effects on the single-particle (s.p.) orbits
well account for most of the tensor-force effects in deformed nuclei:
the tensor force acts repulsively,
it favors sphericity at the $\ell s$-closed magic numbers
while favors deformation at the $jj$-closed ones,
and presence of the unique-parity orbits near the Fermi level
delays deformation in the lower part of the major shell.
The axial HF calculations seem to get rid of the discrepancy
of the prediction of magic numbers
made via the spherical Hartree-Fock-Bogolyubov (HFB) calculations,
at $^{32}$Mg and in the $60\lesssim N\lesssim 70$ Zr nuclei.

On the neutron halos,
a new mechanism `unpaired-particle haloing' is argued,
which accounts for enhancement of neutron halos in odd-$N$ nuclei,
and is exemplified at $^{37}$Mg.
Deformation is important as well,
acting in an intertwined manner with the pairing.
With proper balance of these effects,
the SCMF calculations with M3Y-P6 well describe
the irregular $N$-dependence of the matter radii in the neutron-rich Mg nuclei.
A neutron halo is also predicted for $^{40}$Mg.
Both halos at $^{37,40}$Mg have peanut shape,
resulting from the $p$-wave dominance in them.

Though it is essential to nuclear shell structure,
the $\ell s$-splitting had not been well accounted for at microscopic levels.
Relatively recently, it was indicated
that the $3N$ force may account for the missing part of the $\ell s$-splitting,
based on the chiral effective field theory.
Inspired by this indication,
the density-dependent LS channel is newly introduced in M3Y-P6a,
and applied to the radii of the spherical nuclei via the HFB calculations.
It is found that the observed kinks in the differential charge radii
at $^{48}$Ca, $^{132}$Sn and $^{208}$Pb in the individual isotopes
are reproduced.
In particular, the kink at $^{132}$Sn had been predicted
only by the M3Y-P6a interaction and Fayans' energy-density functional (EDF),
and has been experimentally discovered recently.
Kinks are also predicted for matter radii,
and anti-kinks, \textit{i.e.} inverted kinks,
are predicted at the $\ell s$-closed magic numbers.
An anti-kink in the charge radii has been observed at $^{40}$Ca.
The anti-kinks could be good evidence for the $3N$-force effect,
if confirmed also for other $\ell s$-closed nuclei.

Despite the success in some respects,
there remain many things to do and rooms for improvement.
Extensive applications of the semi-realistic interaction are of interest,
although such extensions might lead to a readjustment of the parameters.
Applications to a larger number of nuclei
with taking deformation into account,
hopefully covering all over the nuclear chart, are desirable.
While calculations have been limited to the axially symmetric cases
keeping the parity, $\mathcal{R}$ and $\mathcal{T}$ symmetries,
extensions releasing the symmetry assumptions are desired:
\textit{e.g.} calculations including triaxial deformation
and violation of the $\mathcal{T}$ symmetry.
It is noted that the $\mathcal{T}$ symmetry cannot be maintained
in odd-$A$ nuclei,
if energy is fully optimized within the HF or HFB states.
These developments will allow examining the effective interaction
via nuclear masses,
whose accurate prediction is vital to astrophysics~\cite{
  Goriely-Chamel-Pearson_2009}.
Applications of the M3Y-type interaction to the RPA calculation
were performed for limited cases:
the $M1$ transition in $^{208}$Pb~\cite{Shizuma-etal_2008},
the $E2$ transitions in $^{78}$Ni~\cite{Nakada_2010b}
and the Sn isotopes~\cite{Nakada_2012}.
Since excitations provide additional information
about the effective interaction,
further applications will be useful,
although care would be needed on the effective mass.
Combination with the quantum-number projections,
\textit{e.g.} the angular-momentum projection (AMP),
and the extension to the generator-coordinate method
may also help to investigate excited states,
as well as improving the description of the ground states
by incorporating the residual correlations.
The AMP is harmonious with the interpretation of the nuclear EDF
as a tool to derive intrinsic states~\cite{Messud_2011,Messud_2013}.
However, the complication in computational treatment
because of the finite-range and the Yukawa form of the interaction
could be an obstacle to these extensive applications.
Further exploration of numerical methods could be a way
to overcome this obstacle.
On the other hand,
the microscopic or semi-microscopic study of nuclear structure
would be facilitated
if an appropriate approximation scheme is developed.
This direction seems to merge
into an \textit{ab initio} nuclear EDF~\cite{Drut-Furnstahl-Platter_2010}.
I refer to Refs.~\refcite{Baldo-Robledo-Schuck-Vinas_2013,NavarroPerez-etal_2018,Shen-etal_2019}
as related works.
Through such extensions,
properties of exotic nuclei can be better understood
and hopefully predicted with excellent precision and high reliability.
Experimental data on them may further clarify
the roles of the nucleonic interaction.

Our understanding of nuclear properties has greatly progressed,
and description in good connection to the bare nucleonic interaction
is being within reach,
thanks partly to the new data in exotic nuclei
and to the development of the theory of the nuclear force itself.
As final words, I hope that this review
will facilitate microscopic and global understanding
of low-energy phenomena of nuclei, particularly the nuclear structure.

\section*{Acknowledgements}

Depending on the topics discussed in this review,
I have collaborated with J.~Margueron, T.~Inakura,
M.~Sato, K.~Sugiura, Y.~Suzuki, S.~Miyahara, Y.~Tsukioka
and K.~Takayama, to whom I would express my gratitude.
I am grateful also to K.~Kat\={o}, H.~Kurasawa, M.~Kohno, K.~Ogawa,
T.~Otsuka, A.~Tamii, M.~Yamagami, T.~Shizuma, N.~Van~Giai, Y.~Yamamoto,
Y.~R.~Shimizu, D.~T.~Khoa, S.~Shlomo, W.~Nazarewicz, M.~Matsuo, T.~Nakatsukasa,
H.~Sagawa, S.-G.~Zhou, A.~Barzakh, T.~Suda, H.~Liang, K.~Iida,
and many other colleagues for useful and enlightening discussions.
I finally thank J.~Meng for recommending me to write this review.

Parts of numerical calculations were performed
on HITAC SR24000 at IMIT, Chiba University,
and CRAY XC40 at YITP, Kyoto University.

\appendix

\section{Center-of-mass corrections to nuclear densities}\label{app:cmcorr}

The translational symmetry is necessarily violated in the SCMF w.f.'s.
Therefore, particular care is needed for influence of the c.m. motion.
While projection was developed~\cite{RodriguezGuzman-Schmid_2004},
it is not easy to implement.
In this Appendix, a method of correcting
the influence of the c.m. motion on the nuclear densities is discussed.

\subsection{Violation of translational symmetry\label{subsec:trans-sym}}

Even though the nuclear Hamiltonian has the translational symmetry,
it is unavoidably violated in the MF w.f.'s.
The translationally symmetric Hamiltonian is separated as
\be H_\mathrm{full}=H+H_\mathrm{c.m.}\,, \label{eq:H-full}
\ee
where $H$ and $H_\mathrm{c.m.}$ represent
the relative and c.m. Hamiltonians, respectively;
see Eq.~(\ref{eq:Hamil}).
Equation~(\ref{eq:H-full}) immediately derives
that energy eigenstates can be expressed by a direct product,
\be |\Psi\ket=|\Psi_\mathrm{rel.}\ket\otimes|\Psi_\mathrm{c.m.}\ket\,.
\label{eq:wf-full}\ee
The variables are well separated
into the relative coordinates and the c.m. coordinate,
$|\Psi_\mathrm{rel.}\ket$ depends on $3(A-1)$ relative coordinates
(apart from the spin and isospin)
in contrast to the $3A$ coordinates
$\vect{r}_1,\vect{r}_2,\cdots,\vect{r}_A$ in $|\Psi\ket$.
As $H_\mathrm{c.m.}=\vect{P}^2/2AM$,
$|\Psi_\mathrm{c.m.}\ket$ is taken to be
$|\Psi_\mathrm{c.m.}\ket\propto e^{i\vect{K}\cdot\vect{R}}$.
Our interest is in $H$ and its eigenstate $|\Psi_\mathrm{rel.}\ket$,
since the c.m. d.o.f. are irrelevant to nuclear structure.

However, it is not easy to express many-body w.f.'s explicitly
by the relative coordinates, unless $A$ is small.
As $A$ nucleons are handled in a democratic manner,
the HF w.f.'s are represented by a single Slater determinant
consisting of the s.p. w.f.'s
$\{\varphi_{k_1}(\vect{r}_1),\varphi_{k_2}(\vect{r}_2),\cdots,
\varphi_{k_A}(\vect{r}_A)\}$.
As long as $\varphi_k(\vect{r})$ is localized,
the total w.f. $|\Phi\ket$ is also localized.
Therefore it is impossible to reproduce
$|\Psi_\mathrm{c.m.}\ket\propto e^{i\vect{K}\cdot\vect{R}}$,
which requires a w.f. spread over the entire space.
The same discussion applies to the HFB case.
Thus the translational symmetry is spontaneously violated in the MF w.f.'s
and the MF w.f.'s are influenced by the c.m. motion.
Remark the uncertainty relation between $\vect{R}$ and $\vect{P}$,
which are canonical conjugate variables.
Therefore, it is impossible to constrain either of $\vect{R}$ or $\vect{P}$
without an infinite size of fluctuation of the other.

A practical question is whether and how the influence of the c.m. motion
can be removed in the observables.
Let us first consider the matter radius and the density distribution.
The nuclear m.s. matter radius is defined as
\be \bra r^2\ket = \frac{1}{A}\Big\bra\Psi_\mathrm{rel.}\Big|
\sum_{i=1}^A (\vect{r}_i-\vect{R})^2\Big|\Psi_\mathrm{rel.}\Big\ket
= \frac{1}{A}\Big\bra\Psi\Big| \sum_{i=1}^A (\vect{r}_i-\vect{R})^2
\Big|\Psi\Big\ket\,.
 \label{eq:r2-full}
\ee
Note that $\vect{R}^2=\sum_{i=1}^A \vect{r}_i^2
+\sum_{i<j}^A \vect{r}_i\cdot\vect{r}_j$ consists
one- and two-body operators.
The nucleon density distribution in nuclei
is a physical quantity carrying significant information of nuclear structure.
It is also a fundamental ingredient of the EDF approaches.
The matter density is defined by
\be \rho(\vect{r}) = \Big\bra\Psi_\mathrm{rel.}\Big| \sum_{i=1}^A
\delta\big(\vect{r}-(\vect{r}_i-\vect{R})\big) \Big|\Psi_\mathrm{rel.}\Big\ket
= \Big\bra\Psi\Big| \sum_{i=1}^A
\delta\big(\vect{r}-(\vect{r}_i-\vect{R})\big) \Big|\Psi\Big\ket\,.
 \label{eq:dens-full}
\ee
Analogously, the point-proton m.s. radius and density are defined as
\be\begin{split}
\bra r^2\ket_p &= \frac{1}{Z}\Big\bra\Psi_\mathrm{rel.}\Big|
\sum_{i\in p} (\vect{r}_i-\vect{R})^2\Big|\Psi_\mathrm{rel.}\Big\ket
= \frac{1}{Z}\Big\bra\Psi\Big| \sum_{i\in p} (\vect{r}_i-\vect{R})^2
\Big|\Psi\Big\ket\,,\\
\rho_p(\vect{r}) &= \Big\bra\Psi_\mathrm{rel.}\Big| \sum_{i\in p}
\delta\big(\vect{r}-(\vect{r}_i-\vect{R})\big) \Big|\Psi_\mathrm{rel.}\Big\ket
= \Big\bra\Psi\Big| \sum_{i\in p}
\delta\big(\vect{r}-(\vect{r}_i-\vect{R})\big) \Big|\Psi\Big\ket\,,
\end{split}\label{eq:r2&dens-p}\ee
and likewise for the point-neutron m.s. radius and density.
It should be noticed that $\rho(\vect{r})$ and $\rho_p(\vect{r})$
satisfy the following relations,
\be\begin{split}
 \int d^3r\,\rho(\vect{r}) &= A\,,\\
 \int d^3r\,r^2\,\rho(\vect{r}) &= A\,\bra r^2\ket\,,
\end{split}\label{eq:integ_rho}\ee
and
\be\begin{split}
 \int d^3r\,\rho_p(\vect{r}) &= Z\,,\\
 \int d^3r\,r^2\,\rho_p(\vect{r}) &= Z\,\bra r^2\ket_p\,.
\end{split}\label{eq:integ_rho_p}\ee

The m.s. radius has often been calculated in the MF approaches by
\be \bra r^2\ket^{(0)} = \frac{1}{A}
\Big\bra\Phi\Big|\sum_{i=1}^A \vect{r}_i^2\Big|\Phi\Big\ket\,,
\label{eq:r2-MF}\ee
where the superscript $(0)$ represents that it is a quantity
with no c.m. correction.
It is reasonable to apply Eq.~(\ref{eq:r2-full}) to $|\Phi\ket$,
leading to the c.m.-corrected matter radius of Eq.~(\ref{eq:r2-matter}).
For $\bra r^2\ket_p$, the c.m. correction gives
\be\begin{split}
\bra r^2\ket_p &= \frac{1}{Z} \Big\bra\Phi\Big|
\sum_{i\in p} (\vect{r}_i-\vect{R})^2\Big|\Phi\Big\ket \\
&= \frac{1}{Z} \Big\bra\Phi\Big|\sum_{i\in p} \vect{r}_i^2\Big|\Phi\Big\ket
- \bra\Phi|\vect{R}^2|\Phi\ket
- \frac{2N}{A}\bra\Phi|(\vect{R}_p-\vect{R}_n)\cdot\vect{R}|\Phi\ket\,,
\end{split}\label{eq:r2_p-MF_corr}\ee
where $Z\vect{R}_p=\sum_{i\in p} \vect{r}_i$ and likewise for $\vect{R}_n$.
This is compared to the c.m. correction customarily applied
to the $E1$ transitions~\cite{Eisenberg-Greiner_2}.

In the MF regime the matter density has been calculated by
\be \rho^{(0)}(\vect{r})
= \Big\bra\Phi\Big| \sum_{i=1}^A \delta(\vect{r}-\vect{r}_i)
\Big|\Phi\Big\ket\,,
\label{eq:dens-MF}\ee
whereas the c.m.-corrected matter density should be obtained by
\be \rho(\vect{r})
 = \Big\bra\Phi\Big| \sum_{i=1}^A
 \delta\big(\vect{r}-(\vect{r}_i-\vect{R})\big) \Big|\Phi\Big\ket\,.
 \label{eq:dens-MF_corr}
\ee
In contrast to the m.s. radius,
the operator $\delta\big(\vect{r}-(\vect{r}_i-\vect{R})\big)$
in Eq.~(\ref{eq:dens-MF_corr}) contains many-body operators,
and is not tractable without approximation.
However, $\rho^{(0)}(\vect{r})$ is not consistent
with $\bra r^2\ket$ in the context of (\ref{eq:integ_rho}),
\be
 \int d^3r\,r^2\,\rho^{(0)}(\vect{r})
 = A\,\bra r^2\ket^{(0)}\ne A\,\bra r^2\ket\,.
 \label{eq:integ_rho-MF}
\ee

\subsection{Center-of-mass correction to density\label{subsec:dens}}

In connection to the matter density,
we shall consider the form factor, which is defined by
\be \rho(\vect{r})
 = \frac{1}{(2\pi)^3}\int d^3q\, e^{-i\vect{q}\cdot\vect{r}}\,F(\vect{q})\,,
 \label{eq:dens&ff}
\ee
yielding
\be F(\vect{q}) = \Big\bra\Psi_\mathrm{rel.}\Big| \sum_{i=1}^A
e^{i\vect{q}\cdot(\vect{r}_i-\vect{R})}\Big|\Psi_\mathrm{rel.}\Big\ket
= \Big\bra\Psi\Big| \sum_{i=1}^A e^{i\vect{q}\cdot(\vect{r}_i-\vect{R})}
\Big|\Psi\Big\ket\,.
  \label{eq:ff-full}
\ee
Similarly. the form factor associated with $\rho_p(\vect{r})$
can be defined as
\be \rho_p(\vect{r})
 = \frac{1}{(2\pi)^3}\int d^3q\, e^{-i\vect{q}\cdot\vect{r}}\,F_p(\vect{q})\,,
 \label{eq:dens&ff_p}
\ee
deriving
\be F_p(\vect{q}) = \Big\bra\Psi_\mathrm{rel.}\Big| \sum_{i\in p}
e^{i\vect{q}\cdot(\vect{r}_i-\vect{R})}\Big|\Psi_\mathrm{rel.}\Big\ket
= \Big\bra\Psi\Big| \sum_{i\in p} e^{i\vect{q}\cdot(\vect{r}_i-\vect{R})}
\Big|\Psi\Big\ket\,.
  \label{eq:ff_p-full}
\ee
The separability of $|\Psi\ket$ in Eq.~(\ref{eq:wf-full}) yields
\be
 \Big\bra\Psi\Big| \sum_{i=1}^A e^{i\vect{q}\cdot\vect{r}_i}
 \Big|\Psi\Big\ket
 = \bra\Psi|e^{i\vect{q}\cdot\vect{R}}|\Psi\ket\,
 \Big\bra\Psi\Big| \sum_{i=1}^A e^{i\vect{q}\cdot(\vect{r}_i-\vect{R})}
 \Big|\Psi\Big\ket\,, \label{eq:exp-sep}
\ee
since $\vect{R}$ and $\vect{r}_i-\vect{R}$ depend
only on the c.m. and the relative coordinates, respectively,
and therefore~\cite{Feshbach}
\be F(\vect{q})
 = \bra\Psi|e^{i\vect{q}\cdot\vect{R}}|\Psi\ket^{-1}\,
 \Big\bra\Psi\Big| \sum_{i=1}^A e^{i\vect{q}\cdot\vect{r}_i}\Big|\Psi\Big\ket\,.
 \label{eq:ff-full2}
\ee
The same algebra gives
\be F_p(\vect{q})
 = \bra\Psi|e^{i\vect{q}\cdot\vect{R}}|\Psi\ket^{-1}\,
 \Big\bra\Psi\Big| \sum_{i\in p} e^{i\vect{q}\cdot\vect{r}_i}\Big|\Psi\Big\ket\,.
 \label{eq:ff_p-full2}
\ee

In order to minimize the influence of $H_\mathrm{c.m.}$ on the energy,
$H\,(=H_\mathrm{full}-H_\mathrm{c.m.})$ is used
for the SCMF calculations throughout this review,
as have been implemented in Ref.~\refcite{Campi-Sprung_1972}.
Although the c.m. part of the MF w.f. $|\Phi\ket$
is not exactly separable as in Eq.~(\ref{eq:wf-full}),
it is expected that contamination of the c.m. motion is small
because the energy is minimized in the SCMF calculations,
and the following approximation is still useful,
\be |\Phi\ket\approx|\Phi_\mathrm{rel.}\ket\otimes
 |\Phi_\mathrm{c.m.}\ket\,. \label{eq:wf-MF}
\ee
For the MF w.f. $|\Phi\ket$,
the c.m.-corrected form factor is defined by
\be F(\vect{q})
 =\Big\bra\Phi\Big| \sum_{i=1}^A e^{i\vect{q}\cdot(\vect{r}_i-\vect{R})}
 \Big|\Phi\Big\ket\,. \label{eq:ff-MF_corr}
\ee
Under the approximation of Eq.~(\ref{eq:wf-MF}),
an equation analogous to Eq.~(\ref{eq:ff-full2}) is obtained,
\be F(\vect{q})
 \approx \bra\Phi|e^{i\vect{q}\cdot\vect{R}}|\Phi\ket^{-1}\,
 \Big\bra\Phi\Big| \sum_{i=1}^A e^{i\vect{q}\cdot\vect{r}_i}\Big|\Phi\Big\ket\,.
 \label{eq:ff-MF_corr1}
\ee
In contrast to the many-body operator $e^{i\vect{q}\cdot(\vect{r}_i-\vect{R})}$,
$e^{i\vect{q}\cdot\vect{r}_i}$ is a one-body operator which is tractable.
In the HF case,
\be
 \Big\bra\Phi\Big| \sum_{i=1}^A e^{i\vect{q}\cdot\vect{r}_i}\Big|\Phi\Big\ket
 = \sum_{i=1}^A \bra\varphi_{k_i}|e^{i\vect{q}\cdot\vect{r}}|\varphi_{k_i}\ket\,.
 \label{eq:exp-1b}
\ee
The expectation value $\bra\Phi|e^{i\vect{q}\cdot\vect{R}}|\Phi\ket$
is determined by $\bra\Phi|\vect{R}^n|\Phi\ket$ ($n=0,1,\cdots$).
At low energy, the c.m. part of the MF w.f. $|\Phi\ket$
should not be very complicated.
It seems sensible to apply the cumulant expansion
to $\bra\Phi|e^{i\vect{q}\cdot\vect{R}}|\Phi\ket$,
deriving
\be \bra\Phi|e^{i\vect{q}\cdot\vect{R}}|\Phi\ket
 = \exp\Big[i\vect{q}\cdot\bra\Phi|\vect{R}|\Phi\ket
 -\frac{1}{2}\big\{\bra\Phi|(\vect{q}\cdot\vect{R})^2|\Phi\ket
 -(\vect{q}\cdot\bra\Phi|\vect{R}|\Phi\ket)^2\big\}
 +\cdots\Big]\,.
 \label{eq:cumulant0}
\ee
As we postulate $\bra\Phi|\vect{R}|\Phi\ket=0$,
Eq.~(\ref{eq:cumulant0}) becomes
\be \bra\Phi|e^{i\vect{q}\cdot\vect{R}}|\Phi\ket
 = \exp\Big[-\frac{1}{2}\bra\Phi|(\vect{q}\cdot\vect{R})^2
 |\Phi\ket + \cdots\Big]\,,
 \label{eq:cumulant1}
\ee
and Eq.~(\ref{eq:ff-MF_corr1}) is further approximated as
\be F(\vect{q})
 \approx e^{\frac{1}{2}\bra\Phi|(\vect{q}\cdot\vect{R})^2|\Phi\ket}\,
 \Big\bra\Phi\Big| \sum_{i=1}^A e^{i\vect{q}\cdot\vect{r}_i}
 \Big|\Phi\Big\ket\,.
 \label{eq:ff-MF_corr2}
\ee
This gives an approximation to $\rho(\vect{r})$,
\be \rho(\vect{r})
 \approx \frac{1}{(2\pi)^3}\int d^3q\, e^{-i\vect{q}\cdot\vect{r}}\,
 e^{\frac{1}{2}\bra\Phi|(\vect{q}\cdot\vect{R})^2|\Phi\ket}\,
 \Big\bra\Phi\Big| \sum_{i=1}^A e^{i\vect{q}\cdot\vect{r}_i}
 \Big|\Phi\Big\ket\,.
\label{eq:dens-MF_corr2}\ee
For the point-proton density, the same approximation yields
\be \rho_p(\vect{r})
 \approx \frac{1}{(2\pi)^3}\int d^3q\, e^{-i\vect{q}\cdot\vect{r}}\,
 e^{\frac{1}{2}\bra\Phi|(\vect{q}\cdot\vect{R})^2|\Phi\ket}\,
 \Big\bra\Phi\Big| \sum_{i\in p} e^{i\vect{q}\cdot\vect{r}_i}
 \Big|\Phi\Big\ket\,.
\label{eq:dens_p-MF_corr2}\ee
Since
\be
 \bra\Phi|(\vect{q}\cdot\vect{R})^2|\Phi\ket
 = \sum_{\alpha,\beta=x,y,z} q_\alpha q_\beta\,
 \bra\Phi|R_\alpha R_\beta|\Phi\ket\,,
\ee
the coefficient $e^{\frac{1}{2}\bra\Phi|(\vect{q}\cdot\vect{R})^2
|\Phi\ket}$ is given by
$Q_{\alpha\beta}=\bra\Phi|R_\alpha R_\beta|\Phi\ket$.

It is remarked that $\rho(\vect{r})$ of Eq.~(\ref{eq:dens-MF_corr2}) satisfies
\be\begin{split}
 \int d^3r\,\rho(\vect{r}) &= A\,, \\
 \int d^3r\,r^2\,\rho(\vect{r}) &= A\,\bra r^2\ket\,,
\end{split}\label{eq:integ_rho-MF_corr}
\ee
as verified by using
$\int d^3r\,e^{-i\vect{q}\cdot\vect{r}}=(2\pi)^3\,\delta(\vect{q})$
and $\int d^3r\,r^2\,e^{-i\vect{q}\cdot\vect{r}}
=-(2\pi)^3\,\nabla_{\vect{q}}^2\,\delta(\vect{q})$.
Equation~(\ref{eq:integ_rho-MF_corr}) exactly holds
partly because the approximation (\ref{eq:wf-MF}) does not influence
the matter density.
While the approximation of Eq.~(\ref{eq:wf-MF}) derives
$\bra\Phi|(\vect{r}_i-\vect{R})\cdot\vect{R}|\Phi\ket
\approx\bra\Phi|(\vect{r}_i-\vect{R})|\Phi\ket\cdot
\bra\Phi|\vect{R}|\Phi\ket$,
for the matter density this term emerges only in the form
$\bra\Phi|\sum_{i=1}^A(\vect{r}_i-\vect{R})\cdot\vect{R}|\Phi\ket$
which vanishes because of $\sum_{i=1}^A(\vect{r}_i-\vect{R})=0$.
This consistency of Eq.~(\ref{eq:integ_rho_p}) does not hold
exactly for $\rho_p(\vect{r})$ and $\bra r^2\ket_p$:
Eq.~(\ref{eq:dens_p-MF_corr2}) derives
\be\begin{split}
 \int d^3r\,\rho_p(\vect{r}) &= Z\,, \\
 \int d^3r\,r^2\,\rho_p(\vect{r})
 &= \Big\bra\Phi\Big| \sum_{i\in p} \vect{r}_i^2\Big|\Phi\Big\ket
 - Z\,\bra\Phi|\vect{R}^2|\Phi\ket \\
 &= Z\bigg[\bra r^2\ket_p + \frac{2N}{A}\bra\Phi|(\vect{R}_p-\vect{R}_n)
   \cdot\vect{R}|\Phi\ket\bigg]\,.
\end{split}\label{eq:integ_rho_p-MF_corr}
\ee
However, the consistency remains to good precision,
as far as $\bra\Phi|(\vect{R}_p-\vect{R}_n)\cdot\vect{R}|\Phi\ket$ is small.

A part of the integration with respect to $\vect{q}$
in Eq.~(\ref{eq:dens-MF_corr2})
can be carried out analytically
when a certain symmetry is maintained in $|\Phi\ket$.
Under the spherical symmetry,
we have $Q_{\alpha\beta}=\frac{1}{3}\delta_{\alpha\beta}
\bra\Phi|\vect{R}^2|\Phi\ket$
and thereby $\bra\Phi|(\vect{q}\cdot\vect{R})^2|\Phi\ket
=\frac{1}{3}q^2\bra\Phi|\vect{R}^2|\Phi\ket$.
The s.p. w.f. $\varphi_k(\vect{r})=R_{n\ell j}(r)\,
[Y^{(\ell)}(\hat{\vect{r}})\,\chi]^{(j)}_m$ yields
\be
 \sum_m \bra\varphi_k|e^{-i\vect{q}\cdot\vect{r}}|\varphi_k\ket
 =(2j+1)\int_0^\infty r^2dr\,j_0(qr)\,\big[R_{n\ell j}(r)\big]^2\,,
\ee
with the zeroth-order spherical Bessel function $j_0(x)$.
Equation~(\ref{eq:dens-MF_corr2}) then becomes
\be \rho(\vect{r})
 \approx \frac{1}{2\pi^2}\int_0^\infty q^2dq\, j_0(qr)\,
 e^{\frac{1}{6}q^2\bra\Phi|\vect{R}^2|\Phi\ket}\,
 \Big\bra\Phi\Big| \sum_{i=1}^A e^{i\vect{q}\cdot\vect{r}_i}
 \Big|\Phi\Big\ket\,. \label{eq:dens-MF_corr3}
\ee
When the axial symmetry around the $z$-axis holds,
$\bra\Phi|(\vect{q}\cdot\vect{R})^2|\Phi\ket
=\frac{1}{3}q^2\big[\bra\Phi|\vect{R}^2|\Phi\ket
+\frac{8\pi}{5}Y^{(2)}_0(\hat{\vect{q}})
\bra\Phi|R^2 Y^{(2)}_0(\hat{\vect{R}})|\Phi\ket\big]$.

One may wonder whether the $q$-integration of Eq.~(\ref{eq:dens-MF_corr2})
is convergent.
Though not fully guaranteed,
we can reasonably expect that it is convergent,
since the momentum distribution of bound nucleons
is limited up to $k_\mathrm{F}\sim 1.34\,\mathrm{fm}$ to good approximation.
Within the GEM with the parameter-set of Eq.~(\ref{eq:basis-param}),
the $q$-integration converges as long as the following condition
is satisfied:
\be \bra\Phi|\vect{R}^2|\Phi\ket
< \frac{3}{4}\,\min\bigg[\mathrm{Re}\Big(\frac{1}{\nu}\Big)\bigg]
= \frac{3}{4\{1+(\pi/2)^2\}}\,\frac{1}{\nu_0}
\approx (1.12\,\mathrm{fm})^2. \label{eq:R-criterion}\ee

The proton and neutron densities $\rho_\tau(\vect{r})$
can be corrected in an analogous manner, as mentioned already.
For the charge form factor $F_c(\vect{q})$,
additional corrections are made
according to Eq.~(20) of Ref.~\refcite{Friar-Negele_1975},
in which the nucleon finite-size effects including the magnetic effects
are incorporated.
The charge density is then calculated by its inverse Fourier transform,
\be \rho_c(\vect{r})
 = \frac{1}{(2\pi)^3}\int d^3q\, e^{-i\vect{q}\cdot\vect{r}}\,F_c(\vect{q})\,,
 \label{eq:dens&ff_c}
\ee

\subsection{Measure of precision of 2nd-order cumulant expansion
  \label{subsec:measure}}

The precision of the above approximations can be examined
in the following manner.
Let us consider the harmonic oscillator (HO) form
for the c.m. Hamiltonian,
\be
 H_\mathrm{c.m.}^{(\mathrm{HO})}=\frac{\vect{P}^2}{2AM}+\frac{AM}{2}
  \sum_{\alpha,\beta=x,y,z} \Xi_{\alpha\beta}R_\alpha R_\beta\,,
 \label{eq:Hcm}
\ee
instead of $H_\mathrm{c.m.}=\vect{P}^2/2AM$.
The symmetric matrix $\mathsf{\Xi}=(\Xi_{\alpha\beta})$ is positive-definite,
whose elements are determined later.
It is apparent that, if we take the $x,y,z$ axes
with which $\mathsf{\Xi}$ is diagonalized,
$(\Xi_{xx}^{1/2},\Xi_{yy}^{1/2},\Xi_{zz}^{1/2})$
corresponds to the frequency parameters.
As an effect of the additional potential,
the exact c.m. w.f. $|\Psi_\mathrm{c.m.}\ket$ of Eq.~(\ref{eq:wf-full})
becomes constrained to a finite space,
not affecting $|\Psi_\mathrm{rel.}\ket$.
At its lowest state $|\Psi_\mathrm{c.m.}\ket$ has a simple Gaussian form as
\be |\Psi_\mathrm{c.m.}^{(\mathrm{HO})}\ket \propto
\exp\Big[-\frac{AM}{2}\sum_{\alpha,\beta}(\mathsf{\Xi}^{1/2})_{\alpha\beta}
  R_\alpha R_\beta\Big]\,.
\ee
It is remarked that
the 2nd-order cumulant expansion of (\ref{eq:cumulant1}) is exact
if the c.m. w.f. has the Gaussian form.

The c.m. part of the MF w.f. $|\Phi\ket$,
$|\Phi_\mathrm{c.m.}\ket$ if $|\Phi\ket$ is separable
as in Eq.~(\ref{eq:wf-MF}),
can be expanded by the HO eigenfunctions.
Since the energy is minimized in the SCMF calculations,
it is expected that the c.m. part of $|\Phi\ket$
tends to be dominated by the lowest state
as $|\Phi_\mathrm{c.m.}\ket\approx|\Psi_\mathrm{c.m.}^{(\mathrm{HO})}\ket$
when the parameters $\{\Xi_{\alpha\beta}\}$ are optimized,
although $H_\mathrm{c.m.}^{(\mathrm{HO})}$ is not used in the SCMF calculations.
The parameters $\{\Xi_{\alpha\beta}\}$ determine
$\bra\Phi|R_\alpha R_\beta|\Phi\ket$ in the lowest eigenfunction
of $H_\mathrm{c.m.}^{(\mathrm{HO})}$.
Conversely, $\Xi_{\alpha\beta}$ can be fixed
from $Q_{\alpha\beta}=\bra\Phi|R_\alpha R_\beta|\Phi\ket$ by
\be
 (\mathsf{\Xi}^{1/2})_{\alpha\beta} = \frac{1}{2AM}
  (\mathsf{Q}^{-1})_{\alpha\beta}\,, \label{eq:Xi}
\ee
in accordance with the approximation (\ref{eq:cumulant1}).
This is reduced to $(\mathsf{\Xi}^{1/2})_{\alpha\beta}
=(3\delta_{\alpha\beta}/2AM)\bra\Phi|\vect{R}^2|\Phi\ket^{-1}$
when $|\Phi\ket$ has the spherical symmetry,
and to $\Xi^{1/2}_{xx}=\Xi^{1/2}_{yy}=(1/AM)\bra\Phi|R^2-Z^2|\Phi\ket^{-1}$,
$\Xi^{1/2}_{zz}=(1/2AM)\bra\Phi|Z^2|\Phi\ket^{-1}
=(3/2AM)\bra\Phi|R^2[1+2\sqrt{(4\pi)/5}\,Y^{(2)}_0(\hat{\vect{R}})]
|\Phi\ket^{-1}$
when $|\Phi\ket$ has the axial symmetry.

With $\Xi_{\alpha\beta}$ of Eq.~(\ref{eq:Xi}),
the value of $\bra\Phi|H_\mathrm{c.m.}^{(\mathrm{HO})}|\Phi\ket$
gives a measure of mixing of excited components
of $H_\mathrm{c.m.}^{(\mathrm{HO})}$.
If $\bra\Phi|H_\mathrm{c.m.}^{(\mathrm{HO})}|\Phi\ket$ is close
to the eigenvalue for the lowest state
$\frac{1}{2}\mathrm{Tr}(\mathsf{\Xi}^{1/2})$,
it proves that both Eqs.~(\ref{eq:wf-MF}) and (\ref{eq:cumulant1})
are fulfilled to good precision.
Note that, owing to Eq.~(\ref{eq:Xi}),
the expectation value of the potential term
of $H_\mathrm{c.m.}^{(\mathrm{HO})}$
is always $\frac{1}{4}\mathrm{Tr}(\mathsf{\Xi}^{1/2})$,
while $\bra\Phi|\vect{P}^2/(2AM)|\Phi\ket
\geq \frac{1}{4}\mathrm{Tr}(\mathsf{\Xi}^{1/2})$.
As examples, the ratio of $\bra\Phi|H_\mathrm{c.m.}^{(\mathrm{HO})}|\Phi\ket$
to $\frac{1}{2}\mathrm{Tr}(\mathsf{\Xi}^{1/2})$
obtained in the spherical HF calculations with M3Y-P6
are tabulated in Table~\ref{tab:cm-corr} for several nuclei.
The $\sqrt{\bra\Phi|\vect{R}^2|\Phi\ket}$ values are also shown
to confirm that the condition (\ref{eq:R-criterion}) is well satisfied.

\begin{table}[pt]
 \caption{Ratio $\bra\Phi|H_\mathrm{c.m.}^{(\mathrm{HO})}|\Phi\ket\big/
   \frac{1}{2}\mathrm{Tr}(\mathsf{\Xi}^{1/2})$,
   $\sqrt{\bra\Phi|\vect{R}^2|\Phi\ket}$,
   $\sqrt{\bra r^2\ket_p}$ from (\protect\ref{eq:r2_p-MF_corr})
   and (\protect\ref{eq:integ_rho_p-MF_corr})
obtained in the spherical HF calculations with M3Y-P6.
 \label{tab:cm-corr}}
\centerline
{\begin{tabular}{crrrr}
\toprule 
Nuclide & Ratio & $\sqrt{\bra\vect{R}^2\ket}~(\mathrm{fm})$ &
\multicolumn{2}{c}{$\sqrt{\bra r^2\ket_p}~(\mathrm{fm})$}\\
&&& (\protect\ref{eq:r2_p-MF_corr}) & (\protect\ref{eq:integ_rho_p-MF_corr}) \\
\colrule 
$^{16}$O & $1.008$ & $0.535$~~~& $2.599$ & $2.600$ \\
$^{24}$O & $1.051$ & $0.493$~~~& $2.696$ & $2.681$ \\
$^{40}$Ca & $1.009$ & $0.375$~~~& $3.389$ & $3.389$ \\
$^{48}$Ca & $1.009$ & $0.341$~~~& $3.415$ & $3.415$ \\
$^{56}$Ni & $1.009$ & $0.318$~~~& $3.674$ & $3.675$ \\
$^{78}$Ni & $1.014$ & $0.287$~~~& $3.906$ & $3.906$ \\
$^{90}$Zr & $1.012$ & $0.269$~~~& $4.193$ & $4.194$ \\
$^{100}$Sn & $1.013$ & $0.256$~~~& $4.397$ & $4.440$ \\
$^{132}$Sn & $1.017$ & $0.235$~~~& $4.649$ & $4.649$ \\
$^{208}$Pb & $1.022$ & $0.199$~~~& $5.429$ & $5.429$ \\
\botrule 
\end{tabular}}
\end{table}

In Table~\ref{tab:cm-corr},
the point-proton r.m.s. radii $\sqrt{\bra r^2\ket_p}$ are also presented.
By comparing the values from Eqs.~(\ref{eq:r2_p-MF_corr})
and (\ref{eq:integ_rho_p-MF_corr}),
it is confirmed that the consistency of Eq.~(\ref{eq:integ_rho_p})
practically holds via the present approximation.

Although one may consider an extension of the cumulant expansion
of Eq.~(\ref{eq:cumulant1}) to a higher-order,
the approximation of Eq.~(\ref{eq:wf-MF}) should also be taken care of.
Without an appropriate prescription improving the latter approximation,
the higher-order cumulant expansion seems to have no advantage.

\bibliography{references}
\bibliographystyle{ws-ijmpe}

\end{document}